\documentclass{emulateapj}

\newcommand{\ha}{H$\alpha$}

\newcommand{\hanii}{H$\alpha+$[N{\sc ii}]}
\newcommand{\nii}{[N{\sc ii}]}
\newcommand{\hi}{H{\sc i}}
\newcommand{\hii}{H\,{\sc ii}}
\newcommand{\msun}{M$\odot$}
\newcommand{\msunyr}{M$\odot$\,yr$^{-1}$}
\newcommand{\ergscm}{erg\,s$^{-1}$\,cm$^{-2}$}
\newcommand{\ergscmas}{erg\,s$^{-1}$\,cm$^{-2}$\,arcsec$^{-2}$}
\newcommand{\magarc}{mag\,arcsec$^{-2}$}
\newcommand{\tn}{\tablenotemark}

\shorttitle{Stellar Disks of Ring galaxies}
\shortauthors{Romano et al.}

\begin{document}

\title{Stellar disks of Collisional Ring Galaxies I. New multiband images, 
Radial intensity and color profiles, and confrontation with N-body simulations}
\author{R. Romano\altaffilmark{1}, Y.D. Mayya\altaffilmark{1} and 
E. I.  Vorobyov\altaffilmark{2}}
\altaffiltext{1}{Instituto Nacional de Astrof\'{\i}sica Optica y Electronica,
       Luis Enrique Erro No. 1, Tonantzintla, Apdo Postal 51 y 216, 72840,
       Puebla, Mexico: rromano@inaoep.mx, ydm@inaoep.mx}
\altaffiltext{2}{
The Institute for Computational Astrophysics, Saint Mary's University, Halifax NS,
B3H 3C3, Canada; and Institute of Physics, South Federal University, 
Stachki 194, Rostov-on-Don, Russia.
}
\accepted{in Astronomical Journal}

\begin{abstract}
We present new multi-band imaging data in the optical ($BVRI$ and \ha)
and near infrared bands ($JHK$) of 15 candidate
ring galaxies from the sample of \citet{app87}. We use these data to obtain
color composite images, global magnitudes and colors of both the ring galaxy
and its companion(s), and radial profiles of intensity and colors.
We find that only nine of the observed galaxies have multi-band
morphologies expected for the classical collisional scenario of ring
formation, indicating the high degree of contamination of the ring galaxy
sample by galaxies without a clear ring morphology.
The radial intensity profiles, obtained by masking the off-centered nucleus,
peak at the position of the ring, with the profiles in the continuum bands
broader than that in the \ha\ line. The images as well as the radial intensity
and color profiles clearly demonstrate the existence of the pre-collisional
stellar disk outside the star-forming ring, which is in general bluer than
the disk internal to the ring.
The stellar disk seems to have retained its size, with the disk outside the
ring having a shorter exponential scale length as compared to the values
expected in normal spiral galaxies of comparable masses.
The rings in our sample of galaxies are found to be
located preferentially at around half-way through the stellar disk.
The most likely reason for this preference is bias against detecting
rings when they are close to the center (they would be confused
with the resonant rings), and at the edge of the disk the gas surface
density may be below the critical density required for star formation.
Most of the observed characteristics point to relatively recent collisions
($<80$~Myr ago) according to the N-body simulations of \citet{ger96}.
\end{abstract}

\keywords{galaxies: photometry --- galaxies: interactions}

\section{Introduction}

Ring galaxies are a class of objects whose optical appearance is dominated
by a ring or a ring-like structure. In one of the earliest discussions
of ring galaxies, \citet{bur59} suggested that these objects
may be the aftermath of a close collision between an elliptical and spiral
galaxy. On the other hand, \citet{fre74} suggested that
the ring is probably the result of a collision between a spiral galaxy and
an intergalactic cloud of \hi. \citet{the76,the77} systematically
studied the basic observational properties of a sample of ring galaxies and
suggested that rings are formed when an intruding galaxy passes nearly
through the center of a normal disk galaxy. \citet{lyn76} used numerical 
simulations and settled the issue regarding their origin.
They demonstrated that ring galaxies 
are formed as a result of an on-axis collision between an intruder galaxy and 
a gas-rich disk-galaxy. In this scenario, the collision sets off an expanding 
ring wave, which in turn triggers star formation in its wake. Due to the 
expanding nature of the wave, the ring of star formation also 
advances successively to larger radii with time. This scenario has stood the 
test of the multi-band data those have become available since then
(Appleton \& Struck-Marcell 1996 and references therein).

Many of the predictions from theoretical models have been tested 
observationally in the Cartwheel, the prototype ring galaxy.
The ring in this galaxy is expanding
\citep{fos77}, and is forming massive stars \citep{hig95}. A radial color 
gradient was also noticed by \citet{mar92}, which was found to be in agreement 
with the sequential ordering of stellar ages \citep{kor01, vor01}. 
\citet{app97} established the color gradient in some other ring galaxies.  
\citet{mar95} carried out \ha\ imaging observations of eight ring galaxies
and found that majority of the star-forming regions are located exclusively
in the ring. 

In the collisional scenario proposed by \citet{lyn76}, a stellar density
wave is set off as the collective response of the stars that were 
present in the pre-collisional disk of the target galaxy. However, we know 
very little about the underlying stellar disk of the target galaxy. 
It is generally believed that the observed rings, especially in the near 
infrared continuum bands, trace the location of the stellar density waves. 
The stellar disk outside the ring should carry important information on the 
nature of the pre-collisional disk. However, such a disk has not been traced
even in the Cartwheel. \citet{app97} tried to infer an outer disk 
using radial profiles of surface brightness and color in four galaxies, and 
found such a disk in two cases (IIHz4 and VIIZw466). Their images were not 
deep enough to study the azimuthal structure of the outer disks.

\citet{kor01} have developed a method to estimate the contribution of the 
newly formed stars in different wavelength bands as the density wave expands 
in the target galaxy. The density wave in the underlying stellar disk
can be recovered by subtracting the contribution of the newly formed stars.
Availability of \ha\ and continuum images are fundamental in achieving
this objective. The continuum images should reach at least 2 magnitudes
deeper than the brightness of the ring in order to register
the stellar disk on either side of the star-forming ring.
With this goal in mind, we carried out new  
imaging observations of a sample of 15 candidate ring galaxies in 
the $BVRIJHK$ broad bands and in the emission line of \ha. 
Some of the candidate galaxies may not have formed by the scenario 
proposed by \citet{lyn76}, and these galaxies were included in our observing
list with the hope that the new observations would allow us to filter out 
contaminating galaxies. 
The data and detailed surface photometric analysis of all sample 
galaxies are presented in this paper. 
In a forthcoming paper (Paper~II), we will discuss the results obtained from
the extraction of the underlying stellar density wave 
in galaxies that are most likely formed by the classical collisional scenario.

Sample and the details of our observations are discussed in \S~2. 
Morphological descriptions and one dimensional profiles of individual galaxies 
are presented in \S~3. In \S~4, we compare the photometric properties of ring 
galaxies with that of normal galaxies. Results obtained from the new data
are discussed in the context of N-body simulations in \S~5. Conclusions
from our study are presented in \S~6. Gray scale maps in the \ha\ 
and $B$-bands and one dimensional surface 
brightness and color profiles for each galaxy are presented in the Appendix.  
Throughout this paper, all distance scaling assumes a value for the
Hubble Constant of 75 km\,s$^{-1}$\,Mpc$^{-1}$.

\section{Observations and Reductions}

Ring galaxies from the sample of \citet{app87} were selected 
for optical and NIR photometric study. A total of 15 galaxies 
that were north of declination $=-15^\circ$ were observed in 
the $BVRIJHK$ broad bands and in the emission line of \ha.
The sample galaxies are listed in Table 1. The equatorial coordinates,
radial velocities (V$_{\rm 0}$), ring diameters and total $B$-band magnitudes
in the table were taken from \citet{app87}.

Three of our program galaxies (IIZw28, IIHz4 and VIIZw466) were common with 
the $BVRJHK$ and \ha\ imaging study by \citet{app97} and \citet{mar95}, while 
three other galaxies (Arp143, NGC985 and NGC5410) were mapped in 
\ha\ by \citet{hig97}, \citet{rod90} and \citet{mar95}, respectively. 
For the remaining galaxies, the present study is the first attempt in obtaining 
uniform 
multi-band digital quality images. Our images reach surface brightness levels 
of $\mu_B\sim25$~\magarc\ (3--4$\sigma$) in the $B$-band and 
0.2--2$\times10^{-16}$~\ergscmas\ (5$\sigma$) in the \ha\ band.

\subsection{Optical imaging in the $BVRI$ and \ha\  bands}

All optical observations  were carried  out with  the {\it Observatorio 
Astrof{\'{\i}}sico Guillermo Haro} 2.1-m  telescope at  Cananea, Mexico. 
We  used a  Tektronics CCD  of $1024\times 1024$ pixel  format at the $f/12$ 
Cassegrain  focus of the telescope with $3\times 3$ pixel binning, resulting 
in an image of $0\farcs6$ pixel$^{-1}$  and a field  of view of 
$3\farcm4\times3\farcm4$. 

Table 2 contains a detailed log of the observations. The observing runs for
each galaxy are given in column 2, followed by the exposure times in the 
broadband filters. These broadband filters correspond to the standard 
Johnson-Cousins $BVRI$ system. The central wavelength of the \ha\ filter for
each galaxy is given in column~7, followed by a column containing the 
exposure times in these filters.
The \ha\ filters were of square-shape with typically a width of 100\AA,
and include both the \nii\ lines flanking the \ha. For five galaxies,
observations were carried out in emission-line free narrow bands to 
facilitate subtraction of the in-band continuum from the \ha-filter images.
The central wavelength of the off-band continuum filters used for these galaxies
are given in column~9. For the rest of the galaxies, the $R$-band images 
were used for the purpose of continuum subtraction. 
Twilight sky  exposures were taken  for flat-fielding purposes.  
Several bias  frames were obtained at the  start and end of each night.   

\subsection{Near infrared imaging}

All near infrared observations were carried out in the $J,H$ and $K$ bands 
with the {\it  Observatorio Astronomico Nacional} 2.1-m telescope
at San Pedro Martir,  Mexico.  The CAMILA instrument \citep{cru94},
that hosts  a NICMOS 3  detector of  256$\times$256 pixel
format,  was  used  in  the   imaging  mode  with  the  focal  reducer
configuration  $f/4.5$. This  results  in a
spatial sampling of $0\farcs85$ pixel$^{-1}$  and a total  field of
view of $3\farcm6\times3\farcm6$.   Each      observation consisted of
a sequence of  object and sky exposures, with  the integration time of
an individual exposure limited by the sky counts,  which was  kept well  
below  the non-linear  regime of  the detector.  A  typical image  
sequence consisted of  10 exposures, six on  the object and  four on the  sky. 
NIR observations could not be carried out for three of the sample galaxies.
Though of lesser sensitivity, we used the 2MASS\footnote{This publication 
makes use
of data products from the Two Micron All Sky Survey, which is a joint project
of the University of Massachusetts and the Infrared Processing and Analysis
Center/California Institute of Technology, funded by the National Aeronautics
and Space Administration and the National Science Foundation.} 
images of these three galaxies for the sake of completeness. 
A  series of twilight and  night-sky images were  taken for  flat-fielding  
purposes. 
Observing runs and exposure times in the $JHK$ bands are given in the last
4 columns of Table~2.

The sky  conditions  for  both  optical  and  NIR  observations  were
generally  photometric and  the seeing  \emph{FWHM} was  in  the range
1$\farcs5$--2$\farcs5$ on different nights. Typical sky brightness
was 21.14, 20.72,  20.18 and 12.27 magnitude arcsec$^{-2}$  in the $B,
V, R$  and $K$  bands, respectively. The  sky brightness in  the $K$
band also includes the background emitted by the warm optics.

\subsection{Image data reduction and Calibration}

The technique used to carry out the image data reduction and calibration
is the same as that followed for the analysis of a sample of lenticular 
galaxies observed during the same runs as the present observations 
\citep{bar05}. The basic data reduction for both the optical and NIR 
images involved subtraction of the bias and sky frames, division by flat 
field frames, registration of  the images  to a common  coordinate system  
and then stacking all  the images of a  given galaxy in each  filter.  
Night-to-night variations of the optical bias frames were negligible, and hence
bias frames  of an entire run  were stacked together  using the median
algorithm to form a master  bias frame, which was then subtracted from
all object frames.  Preparation of the optical flat fields followed
the conventional technique, wherein bias subtracted flats were stacked
and the  resultant frame was  normalized to the mean value in a pre-selected
box near the center of the frame, to  form a
master flat  in each  filter.  Bias subtracted  images of  the program
galaxies   were  divided  by   the  normalized   flat  field   in  the
corresponding filter.  The optical  images suffered from a stray light
problem  that resulted  in a  gradient  in the  sky background,  which
roughly ran through one of the diagonals of the CCD chip. The gradient
was found to be stable throughout  each run and the mean counts scaled
linearly with exposure time.  After several experiments, we found that
the best way to  get rid of the gradient was to  subtract a mean blank
sky  image from  the data  images.   For this  purpose, special  blank
fields were observed  in each filter with exposure  times matching the
typical exposure times of the object frames. 
The adopted procedure eliminated any systematic gradient in the sky 
background, but still resulted in non-negligible residual sky values,
which were taken into account in the error estimation (see \S~2.5).

For the $JHK$  images, a bias frame taken immediately before an object
exposure was subtracted as part of the data   acquisition. A master NIR
flat field in each filter for each night of  observing was prepared  as 
follows. The
night-sky flats  were first stacked  and then subtracted  from stacked
twilight flats.  The frames obtained in this fashion for each run were
then combined and normalized to  the mean value of the resultant frame
to  form a  master   flat.   The sky  frames of  each sequence  of
observations were combined and the resultant image was subtracted from
each of the object frames to get a sky-subtracted image. Flat fielding
was done  by dividing the sky  subtracted images of the  object by the
normalized master flat.  The resulting images were aligned to a common
co-ordinate system using common stars  in the frames and then combined
using  the median  operation.  Only  good  images (as  defined in  the
CAMILA manual  --- see \citet{cru94}) were used  in the combination.
The resulting  combined images were aligned to
corresponding images from the Digitized  Sky Survey (DSS).  As a final
step of  the reduction procedure, the mutually  aligned optical images
were  aligned to the  NIR image  coordinate system.   The transformed
star positions  in the  images agreed to  within $0\farcs2$  as judged
from the coordinates of common stars.

All image  reductions were carried  out using the Image  Reduction and
Analysis  Facility  (IRAF\footnote {IRAF  is  distributed by  National
Optical Astronomy Observatories, which are operated by the Association
of  Universities for  Research in  Astronomy, Inc.,  under cooperative
agreement  with  the National  Science  Foundation.})   and the  Space
Telescope  Science Data  Analysis System  (STSDAS). 

\subsection{Photometric calibration}

Dipper  Asterism stars  in the  M67  field were  observed to  enable
accurate  photometric calibration  of our  optical  observations.  The
stars in this field span a wide  color range ($-0.05 < B - V < 1.35$),
that includes the range of  colors of the program galaxies, and hence
are  suitable for  obtaining  the transformation  coefficients to  the
Cousins $BVRI$  system defined by \citet{bes90}.  The transformation
equations used are:
\begin{eqnarray}
B & = b_0 + \alpha_B + \beta_B (b_0-v_0),  \\
V & = v_0 + \alpha_V + \beta_V (b_0-v_0), \\
R & = r_0 + \alpha_R + \beta_R (v_0-r_0),  \\
I & = i_0 + \alpha_I + \beta_I (v_0-i_0), \\
\log F({\rm H}\alpha) & = \log (C_{\rm H}) + \log (\alpha_{\rm H}) - 0.4\beta_{\rm H} (v_0-r_0),
\end{eqnarray}
where $B, V, R$ and $I$ are standard magnitudes, $b_0, v_0, r_0$ and $i_0$ are
the extinction corrected instrumental magnitudes, $\alpha_B, \alpha_V, 
\alpha_R$ and $\alpha_I$ are  the zero points  and $\beta_B, \beta_V, \beta_R$  
and $\beta_I$ the color coefficients  in bands $B, V, R$ and  $I$ respectively. 
Typical extinction coefficients  for the observatory (0.20, 0.11, 0.07 and 0.03 
for $B, V, R$  and $I$  bands respectively) were  used. Considering  that the
objects and the standard stars  were observed as close to the meridian
as possible,  and in none of  the cases the air mass  exceeded 1.3, the
error  introduced   due  to  possible  variation   in  the  extinction
coefficients is  less than 0.02 magnitude.   The coefficients $\alpha$ and
$\beta$  were  obtained by  using  the  $BVRI$  standard magnitudes  of
\citet{che91}.
Calibration coefficients for \ha\ filters were obtained using the 
monochromatic magnitudes at 6660~\AA, obtained using intermediate-band
($\Delta\lambda=480$~\AA) filters, of five M67 standard stars
given by \citet{fan96} for the 2000-February and 2002-February runs. 
For the 2002-October run, spectroscopic standard star BD28+4211 was
observed in all the \ha\ filters. The spectrum of this star from \citet{oke90}
was integrated to obtain the flux inside each of the \ha\ filters. In the 
2001-December run, the photometric standard star PG0231+051 was observed. 
The $R$-band magnitude of the stars in this field were used to calibrate the 
\ha\ filter, following the method described by \citet{may91}. The coefficient 
$\alpha_{\rm H}$ is the conversion factor between the observed count rate 
($C_{\rm H}$) and the flux ($F({\rm H}\alpha)$ in erg\,s$^{-1}$\,cm$^{-2}$).
Typically, the $\alpha_{\rm H}$ values obtained from different observations 
agree within 1--3\%. The color coefficient for this conversion is found to be 
negligible (i.e. $\beta_{H}$=0.0) for all the \ha\ filters. 

The values  of $\alpha$ and $\beta$ for different observing runs are given in 
Table~3. The color coefficients are reasonably small for all filters except
for the $R$-band. This is understandable given that our $R$-band is centered
around 500~\AA\ to the blue and lacks the red tail as compared to that
recommended by \citet{bes90}.
We found negligible variation of the color coefficient during the two-year 
span when most of these observations were carried out, and hence used the same 
$\beta$ values for all the runs.
The stability of $\alpha$ on different nights was checked using at least 
two standard fields from  the Landolt Selected  Areas \citep{lan92}.
The standard fields observed during our runs are PG0231+051, SA110$-$232,
PG2336+004, Rubin\,149 and PG1323$-$086.   
The rms errors on the transformation coefficients represent the variation of 
zero points obtained using all the stars observed in a run that typically 
consisted of five consecutive nights.

The detector  and filter system combination  that we used  for the NIR
observations  is  identical  to  that  used  in  the  observations  of
standards by \citet{hun98}, and hence the color coefficients are
expected to be negligibly small.  We verified this by observing fields
AS17 and AS36, which contain stars spanning a wide range of colors. We
observed at least 2 standard  fields each night, each field containing
more  than  one  star  and  some  fields such  as  AS17  containing  5
stars. NIR zero points are found to be stable not only for a run, but 
also over all the runs required to complete the observations. The resulting 
zero points are: $20.65\pm0.05, 20.40\pm0.05$ and $20.15\pm0.05$ in $J, H$ 
and $K$ bands, respectively.

\subsection{Photometric errors}

We aimed at carrying out surface photometry of galaxies up to a radius where
the $B$-band surface brightness reaches a level of $\mu_B=25$~\magarc.
Unlike normal spiral galaxies, the ring galaxies have a very limited 
dynamic range 
(typically 2--3 magnitudes), with most of the total flux originating in 
low surface brightness regions. Uncertainties in the estimation of sky value 
is the principal source of error at these surface brightness levels both for 
the optical and NIR observations. We measured the typical mean sky value and 
its rms errors in all the filters for a galaxy.
The error $\delta m$ in magnitude for an observed sky-subtracted count rate, 
$I_{\rm g}$, was calculated using the relation:
\begin{eqnarray}
\delta m =& 1.0857{{\delta I_{\rm g}}\over{I_{\rm g}}}, \\
{\rm where} \hspace{2mm} 
\delta I_{\rm g} =& \sqrt{N_{\rm pix}^2\sigma_{\rm res}^2 + 
                 N_{\rm pix}\sigma_{\rm rms}^2 + I_{\rm g}}.
\end{eqnarray}
The three terms in the expression for ${\delta I_{\rm g}}$ correspond to
the error due to the uncertainty in the residual sky value, rms error on the 
sky value, and the error due to the photon noise of the source, respectively.
Boxes of $21\times21$ pixels a side located at five different 
object-free regions around the galaxy of interest were used to measure 
the $\sigma$ values. The $\sigma_{\rm res}$ is obtained as the rms of the 
mean residual values between the boxes, and the $\sigma_{\rm rms}$ is 
the mean of the pixel-to-pixel rms value within these boxes.
For NIR images $\sigma_{\rm res} << 
\sigma_{\rm rms}$, whereas for optical images $\sigma_{\rm res}$
and $\sigma_{\rm rms}$ were comparable. This is due to the error involved 
in subtracting a gradient in the background that was present in all our 
optical images (see \S~2.3). On the other hand, subtraction of a separate 
sky image in the NIR results in practically zero residuals every where in 
the image. The quantity $N_{\rm pix}$ is the number of pixels in the 
aperture (or in the azimuthal bin) used in obtaining $I_{\rm g}$.

The errors on colors involving the optical and NIR bands (e.g. $V-K$) 
were obtained by adding quadratically the errors on the V and $K$ magnitudes.
The rms errors on the sky values in different optical bands are partially 
correlated, and the same is true for the three NIR bands.
Hence errors on the colors involving only optical or NIR colors are, in 
general, smaller than that on the surface brightness values. 

\subsection{Comparison of our photometry with those in the literature}

We searched the literature for previous photometric study of our sample 
ring galaxies or their companions. We carried out photometry of the
common galaxies, using synthetic apertures that were simulated to be 
as close to the
literature apertures as possible. Our photometry is compared with those 
in the literature in Figure~1, along with the estimated errors in the
photometry. The error bars in the horizontal direction denote the errors
in our photometry, whereas the vertical error bars are obtained by
quadratically adding the errors of our and literature photometry.
The literature photometry is taken from: 
Third Reference Catalogue (\citet{deV91}(RC3 henceforth)) and 
\citet{but96} ($B$-band 
photoelectric aperture photometry for seven and $B-V$ colors for six sample 
galaxies; filled circles with error bars), \citet{app97} ($BVRIJHK$ 
photometry for four sample galaxies; error bars without any central symbol),
\citet{maz95} and \citet{bon87} ($BVR$ CCD photometry for two and one galaxies,
respectively; open squares with error bars), and 2MASS ($JHK$ photometry for 
five galaxies; filled circles with error bars). The sources of \ha\  photometry
are \citet{mar95} (5 galaxies) and \citet{hat04} (Arp148).

From the comparison, it can be seen that there is no systematic 
offset between our photometry and those in the literature for $B$, $B-V$,
$J-K$ and \ha\ photometry. The agreement between ours and literature photometry 
in these bands is, in general, within the quoted errors.
In the case of the $K$-band, our photometry agrees well with that of the
2MASS, whereas \citet{app97} measurements are systematically fainter by
0.1--0.5~mag. Similarly, for those galaxies for which both the optical and
NIR photometry are available, we find that literature $B-K$ colors (obtained by
combining $B$ and $K$ measurements of different authors) are systematically 
redder than our colors by 0.0--0.5~mag.
Agreement between the \ha\ fluxes for three galaxies is within 10\% of
each other. The galaxies with large departures are VIIZw466 and Arp148.
The error for Arp148 is most likely because of unspecified correction 
for [NII] fluxes in the literature. 

\section{Multi-band Morphology}

\subsection{The color Atlas}

We obtained the color images of ring galaxies by digitally combining images in 
the $BV$ and $R$ filters. The IRAF external package {\it color} was used for 
this purpose, where the $RGB$ colors were represented by $R$, $V$, and $B$ band 
filter images, respectively.
These color  composite images are shown in Figure~2, where North is up and East 
is to the left. Galaxies from Arp's (1966) catalog are displayed
first, followed by IIHz4 and Zwicky galaxies, and finally galaxies with an 
NGC number. This ordering sequence is maintained in all tables from Table~4 
onwards.
The following characteristics of ring galaxy systems, some of 
which were well-known, can be noticed in these images: \\
(i) The ring is delineated by bright, blue knots.\\
(ii) The companion galaxy (when seen) is redder than the ring galaxy.\\
(iii) The nucleus of the ring galaxy is off-centered with respect
to the ring, and can be easily distinguished because of its relatively
yellowish or reddish color, even when it is seen superposed on the ring.
The direction of displacement of the nucleus is usually towards
the companion.  \\
(iv) The stars in the external parts of the companion galaxy are stripped
in at least four cases (Arp141, Arp143, Arp144, NGC5410).\\

In the following paragraphs, and in the Appendix, we give a detailed
description of the observed characteristics for each sample galaxy.

\subsection{Two-dimensional distribution of ionized gas and stars}

The continuum subtracted \ha\ emission-line, and the $B$-band images of 
all the program galaxies are displayed in grey-scale in the Appendix, one figure for 
each galaxy. The \ha\ emission originates from the ionized gas associated with 
the massive star-forming regions, whereas the $B$-band images trace the stellar 
light of both young and old stars. In the figures, the \ha\ contours are 
superposed on the $B$-band image, which allows us to directly compare the 
location of active star-forming regions with respect to the distribution of 
the disk stars. Both the \ha\ and the $B$-band images are displayed in 
logarithmic scale in order to illustrate the bright knots and faint parts of 
the stellar disk in the 
same image. Two ellipses are drawn for each galaxy, the inner ellipse 
corresponding to the ring and the outer ellipse corresponding to the intensity 
level of $\mu_B=25$~\magarc. Every ring galaxy of our sample contains a 
resolved stellar source, usually the brightest in the $K$-band, which we 
identify as the nucleus. The position of the nucleus is indicated by a small 
circle on the \ha\ image. The diameter of the circle corresponds to $2\farcs5$.  
The field of view of the displayed images includes the candidate companion 
galaxies as well. These are identified by the letters G1, G2 etc.
Description of the individual galaxies is given as notes in the
Appendix. We describe the general characteristics that can be seen in 
these images below: \\
(i) Majority of the \ha\ emission originates in compact knots, whose 
distribution coincides with the continuum ring that had formed the basis for 
them to be classified as ring galaxies. In three galaxies (Arp145, 
Arp291 and NGC985), a ring cannot be traced in \ha\ in spite of a 
ring-like structure 
in the continuum. In these galaxies, the \ha\ emission is detected only 
around the nucleus that is seen superposed on the continuum ring. In two 
other galaxies (Arp142 and Arp144), a complete ring cannot be traced neither 
in the continuum nor in the \ha\ line. The ring-like structure of NGC5410 
resembles a two-armed spirals in a barred galaxy rather than a ring.
\\
(ii) An \ha\ knot usually has an associated continuum knot, and vice versa. 
However, there are noticeable differences ($\sim1$~kpc) in the positions 
of the knots in the two bands. In most cases, the positional shifts are in 
the azimuthal directions. In a few cases, where the shifts are in the radial 
direction, continuum knots lie on the inner side of the \ha\ knot.
\\
(iii) The width of the ring, measured as the Full Width at Half Maximum
of the radial \ha\ intensity profile, is larger than the size of the most 
intense \ha\ knot, and is due to the radial spread of  \ha\ knots 
around the ellipse used in the construction of the radial profile.
\\
(iv) Among the emission features not associated with the ring,
the most common is the nuclear emission. Emission knots and diffuse emission
internal to the ring are seen in Arp141 and NGC2793.
In two of the classical ring galaxies of the sample (Arp147 and
IIHz4), a couple of interesting structures in the ionized gas can be 
traced outside the ring.  \\
(v) All the sample galaxies have stellar disks extending outside the ring.
In five cases (Arp141, Arp143, Arp145, Arp147, and NGC2793), 
the ellipses that best fit the disk and the ring 
have different centers, orientations and ellipticities. \\
(vi) The stellar disk outside the ring shows considerable degree of asymmetry. 
\\
(vii) Ionized gas is detected in the companions of four of our candidate
ring galaxies (Arp142, Arp148, VIIZw466, and NGC5410).

\subsection{One-dimensional surface photometry}

One dimensional (1-d) radial intensity profiles of galaxies are often used to  
analyze various morphological components in galaxies.
The 1-d profiles are obtained by fitting ellipses to the isophotes, and then
averaging the intensities azimuthally along the fitted ellipse. 
In normal galaxies, the resulting intensities are smoothly decreasing with 
radius, with non-axisymmetric sub-structures such as spiral arms or bars, 
if any, appearing as low-amplitude perturbations. 
On the other hand, the ring is an axially symmetric structure, and its 
contribution to the radial intensity profile is significant. 
In fact, in majority of the ring galaxies, intensity increases away from
the center, and peaks at the position of the ring. 
In most ring galaxies, the brightest photometric component,
usually a bulge or a nucleus, is off-centered with respect to the ring.
These special characteristics of ring galaxies make the surface photometric
profiles to depend critically on the details of the method used to obtain them.

\citet{app97} discussed the problems of getting physically meaningful 
profiles in ring galaxies. In their work, they adopted a method of nested 
rings in which they varied the center of the photometric ellipse smoothly 
between the nucleus and the ring center. The ellipticity and position angle 
were also varied to make sure that the ellipses corresponding to successive 
isophotes do not cross each other. The profiles extracted using this
method critically depend 
on the position and the intensity of the off-centered nucleus, which 
complicates their interpretation. Moreover, technically it is impossible to 
obtain such profiles for systems where the off-centered nucleus is seen 
superposed on the ring. 

We aim to obtain a characteristic post-collisional intensity profile of
ring galaxies. Our interest in obtaining such a  profile is two-fold:
(i) to inter-compare the radial intensity profiles of ring 
galaxies, and (ii) to compare the observed profiles with                              
the predictions of an expanding wave model. Both these studies require
the construction of the intensity profiles, centered on the ring-center, and
without the contaminating effect of any surviving sub-structure (e.g. 
nucleus/bulge). 
Clearly, the method of nested rings adopted by \citet{app97} is not suitable 
for our purpose. Hence, we adopted a different method, the details of which 
are described below.
We started out by setting the values of the geometrical parameters of 
the ellipse (center, ellipticity ($\epsilon$) and position angle (PA)) that 
best reproduces the shape of the ring as traced in \ha. In three cases
where the \ha\ emission does not trace a complete ring (Arp145, Arp291 and 
NGC985), we obtained the ellipse parameters using the $B$-band ring. The 
ellipse parameters were chosen interactively in such a way that the majority 
of the ring knots lie on a single ellipse.
The ring galaxies are characterized by two fundamental structures --- a high
surface brightness ring and a low surface brightness stellar disk.
We found that, in the majority of the cases, the stellar disk at 
$\mu_B=25$\,\magarc\  can be reconciled with the same parameters of the
ellipse (fixed-center) that best fits the ring. However, in five galaxies the 
ellipse parameters for the two components were visibly distinct, as noted in 
point (v) of \S~3.2, and the Appendix figures for the corresponding galaxies.
Nevertheless, the differences are mainly noticeable outside the ring, 
whereas our primary interest is in the part interior to the ring. 
Hence, in all cases, we choose the geometrical parameters for the ring to 
obtain the intensity profiles. The Right Ascension and Declination of the 
ellipse center, the semi-major axis, ellipticity and the position 
angle of the major axis of the best-matched ellipse are tabulated in Table~4. 
The last two columns give the $5\sigma$ surface brightness in the \ha\ and
$B$-bands, respectively.

While extracting the intensity profiles, we masked the nucleus even when it 
is seen superposed on the ring. Any foreground star is also masked.
In three galaxies (Arp291, IZw45 and IIZw28),  the nucleus occupies more 
than a third of the ring perimeter, which made the profile shape depend 
heavily on the chosen mask. We hence show the profiles without any nuclear 
mask for these galaxies. Radial intensity profiles were obtained by azimuthally
averaging the intensities of the unmasked pixels in concentric elliptical
annuli around a fixed center. In Arp142 and Arp144, we couldn't define an 
ellipse in either \ha\ or the $B$-band image. The \ha\ image of Arp142 
gives an impression that the ring in this galaxy is twisted and is seen 
nearly edge-on. The plotted ellipses for these galaxies do not represent any 
isophote, instead they are meant only to obtain representative radial 
intensity profiles and total magnitudes in different bands. We include these 
galaxies only in the discussion of global properties.

Normalized intensity profiles in the continuum bands are compared with that
in the \ha\ line for 8 ring galaxies in Figure~3. Six galaxies 
where the ring is not well-defined are excluded from this figure.
In addition, we excluded IZw45 from this figure due to the 
complications in the extraction of a representative radial profile given its 
small angular size and the dominance of the off-centered nucleus. 
The radius is normalized to $R_{B25}$ (the radius where $\mu_{\rm B}$ = 
25~\magarc), whereas the intensities are normalized to their values 
at $R_{H\alpha}$, the radius where the \ha\ profile peaks.
The most striking feature on these plots is that the \ha\ profiles are
sharper than the continuum profiles on either side of the ring.
Two galaxies (Arp141 and NGC2793)
show secondary peaks inside the ring, which are due to the presence of a
few emission knots interior to the ring. 
Continuum profiles show a variety of forms on the inner side of the ring, 
with the extreme cases being NGC2793, where the $B$-band intensity
profiles are brighter by around 0.2~mag in the center,
and Arp147, the empty ring galaxy, where the center is 2~magnitudes
fainter than the ring. On the other hand, the continuum profiles outside
the ring fall linearly on these plots, implying that the stellar disk outside 
the ring is exponential.

\subsection{Integrated magnitudes and colors}

We obtained the total magnitude (including the nucleus) of each galaxy in all 
the bands by integrating the light inside an elliptical aperture of semi-major 
axis length $R_{B25}$. 
The parameters of the elliptical aperture are the same as that used to obtain 
1-d intensity profiles (see Table~4). 
The measured total $B$ magnitudes and optical and NIR colors along with the 
estimated errors for the sample galaxies are shown in Table~5.
The value of $R_{B25}$ is given in column 2.

We also obtained total magnitudes of the companions by integrating over 
polygonal apertures corresponding to the isophote $\mu_{\rm B}$ = 25~\magarc.
The resulting $B$ magnitudes and optical and NIR colors along with the  
estimated errors for the sample galaxies are given in Table~6. 
Photometry was 
obtained for all the suspected companions surrounding a ring galaxy.
In cases where there were bright foreground stars within the apertures used
for measurements, their contribution was subtracted both for the ring 
and the companion galaxies.

\subsection{Global star formation rates in ring galaxies}

The current star formation in galaxies is inferred using a wide variety of 
tracers \citep{ken98}. The presence of star formation in ring galaxies was 
inferred from the far-infrared (FIR) \citep{app87} and the \ha\ emission
\citep{mar95}.
In the current study, we calculate the SFR using the FIR fluxes and 
compare it with that obtained from the \ha\ fluxes. We also estimate the 
thermal radio continuum (RC) emission and compare it with the literature
20-cm continuum fluxes. Such a study of the Cartwheel galaxy has yielded 
a SFR of 18~\msunyr, and the thermal fraction at 20-cm 
of $\sim10$\% \citep{may05}.

In Table~7, we present the quantities related to the current SFR. 
Observed \ha\ fluxes and \ha\ equivalent widths are given in columns 2 and 3,
respectively. The columns 4 and 5 list the observed FIR and 20~cm RC fluxes. 
The latter fluxes were taken from \citet{jes86}, which could be lower limit
given that their observations were sensitive to only the bright regions. 
Observed \ha\ flux was corrected for the galactic extinction, and 
contribution from the [NII] lines (20\% from \citet{bra98}). 
We then used the calibration of \citet{pan03} for estimating SFRs from 
the \ha\ and FIR luminosities. SFRs obtained using 
the \ha\ luminosities were always found to be less than that found using 
the FIR luminosities. In star-forming galaxies, such a difference is 
understood in terms of the extinction by dust, which reduces the  
\ha\ emission, but not the FIR emission \citep{hir03}. 
Under this hypothesis, the FIR luminosity gives a reliable measure of
the total SFR. We hence derived the SFR using the FIR
and calculated the effective visual extinction $A_{\rm v}$ in such a way
that the SFRs derived using the extinction-corrected \ha\ luminosities match 
these values.
The resulting extinction is applied to the \ha\ fluxes, which were then used 
to estimate the thermal RC flux at 20~cm \citep{ost89}. 
The SFR, $A_{\rm v}$ and the estimated thermal flux as a fraction of the
detected flux are given in the last three columns of Table~7. 

The median value of SFR for our sample of galaxies is 7~\msunyr, which is more 
than a factor of two lower than that in the Cartwheel. Two sample galaxies
have SFRs exceeding that of the Cartwheel. One of them is known to have a 
Seyfert nucleus (NGC985), while the emission from the other (Arp148) is most 
likely originates from its companion (see the \ha\ image in Figure~13).
The observed median \ha\ equivalent width is 44~\AA, which is close to 
the median value obtained by \citet{ken87} for around 40 spiral galaxies 
in the \citet{arp66} sample. 
Thus, ring-making collisions lead to similar enhancements of star formation 
as that from other kinds of collisions leading to peculiar morphological 
structures.

The median value of the effective extinction for our sample is
$A_{\rm v}=2.0$~mag, with two galaxies (Arp145 and Arp148) having
$A_{\rm v}>4$~mag. The median value is marginally higher than the values 
found from optical spectroscopy of individual bright HII regions in ring 
galaxies studied by \citet{bra98}. This difference between the effective
global extinction and extinction towards the bright HII regions suggests 
the presence of star-forming regions that are faint or completely 
obscured in the \ha\ images.
There is independent indication of high extinction in Arp145:
the observed colors and magnitudes at diametrically opposite points
suggest that the northeastern segment of this galaxy suffers
high extinction. In Arp148 system, the companion nucleus is the most likely
location of high extinction. 

The median value of the thermal fraction at 20~cm is 43\%, with only three
galaxies (Arp144, IZw45, NGC985) having values less than 15\%, the mean 
value observed in the normal spiral galaxies. Hence, the sample galaxies 
have more than the normal share of the thermal flux at 20~cm. 
For some of the bright knots, \citet{jes86} obtained the spectral
indices combining 20~cm, 6~cm and 2~cm observations. These indices
are flatter than the value expected for a non-thermally dominated region.
Given that the thermal and non-thermal emission are related to the present 
and past star formation activity, respectively, this relative excess in 
our sample galaxies is consistent with enhanced levels of current star 
formation in our sample galaxies, inferred independently from the observed
\ha\ equivalent widths.

The \ha\ emission is detected in companions of four of our sample galaxies.
The \ha\ fluxes (in \ergscm\ and log units)  and the SFRs (in \msunyr) 
in these galaxies are: 
Arp142 ($-13.77$, 0.65), Arp148 ($-13.52$, 1.87), VIIZw466
($-13.26$, 1.90) and NGC5410 ($-13.09$, 0.20). 

\subsection{Sample of well-defined ring galaxies}

A galaxy was included in the sample of ring galaxies listed by \citet{app87}
if it was classified as such historically. These classifications
were based on the visual inspection of photographic plate material. 
The presence of a ring-like structure (or an arc in a few cases), 
accompanied by a companion 
preferentially along the minor axis of the ring within 
a few ring radii, were sufficient conditions to be classified as a ring 
galaxy. Modern digital data allow us to examine structures of the rings
in much more detail as compared to the photographic images. 
We hence used our data-set to define a sub-sample of well-defined ring 
galaxies, that are most likely to be formed under the scenario presented 
by \citet{lyn76}. We considered a galaxy to be a well-defined ring galaxy 
if:\\
(i) a complete ring is traced in both the \ha\ and the continuum images, and\\
(ii) the observed structure cannot be interpreted as a part of a spiral arm 
or an arc.

The first criterion ensured that the selected galaxies have on-going 
star formation in the ring. Nine of the sample galaxies satisfied these 
criteria. 
A morphological description of the remaining six galaxies that are rejected
is given in Table~8. Three of these galaxies are disturbed barred 
spiral galaxies --- when the winding arm meets the other end of the bar, 
the structure resembles a ring. The other two are galaxies with an arc-like 
structure. The sixth galaxy (Arp145) is a marginal case, as the reason for 
its rejection is the absence of an \ha\  ring. 
The continuum ring is traceable, but is broader and of lower contrast with 
respect to those in the well-defined ring sample. It is possible that this 
is a ring galaxy, but we are witnessing the ring at a later phase 
as compared to those classified as well-defined ring galaxies. Alternatively, 
it can be a normal ring galaxy, except that its star-forming ring is hidden 
from the 
optical view, resulting in a non-\ha-emitting, low-contrast redder continuum 
ring (see the description of this galaxy in the Appendix). 

Interaction definitely had a role in creating the presently observed 
structure of the galaxies in Table~8. Probably, the companion only flew
close to the target galaxy without really passing through it.
We exclude the mis-classified ring galaxies (including Arp145) from the 
discussions in the rest of this paper. Hence, we are left with nine galaxies 
that show well-defined rings in both the continuum and \ha. 
Among these galaxies, Arp141 is a unique case where the companion
is seen along the major axis of the ring.

\section{Photometric properties of ring galaxies compared to the spiral galaxies}

Integrated magnitude and color of a galaxy carry important information on the 
total mass and mean age of the dominating population in that galaxy. 
In this section, we compare the photometric properties, both local and 
global, of ring galaxies with those of normal spiral galaxies.

\subsection{Surface brightness-color relation in ring galaxies}

In normal spiral galaxies, intensity decreases and colors become bluer 
as one moves from the center to the outer parts. This systematic radial
change gives rise to a correlation between these two quantities.
On the other hand, neither the intensity nor the color change monotonically 
in a ring galaxy. Hence, it is interesting to know the intensity-color
relation for ring galaxies and compare it to that of the spiral galaxies.
Such a comparison is shown in Figure 4. 
The relation for the early and late-type spiral galaxies
from the sample of \citet{deJ96} is shown by two straight lines:
at a given $\mu_B$, late type galaxies are 0.4~mag bluer in $B-V$ 
as compared to the early types.
 
Ring galaxies distinguish themselves from spiral galaxies in having a 
characteristic double-valued locus resembling a slanted ``$\Lambda$'' in this 
plane. The points on the bluer and redder branches 
correspond to the regions external and internal to the ring, respectively. 
The point where the two branches intersect pertains to the ring, and is 
often the brightest and the bluest. This is quite different from that seen in 
spiral galaxies, where the brightest regions, normally the central 
regions, are the reddest.
In the last panel of the figure, we compare the average color and surface 
brightness of the ring in different galaxies.
Majority of the rings are bluer than even the late-type galaxies at the
same surface brightness levels ($\mu_B\sim$22--23~\magarc).
This suggests that recently formed stars contribute significantly to the
$B$-band light of the ring.

Several interesting aspects of ring galaxies can be noted in Figure~4.
Colors of external parts ($\mu_B > 24$~\magarc\ on the bluer branch)
are comparable to that of spiral galaxies at similar surface brightness. 
Thus the external part of the disk seems to be unaffected
by the collision and has preserved the pre-collisional disk properties.
However, the disks do not show the tendency for the color to become
bluer as the surface brightness decreases, which is most likely due to 
relatively large errors on 
the colors ($\gtrsim0.1$) at magnitudes fainter than $\mu_B > 24$ in the 
present observations. Future high signal-to-noise deep images would help in 
addressing this issue.
The colors of IIZw28 are bluer than the typical colors of late-type 
galaxies, everywhere, including the faint external parts of the disk. 
Among the 86 galaxies in the sample of \citet{deJ96}, we found two galaxies
(UGC06028 and UGC07169), where the sequence of points in the 
color-magnitude 
plane coincides with the parts external to the ring of IIZw28.
This is illustrated in the panel corresponding to IIZw28, where we plot
the profiles corresponding to UGC06028 (or NGC3495) and UGC07169 (dashed 
curves almost coinciding with the observed points outside the ring). 
The galaxies UGC06028 and UGC07169 are classified as PSABb 
and SABc, respectively in RC3 and do not exhibit obvious signs of 
interaction, though both have extended blue multiple arms dominating their 
morphologies in their external parts. 

The color at the ring-center, the point of impact of the intruder, 
(the reddest color on the redder branch) also follows 
the color-surface brightness relation seen in spiral galaxies. 
However, the central surface 
brightness is more than a magnitude fainter than $\mu_B\sim$21.7~\magarc,
the central surface brightness of the disks of spiral galaxies 
\citep{fre70}.
The nuclear colors, on the other hand, compare well with that 
of the central parts of the rings even when they are seen superposed 
on the ring (with the exception of Arp147 and IIZw28). This is illustrated 
by the solid circle that is placed at $\mu_B\sim$21.7~\magarc, the central 
brightness expected for a Freeman disk. 

The angle of intersection of the blue and red branches is different
in different galaxies. The extreme cases are Arp147, where the two branches 
are parallel, and Arp148, where they are almost perpendicular to each other.
The color and intensities of the bluer branch, which corresponds to the 
disk external to the ring, are barely affected by the post-collision
star formation. On the other hand, many physical parameters related to
the expanding wave control the slope of the redder branch.
The most important among them are the velocity of the expanding wave,
metallicity of the ring galaxy, and the fractional contribution of the
underlying disk stars to the observed profiles. A comprehensive analysis
of the observed profiles in terms of these physical parameters will be 
carried out in Paper~II of this series.

In summary, the ring is the bluest and the brightest part of the galaxy.
The disk external to the ring seems to maintain the pre-collisional disk 
properties. On the other hand, the ring-center is found to be 
fainter and bluer than the values expected at the center of the disk of normal
spiral galaxy.  The blue colors
are easily understood in terms of star formation following the collision.
Such a star formation is also expected to increase the disk brightness, 
and hence the observed low brightness implies that a lot of disk stars 
have escaped from the central regions.
We will come back to this issue in \S5.3. 
An alternative scenario to explain the faint and blue inner disks is that
the progenitors of ring galaxies were low surface brightness galaxies, 
rather than normal spiral galaxies.

\subsection{Color-color relation in ring galaxies}

Color-color relation is traditionally used to illustrate the sequential aging
of stellar populations in ring galaxies \citep{mar92}. 
In Figure~5, we show the locus of points for each
ring galaxy in the $B-V$ vs $B-I$ plane. Only the points interior to the
ring are plotted. Note that the points corresponding to successive radii
are connected by a solid line and hence the observed correlation implies
sequential ordering in both the colors, with the colors systematically redder
in the inner parts. Thus, our sample of ring galaxies follows the locus seen 
in other samples of ring galaxies. The nuclei (filled circles) 
of Arp147 and IIHz4, which do not have \ha\ emission, occupy the reddest 
end of the color-color sequence. Star-forming nuclei, on the other hand, have 
colors intermediate between that of the ring-center and the ring.

In Figure~5, we also plot the locus of points for normal early and late
type galaxies. It is easy to see that the sequential ordering of colors is not
a property unique to ring galaxies, instead colors of normal galaxies are
also sequentially ordered in radius. In fact, ring galaxies show smaller range 
of colors, occupying only the blue end of the color-color sequence. Hence, the 
characteristic property that distinguishes ring galaxies from normal galaxies 
is their bluer position in the color-color sequence, rather than the sequence 
itself.

\section{Results and comparison with N-body simulations}

In this study, we have comprehensively demonstrated the existence of a
stellar disk outside the ring in all our sample galaxies. The study has 
enabled us to photometrically characterize both the disk and the ring, and 
compare the properties of the disk of ring galaxies to those of normal 
spiral galaxies. We use this data-set allows us to test in detail some 
of the predictions of collisional scenario of ring formation. We begin
this section by summarizing the results from the N-body simulations that 
can be confronted with our data-set.

\subsection{Testable results from N-body simulations}

The most comprehensive set of models that investigate the formation and
evolution of ring galaxies, and that can be compared with our observational
results comes from the work of \citet{ger96}. Using a 
three-dimensional, combined N-body/hydrodynamical computer code, they 
studied the dependence of ring structure and its temporal development as
a function of the mass ratio of the two colliding galaxies. In their models,
the target (that transforms into a ring galaxy) is 
a gas-rich disk galaxy, and the intruder is a gas-free spheroid galaxy.
They presented figures illustrating the evolution of the surface density
profiles at several epochs after the impact for both the stellar and gaseous 
components in a form that can be directly compared with our observations. 
The target galaxy had properties similar to the Milky Way 
($M_{stellar} = 5\times10^{10}$\msun, $M_{gas} = 0.1\times M_{stellar}$, 
$M_{halo} = 2.5\times M_{stellar}$). Three values for intruder galaxy mass 
were chosen: equal-mass (C-1), one forth (C-4), and one tenth (C-10) of the 
target galaxy total mass. The disk density and morphology
were calculated out to 17.6~kpc, which corresponds to 4.4 disk scale lengths.

In this paragraph, we summarize some of the important results from this
simulation, that can be directly confronted with our observations.
For collisions involving lower mass galaxies (C-4 and C-10), they found that
the stars essentially behave like those in the classical work of 
\citet{lyn76}. However, the classical picture breaks down in detail
in a collision of equal-mass galaxies: e.g. there is bulk motion of material
over distances large compared to the initial disk dimension. Further, a
ring structure is identifiable only for times $<100$~Myr after 
the collision, with the stars and gas expanding to distances twice the 
initial disk size for later times.
In comparison, in experiments C-4 and C-10, the ring has not reached the
edge of the original disk up to the end of the run (140~Myr).
In all the three experiments, an inner ring forms at the same radius at the
same time, but the location of the outer ring is sensitive to the mass ratio.
However, the inner ring is noticeable only after $\sim100$~Myr, by which 
time the outer ring has already diffused in the equal-mass collisions.
The behavior of gas is also qualitatively different in equal-mass collisions,
as compared to the unequal-mass collisions. The gaseous ring lags behind
the stellar ring in C-1, whereas in the other two experiments it is embedded 
inside, with the density maximum lying at the outer edge of the broad stellar 
ring. The contrast of the ring (compared to 
the surface density of the unperturbed disk) is expected to be higher for
collisions involving higher relative mass intruders.
Considerable off-planar structures, both in stars and gas, in the direction 
of the intruder were found in the C-1 experiment. The vertical spread is more
in the center than it is in the ring. In particular, the nucleus
is dislodged from the plane of the ring. These effects were found to be
only mild in C-4 and C-10 experiments. 
The vertical movements of gas and stars make the surface densities
to be very different from the volume densities, with the maximum differences
expected at the central regions. This would also make the observable
surface densities heavily dependent on the line-of-sight projection angle
of the disk of the ring galaxy. 

\subsection{Relevant observational quantities}

Mass of the ring galaxy, and its ratio to that of the intruder galaxy, 
are the most important quantities for comparing observations with the models. 
In most cases, the intruder can be identified with certainty. However,
the identification becomes a non-trivial task when a ring galaxy is 
surrounded by more than one companion galaxy:
e.g. in the case of the Cartwheel, one of the two galaxies along its minor 
axis was long believed to be the intruder, but the detection of a H\,I 
plume connecting the Cartwheel to a fainter galaxy farther out has opened up 
the discussion on the identity of the intruder \citep{hig96}.
Among the 9 well-defined ring galaxies of our sample, VIIZw466 is the
only one having more than one companion.
In this case, we have considered {\bf G2} as the intruder, following the
detection of an H\,I bridge connecting the ring and {\bf G2} by \citet{acs96}.
On the other hand, there is no confirmed companion 
to IIZw28. However, there is a disturbed object within two optical diameters 
of this galaxy, which we have tentatively identified as the intruder galaxy.
We have calculated mass of this candidate companion  
assuming it is at the same distance as the ring galaxy, and use 
this mass as an upper limit to the companion mass.
In the remaining 7 cases, the only companion that is seen within a few 
galaxy diameters is taken as the intruder. The majority of these companion 
galaxies is dominated by the spheroidal component, a fact consistent
with the assumptions of N-body simulations.

The dynamical masses of galaxies are usually determined using 
rotation curves, or in its
absence the \hi\ line widths with single-dish telescopes. Such masses are
available only for six of our ring galaxies, and none of the companion 
galaxies \citep{jes86}. On the other hand, the stellar masses of galaxies
can be determined by combining the $K$-band photometry with an appropriate 
value for the mass-to-light ratio. The use of the $K$-band
ensures that the observed light originates predominantly in stars that
contribute to the stellar mass of galaxies. \citet{Bel03} have presented
empirical relations between mass-to-light ratio and colors of galaxies,
that allows an estimation of the stellar masses of galaxies taking into account
the contamination to the observed light from recently formed stars.
We used their relation between $M/L_K$ and $B-R$ color to determine
stellar masses for our sample of ring galaxies and their companions.

In Figure~6a, we show the location of ring galaxies ({\it filled circles}) and 
their companions ({\it open triangles}) in color-magnitude plane. 
Quantities are corrected for the galactic extinction, but not for the 
internal extinction.  The size of the filled circle is proportional to 
the relative mass of the companion galaxy.
The relation between the $B-R$ color and $M_K$ magnitude for galaxies
of fixed masses is shown by the dashed lines. 
The stellar masses calculated using this relation along with other 
physical quantities for our sample of ring galaxies are presented in Table~9. 
In column~2, the $K$-band absolute 
magnitude $M_{\rm K}$, corrected for the galactic extinction, is given. 
The photometric masses obtained using the relation of \citet{Bel03} are 
tabulated in column~3. The gas masses, taken as the sum of H\,I and H$_2$, 
with a correction factor of 1.4 to take into account the Helium content,
and the dynamical masses are tabulated in columns~4 and 5, respectively. 
The H\,I and dynamical masses were compiled from \citet{jes86} for 5 galaxies 
and \citet{acs96} for one galaxy (VIIZw466).
The H$_2$ masses were taken from \citet{hor95}. All the masses used 
from the literature are homogenized to the distance used in our work.  
It can be seen that the baryonic mass, taken as the sum of the stellar 
and gaseous masses is less than the dynamical mass, in all but one case,
the exception being VIIZw466.
In column~6, we give the ratio of the baryonic masses of the ring galaxy 
to that of the companion. In calculating this ratio, we have neglected 
the presence of any gas in the companion galaxy, which is reasonable
given that the companion galaxies are redder and bulge-dominated.
Other quantities of interest are the ring radius, $R_{\rm H\alpha}$ and disk 
size usually taken as the radius, $R_{B25}$, measured at $\mu_B=$25~\magarc. 
The values of $R_{\rm H\alpha}$ and $R_{\rm H\alpha}$/$R_{B25}$ 
are given in the last two columns of Table~9.

In order to facilitate comparison with \citet{ger96} model, we use the
mass ratios of the ring galaxy to that of its companion to identify
each of our galaxy with one of their C-1, C-4 and C-10 groups.
The galaxies belonging to each group are listed in Table~10. 

\subsection{High mass intruder galaxies and N-body simulations}

The photometrically derived baryonic mass ratios presented in Table~9 
are expected
to represent their dynamical mass ratios, as there is no reason to believe
different ratio of stellar to dark matter in ring and companion galaxies.
The mass of the companion is smaller than that of the ring 
galaxy in four cases (IIHz4, IZw45, IIZw28 and NGC2793), whereas it is
higher in three cases (Arp141, Arp147, Arp148). In Arp143,
the masses are almost equal. In VIIZw466, the mass ratio depends on the
identification of the companion. With the late-type galaxy {\bf G2} as the
companion \citep{acs96}, the ring galaxy is around 2.5 times massive. However
with G1, the compact companion along the minor axis, as the intruder, 
the ring galaxy mass is around half of that of the companion, a value similar
to that found in other three galaxies with massive companions.
Hence at least in three cases, the intruder seems to be more massive than
the target galaxy. This is unlikely to be due to mis-identification of the
companion, as there are no other candidates in the neighborhood. In addition,
in two of the companion galaxies (Arp141 and Arp148), we can see clear signs 
of interaction in the form of tidally stripped stars. Thus, 
the current mass of these ring galaxies is indeed less than 
that of the companion.

It is not clear whether a gas-rich disk galaxy could survive a head-on 
collision with a higher mass,  but compact, companion, a case not
considered in most of the simulations.
Recent simulations by \citet{nam06} show that such collisions result
in large (up to 30\%) amount of mass being lost from the disk galaxy.
Even in the simulations of \citet{ger96},
where the companion mass at the most equaled the target galaxy mass,
considerable fraction of stars from the central part of the ring galaxy were 
found to escape out of the plane in the 1:1 collisions.
If these stars lie outside our photometric apertures, then it is possible that 
we have underestimated the ring galaxy mass. In the following paragraphs,
we discuss this issue.

\citet{ger96} presented projected radial intensity profiles for their on-axis 
3-D simulations by summing stellar contributions perpendicular to the plane.
Thus, the escaped stars still contribute to the radial intensity profile, 
resulting in a profile that is almost similar to that of the pre-collisional 
disk. On the other hand,
the observed intensity profiles at the central parts are qualitatively
different from those of the 3-D simulations: e.g. observed central surface
brightness (e.g. $\mu_B(0)$) is systematically lower than that at the ring,
whereas \citet{ger96} models always suggest higher stellar densities 
at the center than at the ring. Thus the escaped central stars do not seem to 
contribute to the observed radial profiles. The reasons for this may be that 
the collisions in the observed galaxies are slightly oblique, and the disk 
orientation is not exactly perpendicular to the line-of-sight.

If the collisions are oblique, then the off-planar stars, most of which 
belonged to the central regions, will not be seen projected on to the
central regions on the 2-D images. Instead they are more likely to be seen
projected at a larger radius. Can some of the observed light outside the ring
be attributed to these stars? The fact that the colors at the external parts
of our sample galaxies (see the radial color profiles in the appendix)
are typically bluer than that in the central regions, and that they compare
well with the colors of external parts of normal galaxies, suggests that
the escaped stars do not dominate the light outside the ring. Thus, the 
escaped stars seem to be spread out to much larger radii, and our photometric
apertures do not include them, leading to an under-estimation of 
photometrically derived masses.
Deep wide-field infrared band imaging of ring galaxies would be required to 
detect the missing stellar mass.

In summary, we believe that the ring galaxies with apparently higher
mass intruders have lost as much as half of their masses during the passage
of the companion, and that the two galaxies had comparable masses at the
time of collision. 

\subsection{The disk size-absolute magnitude relation}

In Figure~6b, we plot the disk radius, $R_{B25}$, against $M_K$ for our sample
of ring galaxies (filled circles) and for normal spiral galaxies from the 
sample of \citet{deJ96} (thin dots). In three ring galaxies with massive 
companions, an arrow is drawn with the tip of the arrow indicating the 
magnitude of the ring galaxy if it had the same mass as that of the companion. 
As expected, brighter galaxies are larger and vice versa.
What is intriguing, however, is that the relation shown by the ring galaxies 
is exactly the same as that for normal spiral galaxies.
$M_K$ represents the total magnitude from the underlying disk stars and
is not heavily affected by the contribution from recent stars. Hence it is
a good approximation to assume that the present $M_K$ is the same as that
of the pre-collisional galaxy. Thus, the observed relationship implies
that the size of the stellar disk is unchanged because of the passage of the
intruder.
Under the collisional scenario of ring formation, ring galaxies are expected 
to be larger than normal spiral galaxies of the same mass, especially 
in those with large relative companion masses, and late stages of collision. 
Our observations do not show any dependence of size on the relative mass of the 
companion. Hence, the only way to reconcile our observations with the results 
of N-body simulations of \citet{ger96} is that the collisions occurred less 
than 80~Myr ago. 
The absence of an inner ring in any of our sample galaxies is consistent 
with these relatively recent collisions. 

\subsection{The disk scale length-companion mass relation}

The propagation of a circular stellar density wave in the disk of a ring 
galaxy results in the re-distribution of its stars. As a result, the radial
intensity profile deviates from the intrinsic exponential form, showing
a bump at the position of the ring. N-body simulations of \citet{ger96} 
show that the profile outside the ring continues to be exponential,
but steeper than the original profile, especially for galaxies with 
high-mass companions. At a fixed age, the steepness of the profile depends
on the companion mass, with higher the relative mass of the companion, 
steeper is the profile. Outer disk profiles of ring galaxies with 
relatively low-mass companions become steeper at later times as compared to 
the ones with high-mass companions. In order to measure the steepening using 
the observed profiles,  
we need to know first the intrinsic pre-collisional scale length of the disk.
The observed relation between the disk scale length, $R_{\rm D}$, and the
and the baryonic mass of a galaxy for the normal spiral galaxies can be used 
to infer the disk pre-collisional scale length for each galaxy.
In Figure~6c, we plot the disk scale length $R_{\rm D}$
measured outside the ring, against the stellar mass of the ring galaxies 
(filled circles; in 3 galaxies where the estimated ring galaxy mass is
smaller than that of the companion, we draw the arrows indicating the amount
by which masses have to be increased to match that of the companion). The 
expected relation for normal spiral galaxies from \citet{deJ96}
is shown by small dots. 
It can be seen that the derived scale lengths are systematically smaller 
than that for normal spirals of equivalent mass in all the ring galaxies, 
except one (Arp143). Thus, in general, the observed results support the
predictions of the simulations of \citet{ger96}. 
However, the steepness of the profile is not found to depend on the relative 
mass of the companion, which probably indicates that the rings with 
low-mass companions in the sample are on the average older than those
with high-mass companions.

\subsection{The ring size-disk size relation}

The physical size of the ring in our sample of galaxies varies between
2--12~kpc in radius. In \citet{ger96} simulations, rings reach these values
in less than 60~Myr for 1:1 models, and less than 140~Myr for 10:1 models.
However, when normalized to the disk radius ($R_{\rm B25}$),
all the rings are located between 0.3 and 0.7, with a mean value $\approx0.5$.
These are illustrated in Fig.~6d. Considering the possible broad range of the 
expansion velocities and ring ages, it is surprising that the rings   
preferentially occur at around half-way in the optical disk. Reasons 
for this apparent  preference could be the following:  
the first is regarding the bias against classifying a galaxy as a ring galaxy
when the ring is in its very inner part --- such 
rings are often associated with resonance phenomenon, and are excluded 
from the catalogs of ring galaxies. Also, at early times the intruder
would be still seen projected on the disk, inhibiting the detection of
the ring. Secondly, in the very external parts of the galaxies, gas density 
is usually low, and is below the critical density required for star formation. 
\citet{mar01} found the density of HII regions in normal spiral galaxies 
falls sharply slightly inside the optical radius, and the radius of the sharp 
decrease corresponds to the regions where the gas density drops below the 
critical density. Using their data for 32 spiral galaxies, we calculated 
the threshold radius to be 0.8$R_{\rm B25}$, which is shown by a dashed
horizontal line in Figure~6d. 
Another reason for the absence of rings at large disk radii is that the
expanding wave would have decayed so much by the time it reaches these parts 
of the disk that it won't be efficient enough to trigger star formation.
Hence, star-forming rings are not expected to be found at the very external
parts of the optical disks. 
\cite{biz07} implemented these ideas in their numerical model for Arp~10, and
found that the \ha\ surface brightness reaches its maximum value 
when the ring has expanded to a radius of about $0.5 R_{\rm B25}$, the inner
ring radius observed in that galaxy. Hence, both the physical and observational 
selection effects favor the detection of rings when they are located 
at about half-way through the optical disk.

\section{Conclusions}

We have carried out an analysis of the optical and NIR images of 15 
galaxies in the northern hemisphere from the sample of \citet{app87}.
The images are used to establish a variety of characteristics 
common to these systems. These characteristics are summarized below.

All the 15 galaxies studied show evidence for recent star formation in the
form of detection of FIR and \ha\ emission. In general, the SFR derived using 
the FIR is higher than that derived using the \ha, with the differences
suggesting a median visual extinction of $A_v=2.0$~mag.
Arp145 and Arp148 are the two galaxies where the observed FIR and \ha\ fluxes
suggest effective visual extinction in excess of 4~mag. 
The median SFR (7~\msunyr) 
and \ha\ equivalent width (44\,\AA) suggest that their current star formation
activity is similar to that found in other kinds of interacting galaxies.
The observed 20~cm fluxes of the sample galaxies seem to be having
a higher thermal fraction as compared to that in normal spiral galaxies.

All the 15 galaxies studied show morphological signs of having suffered
an interaction in the recent past. However, only nine of the galaxies
have morphologies consistent with a ring-making collision of the type 
described by \citet{lyn76}.

Our sample of ring galaxies shows the sequential color-color relation 
found in other ring galaxies by \citet{mar92}. However, we found that the
relation for ring galaxies is not very much different from that seen for 
the normal spiral galaxies --- and that the characteristic that distinguishes 
the ring galaxies from the normal galaxies in a color-color plot is 
that the ring galaxies occupy only the blue end of the relation for the 
normal spiral galaxies.

We were able to trace the underlying stellar disk well beyond the ring in all 
our sample galaxies. The outer extension of the stellar disk is in general
symmetric around the ring, with noticeable asymmetries present 
in Arp141, Arp143, Arp147 and NGC2793.
The disk colors in the external parts are redder than the ring, but
are in general bluer than that at the ring-center.

The ring-center of our sample galaxies is fainter (at least by a 
magnitude in $B$), and bluer (by around 0.2 mag in $B-V$) as compared to 
the values at the disk centers of normal spiral galaxies. Recent star 
formation in the wave is expected to make the inner disk bluer (as is 
observed) and brighter (contrary to what is observed).
Thus, the observed low surface brightness of the ring-center implies
either considerable fraction of the inner disk stars have splashed out of the
disk or that the pre-collisional galaxy was of low surface brightness.

The outer disk scale lengths of ring galaxies are found to be systematically 
smaller than that of normal galaxies of comparable mass.
On the other hand, the disk radius $R_{B25}$, measured at the 25~\magarc\ on 
azimuthally averaged $B$-band profiles, seems unchanged due to collisions.
Under the N-body simulations of \citet{ger96} these results imply time 
elapsed since the collision less than 80~Myr.

The physical sizes of the ring in our sample varies from 2--12~kpc. However,
the observed rings seem to be systematically located around half-way through
the disk. This is most likely due to the difficulties in identifying 
a ring galaxy in its initial phase of evolution. The outer parts of the
galaxies may not have sufficient gas surface densities to form stars.

\acknowledgments

We thank Vladimir Korchagin for discussions during the initial stages of
this project that helped to plan the observations. We greatly appreciate
the comments of Gerhardt Meurer as a referee, which were extremely useful 
in improving the presentation of our results in this paper.
We thank all the supporting staff at OAGH and OAN-SPM observatories for 
technical help during the observations.
This work is partly supported by CONACyT (Mexico) research grants 
39714-F and 42609-F, and formed the doctoral thesis of RR.
This research has made use of the NASA/IPAC
Extragalactic Database (NED) which is operated by the Jet
Propulsion Laboratory, California Institute of Technology, under
contract with the National Aeronautics and Space Administration. \\



\centerline{APPENDIX}  


\section{NOTES ON INDIVIDUAL GALAXIES}

In this Appendix, we discuss structural properties of each galaxy, based
on their multi-band appearance and radial profiles of intensity and colors. 
The discussions are centered around (i) the morphology of the ionized gas, 
(ii) the underlying stellar disk, (iii) the nature of the 
nucleus, and (iv) the morphology of the candidate companion galaxies.
We present one composite figure for each galaxy (Figures~7--21). 
Each figure has three parts: 
{\it top} part, gray-scale images in \ha\ and $B$-bands; 
{\it bottom left} part, azimuthally averaged radial intensity
and color profiles; and
{\it bottom right}, the intensity and color profiles along two 
representative position angles, illustrated on the gray-scale $B$-band 
image, one of them passing through the companion.

{\it Top:} 
Intensities on the gray-scale images are shown on a logarithmic scale, with 
the specific purpose of illustrating the faint ionized gas on the \ha\ images,
and the disk external to the ring on the $B$-band images. Some of the
bright structures that are saturated on these plots can be visualized 
clearly on the color composite images in Figure 2.
The contours of \ha\ intensities are superposed on the $B$-band image, 
with the outermost contour corresponding to $5\sigma$ levels (see the 
column~7 of Table~4), and the subsequent contours brighter by factors of 4. 
Best-matched ellipses to the ring (green; inner ellipse) and the
disk (red; outer ellipse) are shown. A small open circle superposed on 
the \ha\ image, indicates the position of the nucleus. The most likely 
intruder galaxy is identified by {\bf G1}. When a second companion galaxy 
is seen, it is identified by {\bf G2}. The orientation, as well as the 
image scale are indicated on the panel corresponding to the \ha\ image.
The two arrows on the $B$-band image indicate the direction of the cuts
corresponding to the plots on the bottom right.

{\it Bottom left:} 
Azimuthally averaged radial profiles of intensity and colors, along 
with errors on them. Azimuthal averaging is done along the ellipse that
best represents the ring (displayed in green on the $B$-band image; see 
Table~4 for the ellipse parameters).
The ring radius (the position of the peak in the \ha\ profile), and disk 
radius (the position where $\mu_B=25$~\magarc) 
are shown by the inner and outer vertical dotted lines, respectively. 
Note that the nucleus and companions were masked in obtaining these radial 
profiles (see text for details). In three cases 
(IZw45, IIZw28 and Arp291), masking could not be implemented in a meaningful 
way because the nuclear light affects more than half of the ring.
We hence display the profiles taken on the unmasked images for these galaxies.

{\it Bottom right:} 
Cuts in intensity and color along two representative position angles.
One of these cuts (shown by the red arrow on the $B$-band image) is carefully
chosen to pass through the faint parts
of the ring (i.e. avoiding any bright \hii\ regions), with the intention
of illustrating the characteristics of the faint diffuse parts of the ring,
that do not contribute much to azimuthal profiles. 
The top three panels show the corresponding profiles.
The radius of the ring as measured on the azimuthally averaged \ha\ profile
is shown by the vertical dotted lines on these plots.
Note that the cuts do not necessarily pass through the ring-center, and 
the origin of the x-axis is chosen to be the point closest to the ring-center,
and the distance increases towards the tip of the arrow.
The second cut passes through the ring and the nucleus of the companion. 
The $B$-band and $B-V$ color profiles corresponding to this cut are 
shown by the solid and dotted lines, respectively, in the bottom most graph.
These intensity and color cuts allow us to appreciate the actual gradients 
without the effects of azimuthal averaging. Note that the ring is relatively
blue, and emits in the \ha\ line even in the parts where there is no 
bright \hii\ regions. The last plot also enables us to compare the relative 
colors, sizes, and separation of the disk, ring and the companion. 

In Figure~22, we show the gray-scale maps of Arp143 in $BVR$H$\alpha IJHK$
bands. Gray-scale maps for the rest of the galaxies are available 
at the website http://www.inaoep.mx/$\tilde{\ }$ydm/rings/. 
Fits files in all these bands will also be made available at this site.
We have obtained the $U$-band images for three galaxies (Arp147, IIHz4, 
NGC2793) which also form part of the gray-scale maps and our dataset. 
\\

{\bf Arp141:}
The ring in this galaxy is defined by compact knots of \ha\ emission.
Several regions of extended emission are seen inside this ring, which
also have corresponding continuum emission. The underlying disk is circular, 
whereas the ring is clearly elliptical (eccentricity=0.16). 
The ring on the eastern side almost coincides with the edge of the
disk, whereas it is only half way through the disk on the western side. 
The nucleus is seen outside the ring, along the line joining the ring with 
the companion.
The companion galaxy has tidal streams, and is clearly participating
in the collision. Note that the companion is seen along the
major axis of the ring --- a unique case among the known ring galaxies.

{\bf Arp142:}
No ring structure can be defined in this arc-like galaxy. 
Hence, the plotted ellipse is not a match to the ring, and is used only 
to obtain azimuthally averaged profiles. Note that the nucleus is 
located at the northern part of the arc, from where most of the \ha\ emission  
originates. The position of the 20~cm radio continuum emission mapped by 
\citet{mol92} coincides with the bright \ha\ emitting zone. 
At longer wavelengths, a two-armed spiral structure is clearly seen 
emanating from the nucleus. Faint \ha\ emission is 
detected from the edges of the arc seen in the continuum bands. There is 
a second candidate for the intruder ({\bf G2}), which shows faint \ha\ 
emission, and is bluer than the ring galaxy.
The bright object along the major axis of {\bf G2} is a foreground star.

{\bf Arp143:}
Ring is very well defined by the distribution of \hii\ regions, with
most of the \ha\ emission originating from the northwestern arc.
The \ha\ emission is also detected from the nucleus.
The stellar disk can be clearly traced outside the ring, and has similar
ellipticity as the ring. However, the ring is offset both in center and
P.A. with respect to the disk. The nucleus is not exactly located at the 
ring center and is pulled towards the companion.
The farther side of the companion shows tidal streams. There is 
considerable azimuthal asymmetry in the disk brightness outside the ring 
in all the bands. 
Our multi-band images confirm the sharp decrease of 
intensity on the western side of the galaxy that was reported by \citet{app92}.

{\bf Arp144:}
A clear ring is not seen either in \ha\ or in continuum bands and the
definition of the ellipse was based on the partial ring that can be traced 
in the \ha. The companion galaxy shows signs of interaction,
in the form of stripped stars on the side opposite to the ring galaxy.
The system does not appear to belong to the classical \citet{lyn76} 
scenario, and is often considered to be the result of stripping of
the gaseous disk from a spiral galaxy during the collision with an 
intergalactic HI cloud \citep{hig88} following a proposal by \citet{fre74}.

{\bf Arp145:}
The ring is broad in the continuum, especially on the southwestern side, 
where it is also considerably brighter and bluer and than the 
opposite side. The \ha\ emission is weak, and is confined to the nucleus and 
a few other knots in the ring. A visual extinction value of $>4.0$~mag is 
inferred for this galaxy by comparing the SFR derived by \ha\ and FIR. 
Color profiles suggest that most of this dust lies on the northeast side of 
the ring. Two bright objects to the southwest of the nucleus are foreground 
stars.

{\bf Arp147:}
This classical empty ring galaxy is the brightest \ha\ emitter in our
sample (leaving out NGC985, where the Seyfert nucleus is responsible for 
the emission), with the emitting compact knots more or less uniformly 
distributed in the ring. The space between the knots in the ring is also 
filled with diffuse \ha\ emission. Note that there is no compact \ha\ knot 
associated with the nucleus, which is extremely red and gives a crescent 
shape to the ring in the $K$-band. Several filamentary structures on the 
northeastern segment can be noticed in the \ha\ image.
The observed crescent structure and the displaced nucleus have been 
reproduced by \citet{ger92}, using stellar 
and gas dynamical models, with {\bf G1} as the intruder. 
The \ha\ emission associated with the companion {\bf G1}, and a foreground
star to its southeast, is not significant given the errors expected 
in the subtraction of continuum image from these bright continuum sources.

{\bf Arp148:}
\ha\ emission is uniformly distributed in the ring, with the only knot
coinciding with the brightest knot on the $K$-band image. We identify
this knot as the nucleus. The \ha\ emitting knot eastward of the ring
is most likely associated with the companion. The companion is elongated 
along the east-west direction, splitting into two objects separated by 
7\arcsec\ in the $B$-band. The $K$-band and \ha\ fluxes are maximum 
between the two $B$-band knots. A visual extinction value of $>5.0$~mag is 
inferred for this system by comparing the SFR derived by \ha\ and FIR.
The observed multi-band morphology of this galaxy suggests that most of the
dust is associated with the \ha\ emitting region of the companion. 
The double lobe structure of the companion at shorter wavelengths is 
due to the obscuration caused by this dust cloud.

{\bf Arp291:}
\ha\ emission in this galaxy is not associated with the ring that is seen
in the continuum bands. Most of the \ha\ emission comes from regions 
surrounding the nucleus, and extends along a filament along the minor axis
of the ring. The \ha\ emission at the nucleus itself is weak.
A faint spur seen on the northeastern side of the nucleus on the continuum 
images gives this galaxy an appearance of a tightly-wound one-armed spiral,
 and hence this galaxy may not be a classical ring galaxy.
There is no spectroscopically confirmed companion for this galaxy. We 
identify the most likely companion by {\bf G1}.

{\bf IIHz4:}
Ring is very well defined both on the \ha\ and continuum images of this
classical double ring galaxy. The faint second ring can be seen to the
north of the main ring on the continuum images, and is denoted by {\bf R2}.
\ha\ emission is not detected from the second ring. 
Nucleus of both the rings is identified.
There is an \ha\ emitting region without noticeable continuum emission,
outside the ring, just south of it. 
A bright blue star is responsible for the glare seen to the north of the 
ring galaxy. This glare didn't permit us to show the second ring on our
$B$-band image and hence the $V$-band image is displayed.

{\bf IZw45:}
\ha\ emission traces most of the ring, with the brightest \ha\ emitting
region coinciding with the nucleus. The continuum ring is slightly flatter
and bigger than the \ha\ ring. The continuum images show a bridge that
connects the nucleus of the ring galaxy to the companion.
This is the only galaxy where the colors at the center of the ring are
bluer than at the ring itself. This is most likely due to the contamination
of the radial profiles by the presence of the nucleus that extends all the 
way to the center of the ring..

{\bf IIZw28:}
The ring can be traced both in the \ha\ and continuum bands of this galaxy.
Most of the \ha\ emission comes from the northern segment that 
includes the nucleus. There is no spectroscopically confirmed companion
to this galaxy. We identify the most likely companion by {\bf G1}.

{\bf VIIZw466:}
Most of the \ha\ emission originates from the western half of the ring,
with hardly any emission detected from the red nucleus. Comparable \ha\ 
emission is also detected from the edge-on companion {\bf G2}. 
This detection confirms the earlier suggestion of star formation activity in 
this galaxy by \citet{app99} based on mid infrared emission. 
\citet{acs96} had confirmed {\bf G2} as the intruder galaxy responsible
for the ring in VIIZw466.
The object to the northwest of {\bf G1} is a background galaxy.

{\bf NGC985:}
The nucleus of this ring galaxy is known for its Seyfert activity 
\citep{rod90}, and is bright on both the \ha\ and continuum images. 
The \ha\ emission traces only
the western segment of the ring. No companion can be identified within a few
disk diameters of this galaxy.

{\bf NGC2793:}
The ring in this galaxy is traced by the \ha\ emitting knots.
The nucleus is located at the southern end of a bar-like elongation.
Patchy \ha\ emission is detected all along the bar.
A compact knot of emission appears on the southern part of the disk outside
the ring. Faint diffuse \ha\ emission can be detected throughout the internal
part of the ring.

{\bf NGC5410:}
In morphology, this galaxy resembles a disturbed barred spiral galaxy, 
rather than a ring galaxy. \ha\ emission is detected from throughout
the galaxy and its companion. The companion shows clear signs of 
interaction.


\clearpage

\begin{center}
\begin{deluxetable}{llcccccc}
\tabletypesize{\scriptsize}
\tablecaption{General properties of the sample galaxies}
\tablehead{
\colhead{Galaxy} & \colhead{Other names}     & \colhead{$\alpha (2000)$} 
                 & \colhead{$\delta (2000)$} & \colhead{V$_0$}
                 & \colhead{Scale}     & \colhead{Dia} 
                 & \colhead{$B_{T}$} \\  
\colhead{}       & \colhead{}                & \colhead{$^h$~~$^m$} 
                 & \colhead{$^{o}$~~$^{\prime}$} & \colhead{km\,s$^{-1}$}  
                 & \colhead{pc\,arcsec$^{-1}$} & \colhead{\arcsec}
                 & \colhead{mag} \\ 
}
\startdata
Arp144   &  NGC7828/9     & 00 06.49 & $-$13 25.25 & 5888   &  402  &   $37.9$  & 14.8 \\ 
Arp145   & VZw229, A0220+41A/B, UGC1840
                           & 02 23.14 & +41 22.33  &  5357  &  371 &    63.2    & 14.3 \\ 
NGC985   &Mrk1048, VV285
                           & 02 34.63 & $-$08 47.26  & 12948  & 897 &    35.0     & 14.1 \\ 
Arp147   & IZw11, IC298A/B, VV787
                           & 03 11.31 & +01 18.95  &  9660  & 669   &    16.3     & 15.7 \\ 
IIZw28   &A0459+03, VV790B
                           & 05 01.73 & +03 34.52  &  8572  & 594   &    14.0     & 15.6 \\ 
Arp141   & UGC3730, VV123
                           & 07 14.35 & +73 28.52  &  2874  & 199 &    75.9     & 14.5 \\ 
Arp143   & NGC2444/5, UGC4016/7, VV117
                           & 07 46.87 & +39 00.60  &  4017  & 278  &    84.5     & 13.6 \\ 
IIHz4    & A0855+37        & 08 58.55 & +37 05.19  & 12862  & 891  &    29.0     & 15.7 \\ 
NGC2793  & UGC4894         & 09 16.79 & +34 25.79  &  1644  & 114  &    56.0     & 14.2 \\ 
Arp142   & NGC2936/7, UGC5130/1, VV316
                           & 09 37.75 & +02 45.52  &  6794  & 471  &    81.0     & 14.0 \\ 
Arp291   & UGC5832,VV112
                           & 10 42.81 & +13 27.60  &  1084  &  75  &    58.0     & 14.3 \\ 
Arp148   & Mayall, A1101+41& 11 03.90 & +40 51.00  & 10363  & 718 &    20.3     & 15.1 \\ 
VIIZw466 & A1229+66        & 12 31.94 & +66 24.68  & 14341  & 993 &    22.0     & 15.6 \\ 
IZw45    & NGC4774         & 12 53.11 & +36 49.10  &  8314  & 576  &    23.0     & 14.9 \\ 
NGC5410  &UGC8931, VV256
                           & 14 00.91 & +40 59.30  &  3785  & 262  &    60.0     & 14.4 \\ 
\enddata
\end{deluxetable}
\end{center}

\begin{center}
\begin{deluxetable}{llrrrrclcrlrrr}
\tabletypesize{\scriptsize}
\tablewidth{0pc}
\tablecaption{Observational log of the sample galaxies\tablenotemark{a}}
\tablehead{
\colhead{Galaxy} & \colhead{Run}
                 & \colhead{B} & \colhead{V} & \colhead{R} & \colhead{I} & \colhead{$\lambda(\rm H\alpha$)}
                 & \colhead{H$\alpha$} & \colhead{$\lambda(\rm H\alpha\rm c$)} & \colhead{H$\alpha\rm c$} 
                 & \colhead{Run} & \colhead{J} & \colhead{H} & \colhead{K} \\
\colhead{} & \colhead{Mon-Yr} & \colhead{Sec} & \colhead{Sec} & \colhead{Sec} & \colhead{Sec} 
& \colhead{nm} & \colhead{Sec} & \colhead{nm} & \colhead{Sec} 
& \colhead{Mon-Yr} & \colhead{Sec} & \colhead{Sec} & \colhead{Sec} \\
}
\startdata
Arp144 & Dec-01 &1800 & 600& 600& 360 & 668&3600&\nodata  &\nodata  & Dec-00 & 840 & 600 & 630  \\
Arp145 & Oct-02 &1200 & 900& 600& 600\tn{b} & 668&3600&662      &2400     &Dec-99  &1080 &720&1200 \\
NGC985 & Feb-02 &1800 &1200& 900& 540 & 686&2700\tn{b}&674      &1800     &Oct-01  &1080 &480& 600 \\
Arp147 & Dec-01 &1800 &1800&1200& 600 & 674&5400\tn{b}&662      &3600     &Dec-99  & 720 & 600&1200  \\
IIZw28 & Feb-00 &600  &1200& 300& 300 & 674&5400\tn{b}&\nodata  &\nodata  & Mar-00 & 960 & 840 & 660  \\
Arp141 & Feb-00 &900  & 600& 300& 600 & 662&3600&\nodata  &\nodata  & 2MASS  &\nodata & \nodata& \nodata \\
Arp143 & Feb-00 &900  & 600& 900&1200 & 662&3600&\nodata  &\nodata  & Dec-99 &1080 & 900& 1500\\
IIHz4  & Feb-00 &1200 &300 & 900& 600 & 686&3600&\nodata  &\nodata  & Mar-00 & 960 & 780 & 630  \\
NGC2793& Feb-00 &600  & 300& 900& 300 & 656&3600&\nodata  &\nodata  & Dec-99 &2160 & 540& 390\\
Arp142 & Feb-00 &600  & 300& 600& 300 & 668&3600&\nodata  &\nodata  & Dec-99 &1080 & 600& 675\\
Arp291 & Feb-00 &600  & 480& 300& 300 & 656&3600&\nodata  &\nodata  & Dec-99 &600  & 660& 450\\
Arp148 & Dec-01 &1800 &1200& 600& 600 & 680&3600&668      &1800     & Mar-02 &1020 & 900& 1050\\
VIIZw466& Feb-00&600  &600 & 600& 300 & 686&3600&\nodata  &\nodata  & 2MASS  &\nodata & \nodata& \nodata \\
IZw45  & Feb-00 &900  &900 & 600& 600 & 674&3600&\nodata  &\nodata  & 2MASS  &\nodata & \nodata & \nodata \\
NGC5410& Feb-02 &1800 &1200&1200& 600 & 662&7200&674      &4800     & Mar-02 &1440 & 1800& 1320\\
\enddata
\tablenotetext{a}{Columns 3--6, 8, 10 and 12--14 contain the net exposure times in seconds 
on the objects in the indicated filters. The central wavelength of the on and off-line
\ha\ filters are given under the columns 7 and 9, respectively. The month and year of the 
optical and NIR observations are listed under the columns 2 and 11, respectively. For 3 
galaxies, the NIR data were taken from the 2MASS atlas image database.}
\tablenotetext{b}{The observing run for this filter is different from the rest of
the filters: Feb-00 (Arp145-I), Dec-01 (NGC985-\ha\ and Arp147-\ha), Dec-00 (IIZw28-\ha).}
\end{deluxetable}
\end{center}

\begin{center}
\begin{deluxetable}{lcccccccccc}
\tabletypesize{\scriptsize}
\tablewidth{0pc}
\tablecaption{Transformation coefficients for optical observations\tablenotemark{a}} 
\tablehead{
\colhead{Quantity} &\colhead{B} & \colhead{V} & \colhead{R} & \colhead{I} 
                      & \colhead{H$\alpha$656} & \colhead{H$\alpha$662} & \colhead{H$\alpha$668}
                      & \colhead{H$\alpha$674} & \colhead{H$\alpha$680} & \colhead{H$\alpha$686} \\  
}
\startdata
Feb-00 (M67-DA) &     &       &       &       &      &      &       &       &       &        \\
$\alpha$     & 23.30  & 23.74 & 22.47 & 23.35 & 9.82 & 8.72 & 8.54  & 8.50  & 8.51  & 9.22   \\
 rms         &  0.01  &  0.01 & 0.01  &  0.01 & 0.02 & 0.03 & 0.03  & 0.03  & 0.03  & 0.03   \\ 
Dec-01 (PG0231+051) & &       &       &       &      &      &       &       &       &        \\
$\alpha$     & 23.04  & 23.61 & 22.40 & 23.19 & 8.92 & 7.95 & 7.94  & 7.75  & 7.85  & 8.26   \\
 rms         &  0.04  &  0.03 &  0.07 &  0.04 & 0.01 & 0.01 & 0.02  & 0.02  & 0.01  & 0.02   \\ 
Feb-02 (M67-DA) &     &       &       &       &      &      &       &       &       &        \\
$\alpha$     & 22.97  & 23.52 & 22.38 & 23.22 & 10.23& 9.10 & 8.97  & 8.94  & 9.04  & 9.63   \\
 rms         &  0.03  &  0.04 & 0.05  &  0.04 &  0.03& 0.03 & 0.02  & 0.02  & 0.02  & 0.01   \\ 
Oct-02 (PG0231+051) & &       &       &       &      &      &       &       &       &        \\
$\alpha$     & 22.60  & 23.21 & 22.15 & 23.35 & 13.01& 13.04& 13.23 & 13.16 & 13.25 &\nodata \\
 rms         &  0.03  &  0.02 & 0.05  &  0.01 & 0.03 &  0.03&  0.03 & 0.03  & 0.03  & \nodata \\ 
             &        &       &       &       &      &      &       &       &       &        \\
$\beta$      &$-$0.11 &  0.06 &$-$0.50&  0.03 & 0.00 & 0.00 & 0.00  &  0.00 &  0.00 &  0.00   \\ 
\enddata
\tablenotetext{a}{See section 2.4 for the definition of $\alpha$ and $\beta$. Values of 
$\alpha$ for H$\alpha$ filters are in units of $10^{-16}$~erg\,s$^{-1}$\,cm$^{-2}$/(count s$^{-1}$).
}
\end{deluxetable}
\end{center}

\begin{center}
\begin{deluxetable}{lrrrrrll}
\tabletypesize{\scriptsize}
\tablecaption{Parameters of the best-matched ellipses to the rings}
\tablehead{
\colhead{Galaxy} & \colhead{$\alpha$ (J2000)} & \colhead{$\delta$ (J2000)}  
                 & \colhead{$a$}              & \colhead{$\epsilon$} 
                 & \colhead{PA} 
           & \colhead{$I_{\rm H\alpha}(5\sigma)$\tablenotemark{a}} 
           & \colhead{$\mu_B(5\sigma)$\tablenotemark{a}} \\
\colhead{}       & \colhead{$\circ$}  & \colhead{$\circ$} & \colhead{\arcsec}
                 & \colhead{}         & \colhead{$\circ$} &  \colhead{}
}
\startdata
 Arp141     & 108.5953 &   73.46458 & 26  & 0.16 &$-$17 & 0.194  & 24.94 \\
 Arp142     & 144.4309 &   2.755613 & 38  & 0.62 &   40 & 1.092  & 24.72 \\
 Arp143     & 116.7312 &   39.01426 & 36  & 0.32 &   60 & 0.762  & 24.62 \\
 Arp144     & 1.6132 &$-$13.41494 & 16  & 0.65 &$-$45   & 1.717  & 24.35\\
Arp145\tn{b}&  35.7846 &   41.37402 & 22  & 0.11 &   35 & 0.784  & 23.75\\
 Arp147     &  47.8264 &    1.31554 &  7  & 0.05 &   90 & 1.102  & 24.61\\
 Arp148     & 165.9702 &   40.85075 &  9  & 0.23 &   28 & 0.982  & 24.58\\
Arp291\tn{b}& 160.7041 &   13.45734 & 27  & 0.35 &$-$81 & 1.525  & 24.62\\
 IIHz4      & 134.6393 &   37.08633 & 14  & 0.22 &$-$77 & 1.123  & 24.83\\
 IZw45      & 193.2774 &   36.81840 & 10  & 0.42 &   90 & 2.147  & 24.74\\
 IIZw28     &  75.4244 &    3.57377 &  6  & 0.10 &   60 & 1.580  & 24.51\\
 VIIZw466   & 188.0200 &   66.40405 & 11  & 0.28 &    0 & 1.659  & 24.38\\
NGC985\tn{b}&  38.6558 & $-$8.78552 & 14  & 0.20 &   72 & 1.202  & 24.82\\
 NGC2793    & 139.1924 &   34.43203 & 19  & 0.17 &   90 & 1.944  & 24.37\\
 NGC5410    & 210.2242 &   40.98832 & 25  & 0.49 &   60 & 0.668  & 24.75\\
\enddata
\tablenotetext{a}{Limiting (5$\sigma$) \ha\ intensity in units 
        of $10^{-16}$\,erg\,cm$^{-2}$\,s$^{-1}$\,arcsec$^{-2}$ (column 7)
and $B$-band surface brightness in mag\,arcsec$^{-2}$ (column 8) of the
azimuthally averaged profiles.}
\tablenotetext{b}{Ellipse parameters based on the $B$-band ring.}
\end{deluxetable}
\end{center}

\begin{center}
\begin{deluxetable}{llccccccc}
\tabletypesize{\scriptsize}
\tablewidth{0pc}
\tablecaption{Photometry of the sample galaxies\tablenotemark{a}}
\tablehead{
\colhead{Galaxy} & \colhead{R$_{B25}$[\arcsec]} & \colhead{$B$} & \colhead{$B-V$} & \colhead{$B-R$} & \colhead{$V-I$}
                 & \colhead{$V-K$} & \colhead{$J-K$} & \colhead{$H-K$}
}
\startdata
Arp141 &  43 & 14.39 $\pm$0.07 &  0.54 $\pm$0.01 &  1.02$\pm$ 0.01 &  0.91 $\pm$0.01 &  2.70 $\pm$0.06 & 0.77$\pm$ 0.02 & 0.20$\pm$0.04 \\
Arp142 &  68 & 14.18 $\pm$0.07 &  0.76 $\pm$0.02 &  1.69$\pm$ 0.02 &  1.24 $\pm$0.01 &  3.67 $\pm$0.05 & 1.05$\pm$ 0.01 & 0.28$\pm$0.01 \\
Arp143 &  70 & 13.30 $\pm$0.08 &  0.57 $\pm$0.02 &  1.22$\pm$ 0.02 &  0.99 $\pm$0.01 &  2.93 $\pm$0.06 & 0.87$\pm$ 0.01 & 0.28$\pm$0.01 \\
Arp144 &  43 & 14.09 $\pm$0.03 &  0.50 $\pm$0.01 &  1.03$\pm$ 0.01 &  1.07 $\pm$0.01 &  3.07 $\pm$0.02 & 0.90$\pm$ 0.01 & 0.27$\pm$0.01 \\
Arp145 &  40 & 14.41 $\pm$0.31 &  0.82 $\pm$0.18 &  1.30$\pm$ 0.15 &  1.11 $\pm$0.12 &  2.89 $\pm$0.13 & 0.80$\pm$ 0.01 & 0.42$\pm$0.01 \\
Arp147 &  17 & 15.99 $\pm$0.05 &  0.42 $\pm$0.03 &  0.74$\pm$ 0.01 &  0.97 $\pm$0.02 &  2.87 $\pm$0.03 & 1.07$\pm$ 0.01 & 0.43$\pm$0.01 \\
Arp148 &  16 & 16.04 $\pm$0.03 &  0.45 $\pm$0.01 &  0.91$\pm$ 0.01 &  1.03 $\pm$0.02 &  3.13 $\pm$0.02 & 1.05$\pm$ 0.02 & 0.46$\pm$0.02 \\
Arp291 &  43 & 14.23 $\pm$0.07 &  0.45 $\pm$0.01 &  0.79$\pm$ 0.08 &  0.79 $\pm$0.02 &  2.69 $\pm$0.06 & 0.81$\pm$ 0.02 & 0.12$\pm$0.01 \\
IIHz4  &  20 & 16.00 $\pm$0.05 &  0.90 $\pm$0.02 &  1.57$\pm$ 0.04 &  1.30 $\pm$0.01 &  3.36 $\pm$0.03 & 0.85$\pm$ 0.01 & 0.18$\pm$0.01 \\
IZw45  &  23 & 14.95 $\pm$0.04 &  0.68 $\pm$0.02 &  0.88$\pm$ 0.04 &  0.90 $\pm$0.01 &  3.06 $\pm$0.03 & 1.04$\pm$ 0.02 & 0.37$\pm$0.02 \\
IIZw28 &  14 & 15.62 $\pm$0.04 &  0.35 $\pm$0.02 &  0.75$\pm$ 0.03 &  0.73 $\pm$0.04 &  2.30 $\pm$0.03 & 0.80$\pm$ 0.03 & 0.23$\pm$0.03 \\
VIIZw466& 17 & 15.74 $\pm$0.09 &  0.35 $\pm$0.05 &  0.68$\pm$ 0.02 &  0.80 $\pm$0.02 &  2.78 $\pm$0.05 & 0.97$\pm$ 0.03 & 0.03$\pm$0.02 \\
NGC985 &  28 & 14.42 $\pm$0.03 &  0.60 $\pm$0.01 &  1.06$\pm$ 0.04 &  1.21 $\pm$0.07 &  3.43 $\pm$0.03 & 1.21$\pm$ 0.01 & 0.39$\pm$0.01 \\
NGC2793&  37 & 13.77 $\pm$0.04 &  0.38 $\pm$0.01 &  0.64$\pm$ 0.01 &  0.75 $\pm$0.02 &  2.39 $\pm$0.05 & 0.96$\pm$ 0.01 & 0.35$\pm$0.01 \\
NGC5410&  54 & 14.00 $\pm$0.04 &  0.41 $\pm$0.01 &  0.81$\pm$ 0.01 &  0.90 $\pm$0.02 &  2.54 $\pm$0.03 & 0.78$\pm$ 0.01 & 0.28$\pm$0.02 \\
\enddata
\tablenotetext{a}{Magnitudes inside the R$_{B25}$ ellipse which are shown in 
the figures in the Appendix. Stars, if any, are masked in doing photometry.
}
\end{deluxetable}
\end{center}

\begin{center}
\begin{deluxetable}{lccccccr}
\tabletypesize{\scriptsize}
\tablewidth{0pc}
\tablecaption{Photometry of the companions of sample galaxies}
\tablehead{
\colhead{Galaxy} &  \colhead{$B$} & \colhead{$B-V$} & \colhead{$B-R$} & \colhead{$V-I$}
                 & \colhead{$V-K$} & \colhead{$J-K$} & \colhead{$H-K$} \\
}
\startdata
Arp141-G1 & 14.41 $\pm$0.08 & 0.86 $\pm$0.03 & 1.45 $\pm$0.02 & 1.24 $\pm$0.01 & 3.48 $\pm$0.04 & 1.21 $\pm$0.01 &  0.03$\pm$ 0.04 \\ 
Arp142-G1 & 14.89 $\pm$0.09 & 0.91 $\pm$0.03 & 1.92 $\pm$0.04 & 1.33 $\pm$0.01 & 3.63 $\pm$0.06 & 0.95 $\pm$0.01 &  0.23$\pm$ 0.01 \\ 
Arp142-G2 & 16.88 $\pm$0.16 & 0.40 $\pm$0.02 & 1.07 $\pm$0.02 & 1.24 $\pm$0.03 & 2.22 $\pm$0.19 & 0.58 $\pm$0.07 &  0.32$\pm$ 0.01 \\ 
Arp143-G1 & 14.29 $\pm$0.11 & 1.05 $\pm$0.06 & 1.88 $\pm$0.07 & 1.25 $\pm$0.01 & 3.42 $\pm$0.05 & 0.98 $\pm$0.01 &  0.19$\pm$ 0.01 \\ 
Arp144-G1 & 14.65 $\pm$0.08 & 0.62 $\pm$0.03 & 1.12 $\pm$0.02 & 1.07 $\pm$0.01 & 2.89 $\pm$0.05 & 0.81 $\pm$0.01 &  0.13$\pm$ 0.01 \\ 
Arp145-G1 & 14.85 $\pm$0.14 & 1.19 $\pm$0.10 & 1.80 $\pm$0.10 & 1.30 $\pm$0.02 & 3.38 $\pm$0.03 & 0.92 $\pm$0.01 &  0.29$\pm$ 0.01 \\ 
Arp147-G1 & 16.39 $\pm$0.04 & 0.88 $\pm$0.03 & 1.59 $\pm$0.02 & 1.25 $\pm$0.01 & 3.56 $\pm$0.01 & 1.07 $\pm$0.01 &  0.37$\pm$ 0.01 \\ 
Arp148-G1 & 15.36 $\pm$0.02 & 0.54 $\pm$0.01 & 1.06 $\pm$0.01 & 1.10 $\pm$0.01 & 3.50 $\pm$0.01 & 1.13 $\pm$0.01 &  0.48$\pm$ 0.02 \\ 
Arp291-G1 & 17.90 $\pm$0.24 & 0.70 $\pm$0.09 & 1.06 $\pm$0.14 & 1.01 $\pm$0.02 & 3.12 $\pm$0.16 & 0.94 $\pm$0.06 &  0.23$\pm$ 0.01 \\ 
IIHz4-R2\tablenotemark{a}      
          & 16.34 $\pm$0.12 & 0.51 $\pm$0.02 & 1.17 $\pm$0.06 & 1.95 $\pm$0.08 & 2.99 $\pm$0.11 & 0.53 $\pm$0.02 &$-$0.05$\pm$ 0.01\\ 
IZw45-G1  & 17.27 $\pm$0.04 & 0.59 $\pm$0.01 & 0.87 $\pm$0.04 & 0.65 $\pm$0.02 & 2.16 $\pm$0.08 & 0.52 $\pm$0.09 &  0.02$\pm$ 0.01 \\ 
IIZw28-G1?& 20.18 $\pm$0.33 & 1.46 $\pm$0.73 & 1.38 $\pm$0.25 & 1.06 $\pm$0.18 & 3.54 $\pm$0.13 & 1.40 $\pm$0.10 &  0.62$\pm$ 0.02 \\ 
VIIZw466-G1& 16.42 $\pm$0.08& 1.02 $\pm$0.06 & 1.74 $\pm$0.04 & 1.25 $\pm$0.01 & 3.28 $\pm$0.02 & 0.89 $\pm$0.02 &  0.16$\pm$ 0.02 \\ 
VIIZw466-G2& 16.93 $\pm$0.14& 0.64 $\pm$0.10 & 1.02 $\pm$0.03 & 1.02 $\pm$0.01 & 2.69 $\pm$0.06 & 0.66 $\pm$0.05 &  0.32$\pm$ 0.02 \\ 
NGC2793-G1& 17.07 $\pm$0.10 & 0.65 $\pm$0.02 & 1.17 $\pm$0.05 & 0.87 $\pm$0.02 & 2.42 $\pm$0.08 & 0.62 $\pm$0.03 &$-$0.19$\pm$ 0.01\\ 
NGC5410-G1& 15.52 $\pm$0.07 & 0.40 $\pm$0.01 & 0.78 $\pm$0.01 & 0.90 $\pm$0.03 & 2.73 $\pm$0.06 & 1.06 $\pm$0.03 &  0.53$\pm$ 0.01 \\ 
\enddata
\tablenotetext{a}{Major contribution to the photometry comes from the nucleus of R2.}
\end{deluxetable}
\end{center}

\begin{center}
\begin{deluxetable}{lrrrrrrc}
\tabletypesize{\scriptsize}
\tablewidth{0pc}
\tablecaption{Star formation properties of ring galaxies}
\tablehead{
\colhead{Galaxy} & \colhead{$\log f$(\hanii)}  
                 & \colhead{EW(\hanii)}      
                 & \colhead{$\log f$(FIR)}    
                 & \colhead{f(20cm)}   
                 & \colhead{SFR(FIR)} & \colhead{A$_{\rm v}$} 
                 & \colhead{frac(th)} \\ 
\colhead{}       & \colhead{erg\,cm$^{-2}$\,s$^{-1}$}
                 & \colhead{\AA}   
                 & \colhead{W\,m$^{-2}$}   
                 & \colhead{mJy}  
                 & \colhead{\msun\,yr$^{-1}$} & \colhead{mag}
                 & \colhead{}  
}
\startdata
Arp141  &  $-12.91 \pm$  0.05 & 12.8  $\pm$  2.8 &$<-13.78$ &$<0.54$&  0.26 &  1.03 &  0.49 \\ 
Arp142  &  $-12.47 \pm$  0.07 & 20.9  $\pm$  7.2 & $-12.91$ &  2.30 & 10.74 &  2.50 &  0.86 \\ 
Arp143  &  $-12.14 \pm$  0.05 & 23.0  $\pm$  8.0 & $-12.73$ &  4.15 &  5.60 &  1.91 &  0.70 \\ 
Arp144  &  $-12.64 \pm$  0.09 & 19.5  $\pm$  4.6 & $-12.96$ & 13.20 &  6.86 &  2.88 &  0.13 \\ 
Arp145  &  $-13.45 \pm$  0.20 &  1.1  $\pm$  0.4 & $-13.30$ &  1.61 &  2.69 &  4.39 &  0.50 \\ 
Arp147  &  $-12.61 \pm$  0.04 &175.4  $\pm$ 15.8 & $-13.31$ &  2.00 &  8.58 &  1.40 &  0.39 \\ 
Arp148  &  $-12.92 \pm$  0.09 & 37.6  $\pm$  4.8 & $-12.49$ & 23.80 & 65.06 &  5.47 &  0.22 \\ 
Arp291  &  $-12.49 \pm$  0.12 & 55.3  $\pm$ 24.2 & $-13.59$ &\nodata&  0.06 &$<0.75$& \nodata \\ 
IIHz4   &  $-13.09 \pm$  0.11 & 26.8  $\pm$  5.7 &$<-13.54$ &  0.74 &  8.95 &  2.41 &  0.62 \\ 
IZw45   &  $-12.43 \pm$  0.04 & 90.1  $\pm$ 23.3 & $-13.03$ &  9.99 & 12.10 &  2.01 &  0.15 \\ 
IIZw28  &  $-12.79 \pm$  0.05 & 81.5  $\pm$ 11.9 &$<-13.56$ &  1.66 &  3.82 &  1.30 &  0.27 \\ 
VIIZw466&  $-13.14 \pm$  0.08 & 49.5  $\pm$ 19.7 &$<-13.69$ &  0.83 &  7.80 &  2.16 &  0.39 \\ 
NGC985  &  $-12.23 \pm$  0.02 & 77.2  $\pm$ 10.1 & $-13.16$ & 10.00 & 21.61 &  0.85 &  0.11 \\ 
NGC2793 &  $-12.39 \pm$  0.08 & 43.7  $\pm$ 14.5 & $-13.32$ &$<0.60$&  0.24 &  0.91 &  1.27 \\ 
NGC5410 &  $-12.39 \pm$  0.07 & 44.2  $\pm$  6.8 & $-13.30$ &  1.73 &  1.34 &  0.97 &  0.46 \\ 
\enddata
\end{deluxetable}
\end{center}

\begin{center}
\begin{deluxetable}{ll} 
\tablecaption{Mis-classified candidate ring galaxies}
\tablehead{ 
\colhead{Galaxy} & \colhead{Reasons for rejection}               \\
}
\startdata
Arp142 & only an arc is traced both in the continuum and \ha\ images \\
Arp144 & only broken arcs are traced in both the continuum and \ha\ images \\
Arp145 & diffuse ring present in the continuum, but not traced in \ha \\
Arp291 & ring resembles a one-armed spiral emanating at the end of a bar \\
NGC985 & ring resembles a one-armed spiral emanating at the end of a bar \\
NGC5410 & ring resembles a two-armed spiral structure emanating from the
ends of a bar.\\
\enddata
\end{deluxetable}
\end{center}

\begin{center}
\begin{deluxetable}{lrrrcrrr}
\tabletypesize{\scriptsize}
\tablecaption{Derived properties of ring galaxies}
\tablehead{
\colhead{Galaxy} & \colhead{$M_{\rm K}$}               
                 & \colhead{$M_\ast$}  
                 & \colhead{$M_{\rm gas}$}   
                 & \colhead{$(M\ast + M_{\rm gas})/M_{\rm dyn}$}   
                 & \colhead{$M_{\rm R}/M_{\rm C}$}   
                 & \colhead{$R_{\rm H\alpha}$} 
                 & \colhead{$R_{\rm H\alpha}/R_{\rm B25}$}  \\
\colhead{}       & \colhead{mag}  & \colhead{$10^{10}$\msun} 
                 & \colhead{$10^{10}$\msun}
                 & \colhead{}          & \colhead{}
                 & \colhead{kpc}          & \colhead{}
}
\startdata
   Arp141 & $-$21.78 &   0.83 &  0.49 &  0.99 &  0.52 &  5.2 &  0.61  \\
   Arp143 & $-$23.86 &   6.05 &  1.34 &  0.63 &  1.01 &  9.9 &  0.51  \\
   Arp147 & $-$22.88 &   2.11 &  1.04 &  0.31 &  0.57 &  4.8 &  0.46  \\
   Arp148 & $-$23.23 &   3.10 &  1.54 &  0.54 &  0.50 &  6.2 &  0.58  \\
   IIHz4  & $-$24.44 &  11.57 &  1.34 &\nodata&  3.49 & 12.1 &  0.70  \\
   IZw45  & $-$24.02 &   6.27 &  0.25 &  0.47 & 22.00 &  4.9 &  0.37  \\
   IIZw28 & $-$22.34 &   1.28 &  0.44 & \nodata&$>8.22$&  3.5 &  0.44  \\
  VIIZw466\tablenotemark{a}& $-$23.80 &   4.83 &  0.65 &  1.18 &  2.49 & 11.0 &  0.63  \\
  NGC2793 & $-$20.71 &   0.28 &  0.12 &\nodata& 19.31 &  2.1 &  0.50  \\
\enddata
\tablenotetext{a}{Taking G2 as the companion. $M_R/M_C=0.53$ if G1 is the 
companion.}
\end{deluxetable}
\end{center}

\begin{center}
\begin{deluxetable}{ccc}
\tablewidth{0pc}
\tablecaption{Ring galaxies separated based on the relative mass of the companion} 
\tablehead{ 
\colhead{ $M_{\rm R}/M_{\rm C} < 2.5$ } & 
\colhead{$2.5 <M_{\rm R}/M_{\rm C} < 6.3$} & 
\colhead{ $M_{\rm R}/M_{\rm C}>6.3$} \\
\colhead{1:1} & \colhead{1:4} & \colhead{1:10} \\
}
\startdata
Arp141 & IIHz4  & IZw45 \\
Arp143 & \nodata & NGC2793 \\
Arp147 & \nodata  & IIZw28    \\
Arp148 & \nodata  & \nodata \\
VIIZw466 & \nodata  & \nodata \\
\enddata
\end{deluxetable}
\end{center}

\clearpage

\begin{figure}
\epsscale{1.00}
\plotone{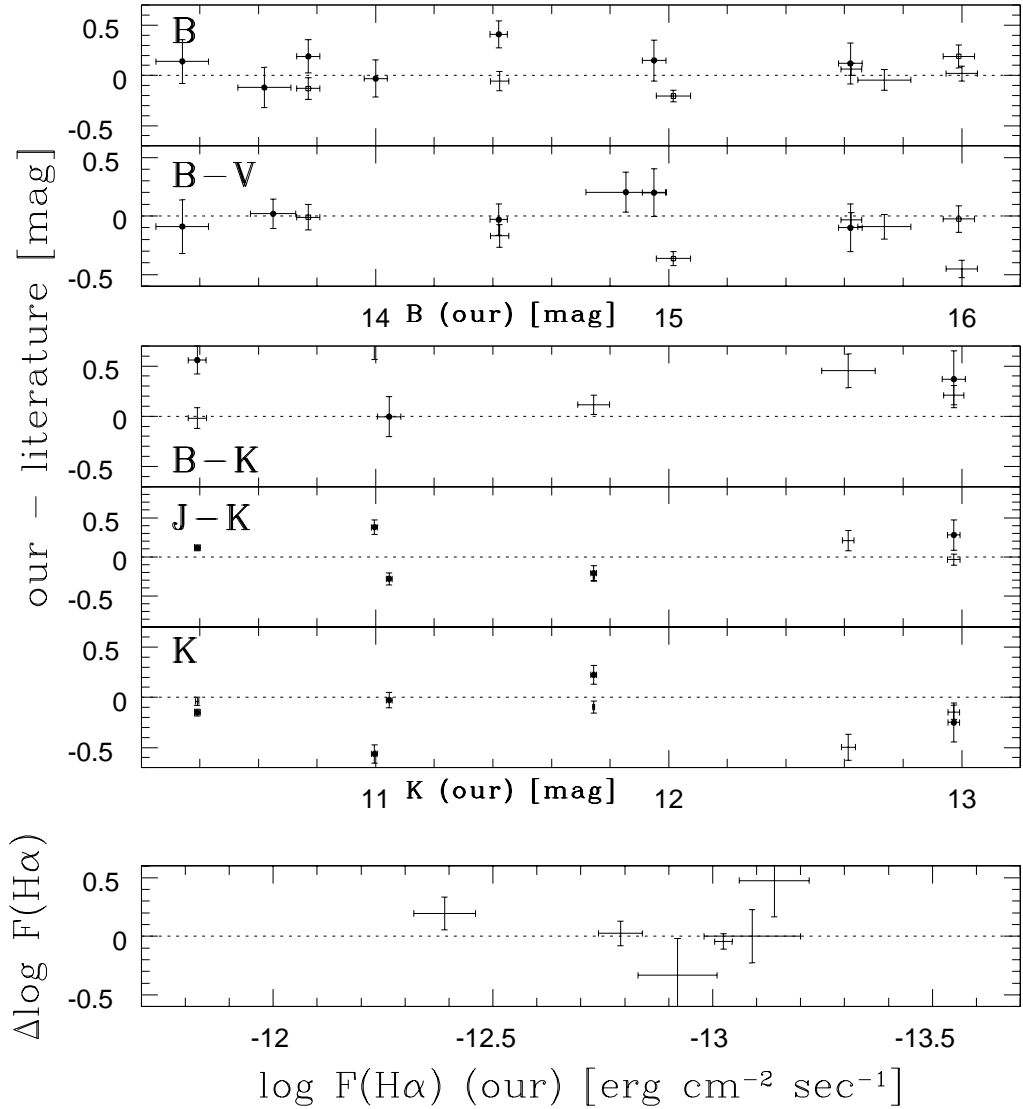}
\caption{Our photometry of ring galaxies in different bands compared to
that in the literature. The principal sources of photometry in $B,V$ (RC3; Buta 1996), 
and $J,H,K$-bands (2MASS) are denoted by {\it filled circles with error bars}.
Other sources of photometry are \citet{maz95} ($B,V$ for two galaxies denoted 
by {\it open squares with error bars}), \citet{app97} ($B,V,J,H,K$ for five
galaxies {\it error bars without a central symbol}). \ha\ photometry 
is from \citet{mar95} for five galaxies, and \citet{hat04} for Arp148.
}
\end{figure}

\begin{figure}
\epsscale{1.00}
\plotone{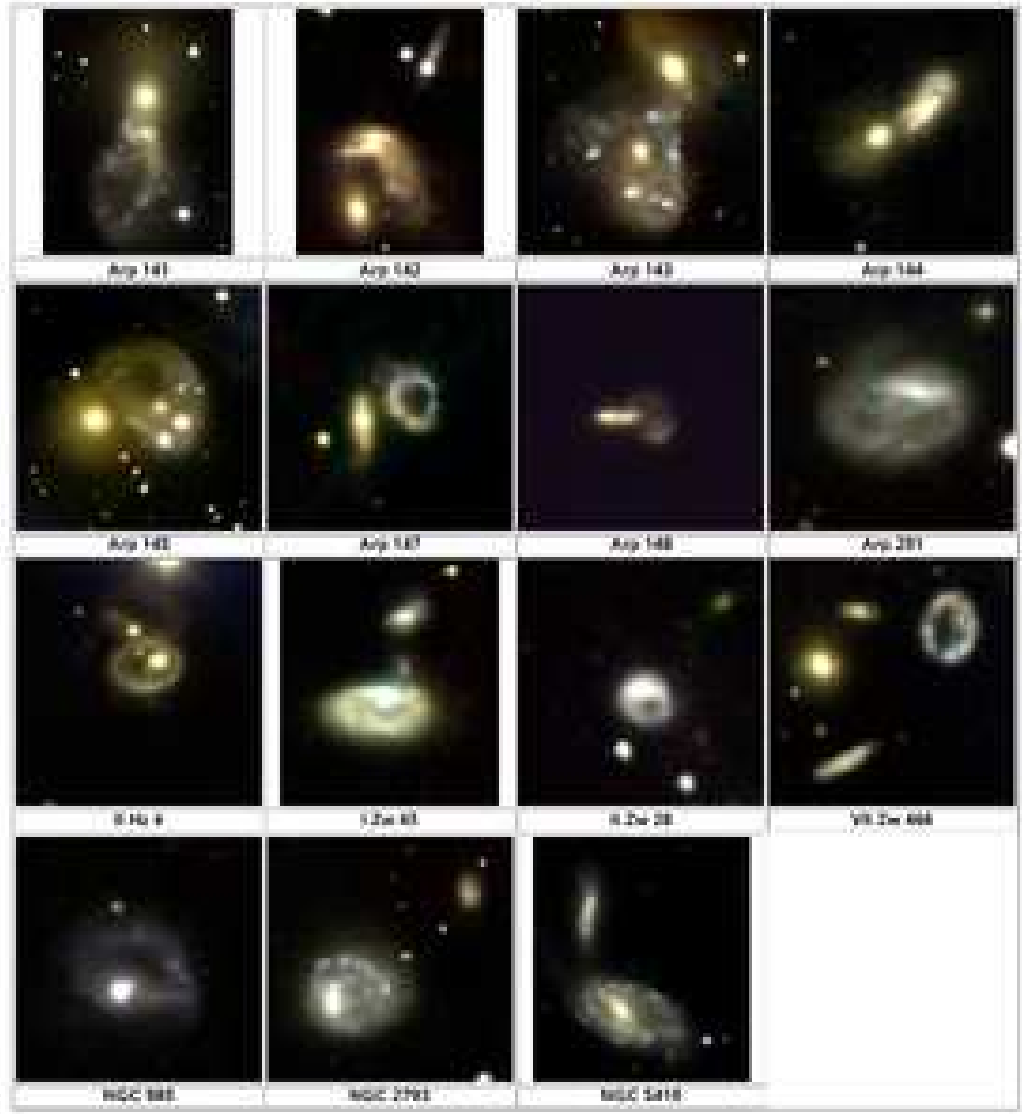}
\caption{Color composite images of ring galaxies obtained by digitally 
combining images in $B$ (blue), $V$ (green) and $R$-band (red) filters. 
Intensities in each of the bands are logarithmically mapped to illustrate
colors of both bright and faint regions. Note that the rings stand out 
on these images as distribution of blue knots. 
The off-centered nuclei and companions are relatively redder.
}
\end{figure}

\begin{figure}
\epsscale{1.20}
\plotone{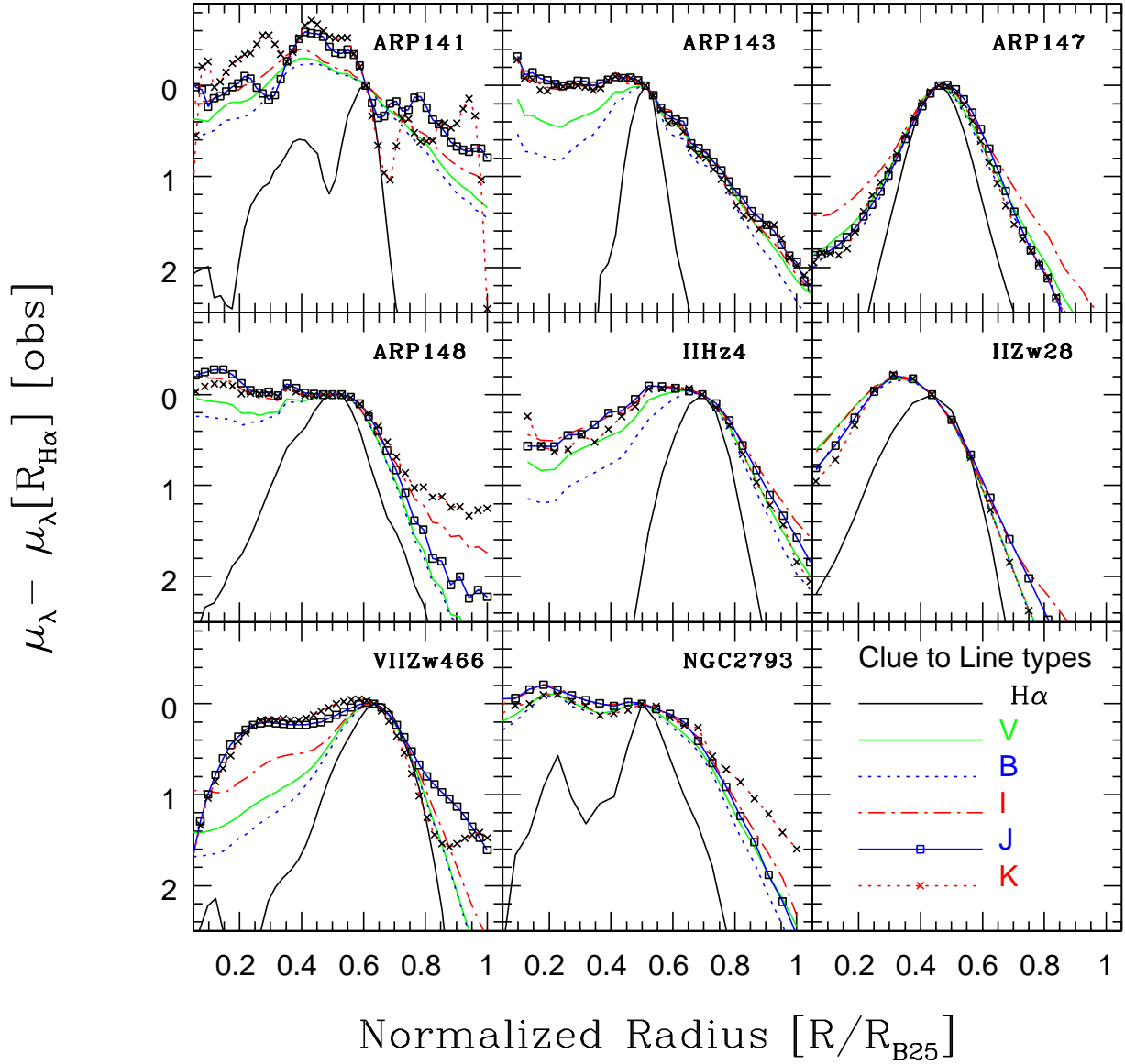}
\caption{Radial surface brightness profiles in $B$ (blue dotted line), $V$ 
(green solid line), $I$ (red dash-dotted line), $J$ (square symbols), $K$ 
(cross symbols), and H$\alpha$ (black solid line) bands 
($J$ and $K$ absent for Arp141 and IZw45 and $K$ absent for 
VIIZw466). Intensities are normalized to their values at the position 
of the ring, whereas the radius is normalized to the disk radius. 
H$\alpha$ profile is the narrowest and the most symmetric of all.
Relative intensity gradients on the inner and outer side of the ring as well
their variations from one galaxy to another can be directly appreciated on
these plots.
}
\end{figure}

\begin{figure}
\epsscale{1.20}
\plotone{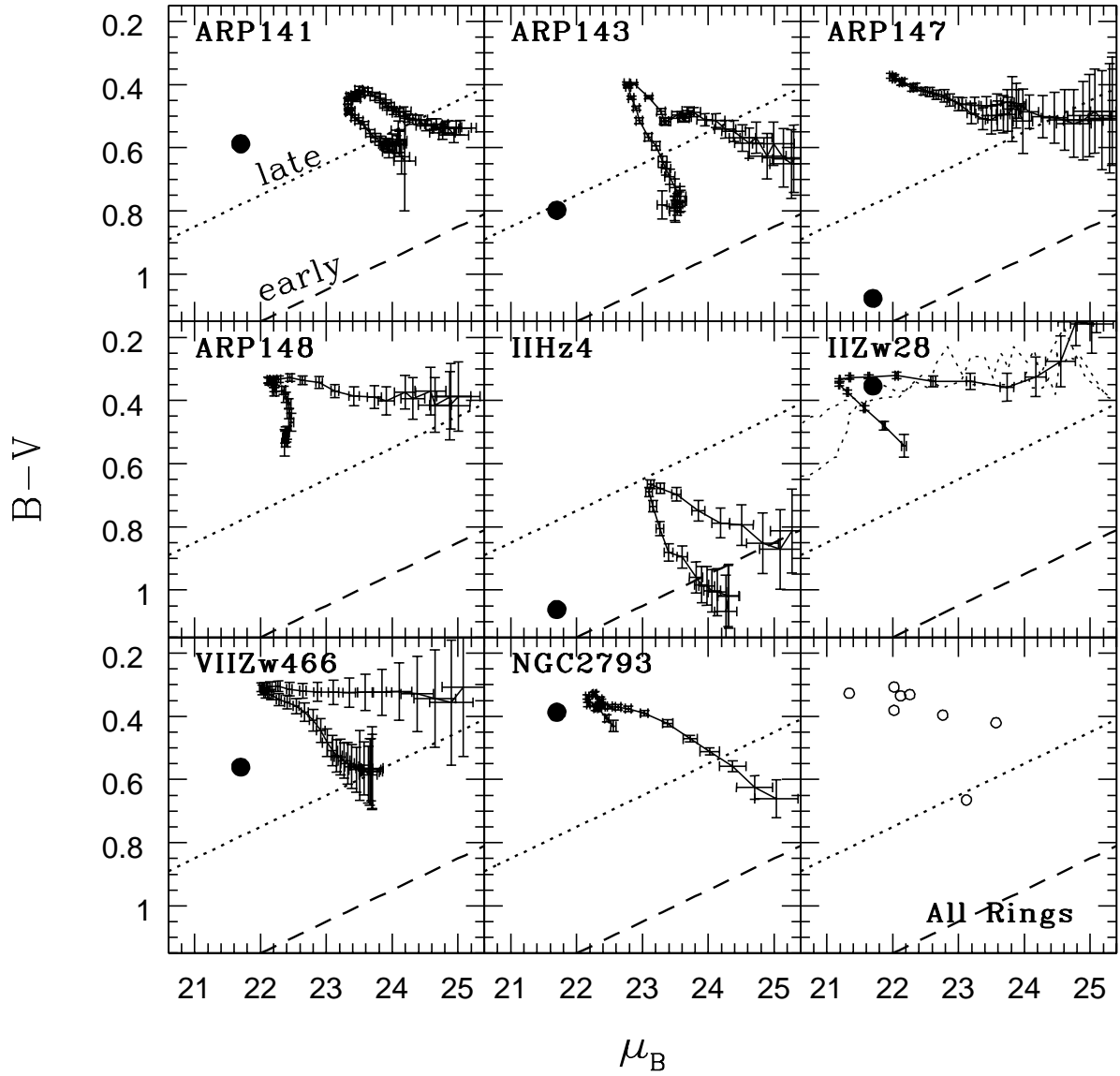}
\caption{Ring galaxies compared to normal late (dotted line) and early 
type (dashed line) galaxies in color vs surface brightness plot. Points 
belonging to successive radial bins are connected by the solid line.
Colors of the off-centered nucleus are indicated by the solid
circle, placed at $\mu_B=21.7$~\magarc, central surface brightness of 
Freeman disks. The ring galaxies trace an inclined $\Lambda$ on this plot, 
with the upper and lower branches correspond to the outer and inner profiles, 
respectively.
}
\end{figure}

\begin{figure}
\epsscale{1.20}
\plotone{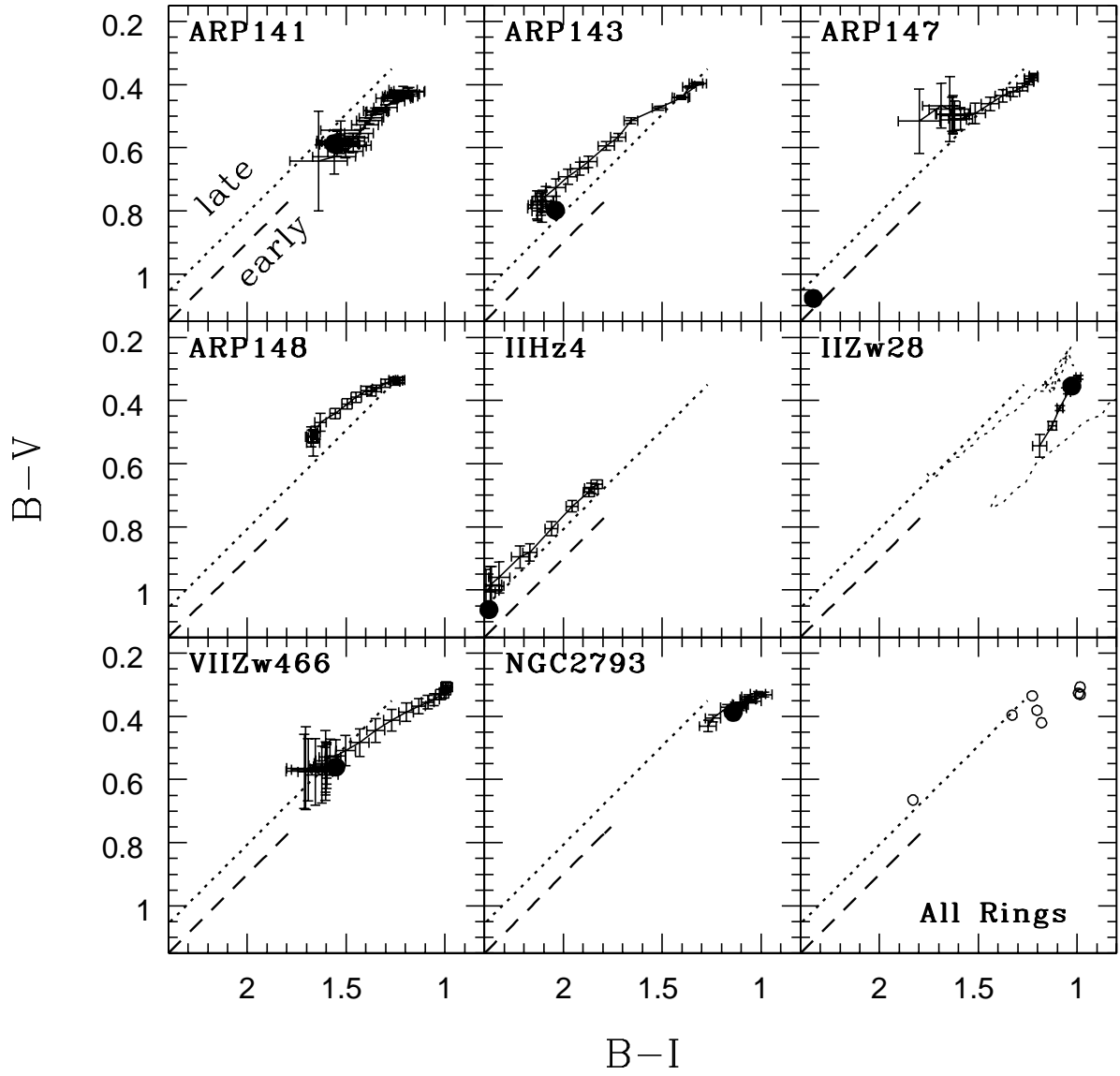}
\caption{Colors interior to the ring are compared to that in normal spiral
galaxies in color-color plots.
Meaning of the symbols is the same as in Figure 4. Ring galaxy colors are 
sequentially ordered with radius. Note that the ring galaxies occupy 
a smaller range of colors as compared to normal spiral galaxies.
}
\end{figure}

\begin{figure}
\epsscale{1.20}
\plotone{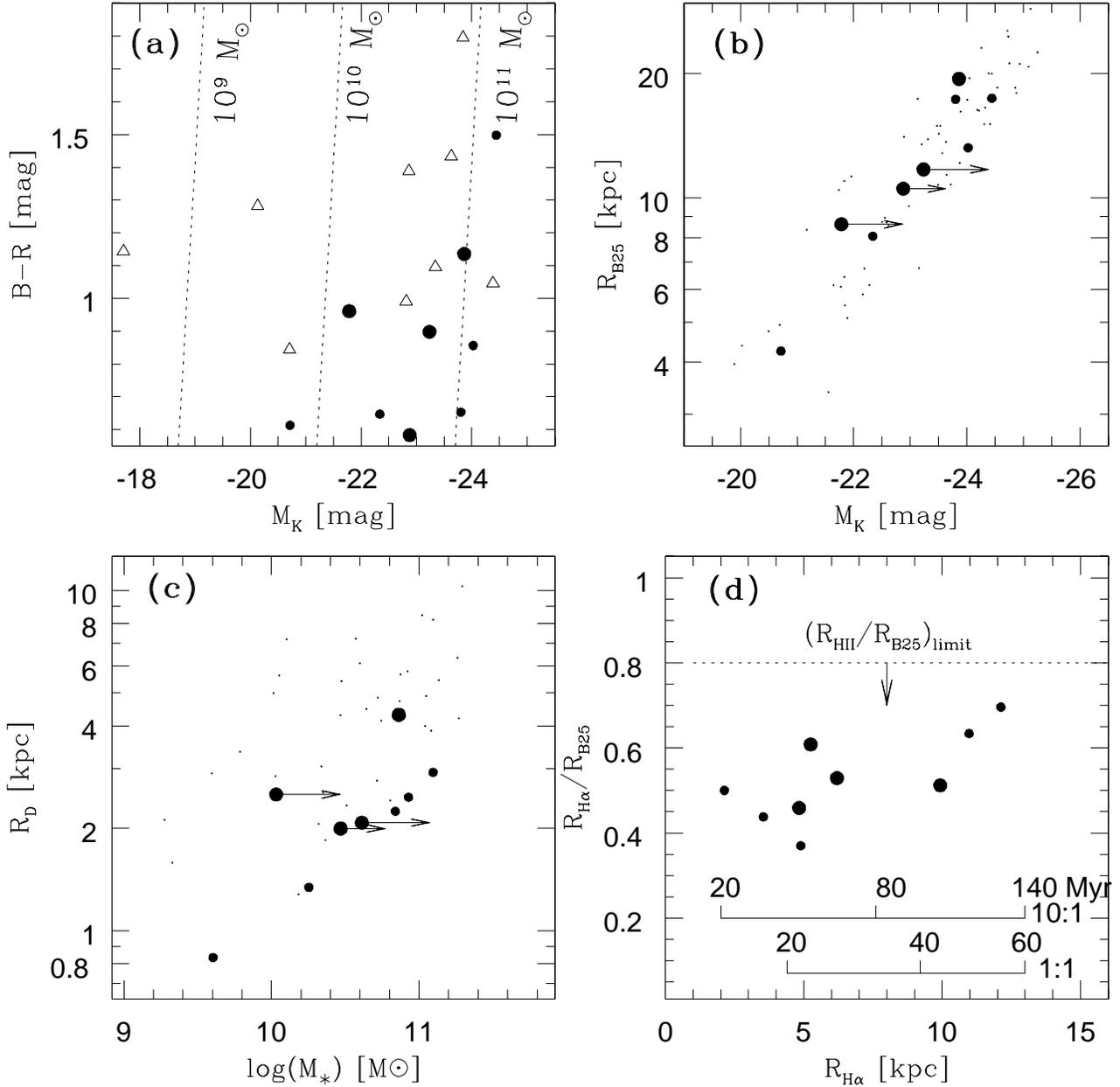}
\caption{
Observed relations for ring galaxies ({\it filled circles}; 
5 galaxies with high mass companions are shown by bigger circles). 
(a) Global $B-R$ vs $M_K$ of ring galaxies and their companions ({\it open 
triangles}). 
The empirical relation of \citet{Bel03} for star-forming galaxies is shown 
by dashed lines for three fixed masses. 
(b) Disk radius, $R_{B25}$ vs $M_K$ relation. 
Points with arrows indicate the maximum amount by which the mass has to be 
increased to match the mass of the companion.
Ring galaxies follow the same relation as that for normal disk galaxies of
\citet{deJ96}, which is shown by the dots. 
(c) Disk scale length external to the ring, $R_{\rm D}$ vs $\log M_\ast$.
For a given disk mass, ring galaxy scale lengths are systematically smaller 
than that for normal galaxies of \citet{deJ96}, which is shown by the dots.
(d) $R_{{\rm H}\alpha}/R_{B25}$ vs $R_{{\rm H}\alpha}$. 
The time required to reach the observed ring sizes are shown for 1:1 (massive
companions) and 10:1 (low-mass companions) models of \citet{ger96}.
The dashed horizontal line shows the mean radius from a sample of 
\citep{mar01} beyond which there are no detectable HII regions in normal 
spiral galaxies. 
}
\end{figure}

\clearpage

\begin{figure}
\epsscale{0.75}
\plotone{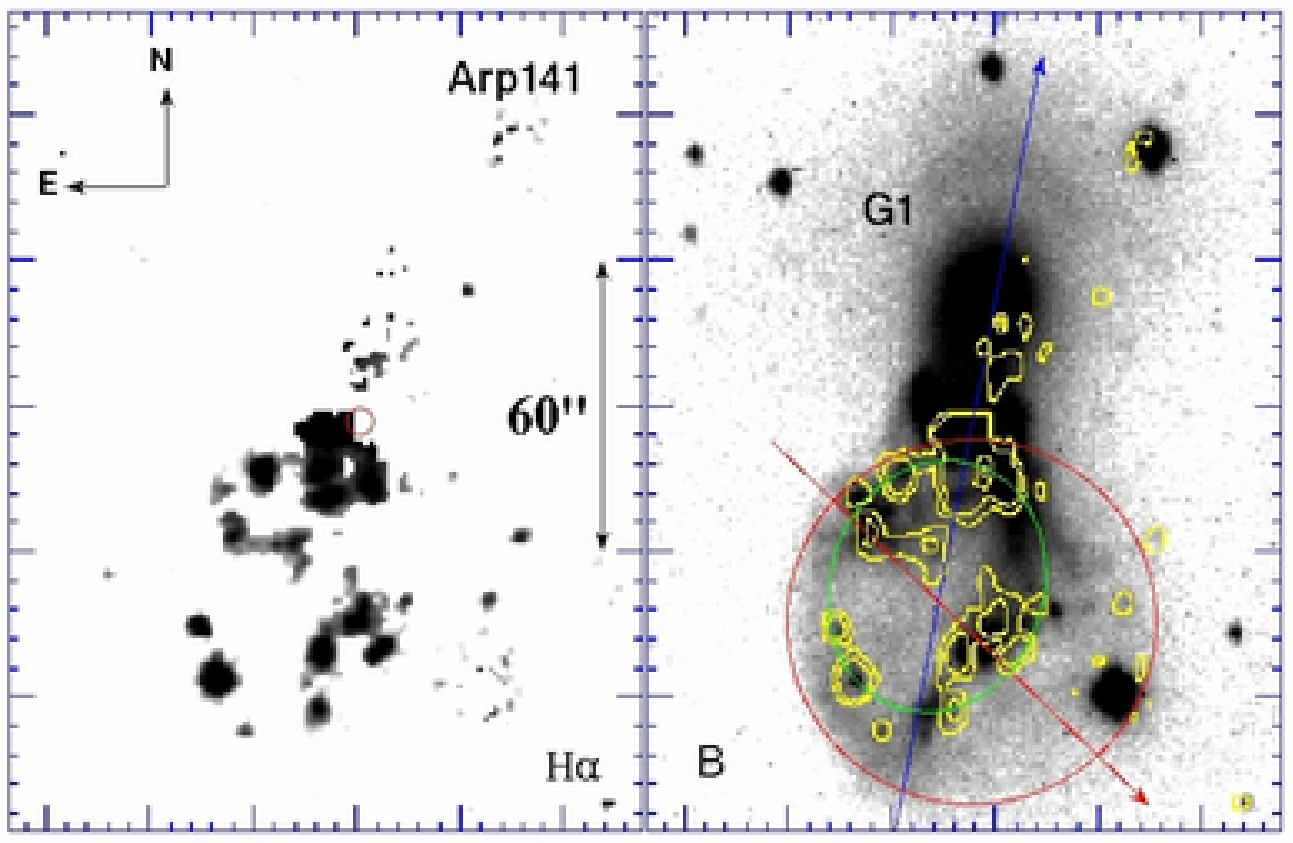}
\centerline{\hspace*{1.5cm}\includegraphics[width=14cm]{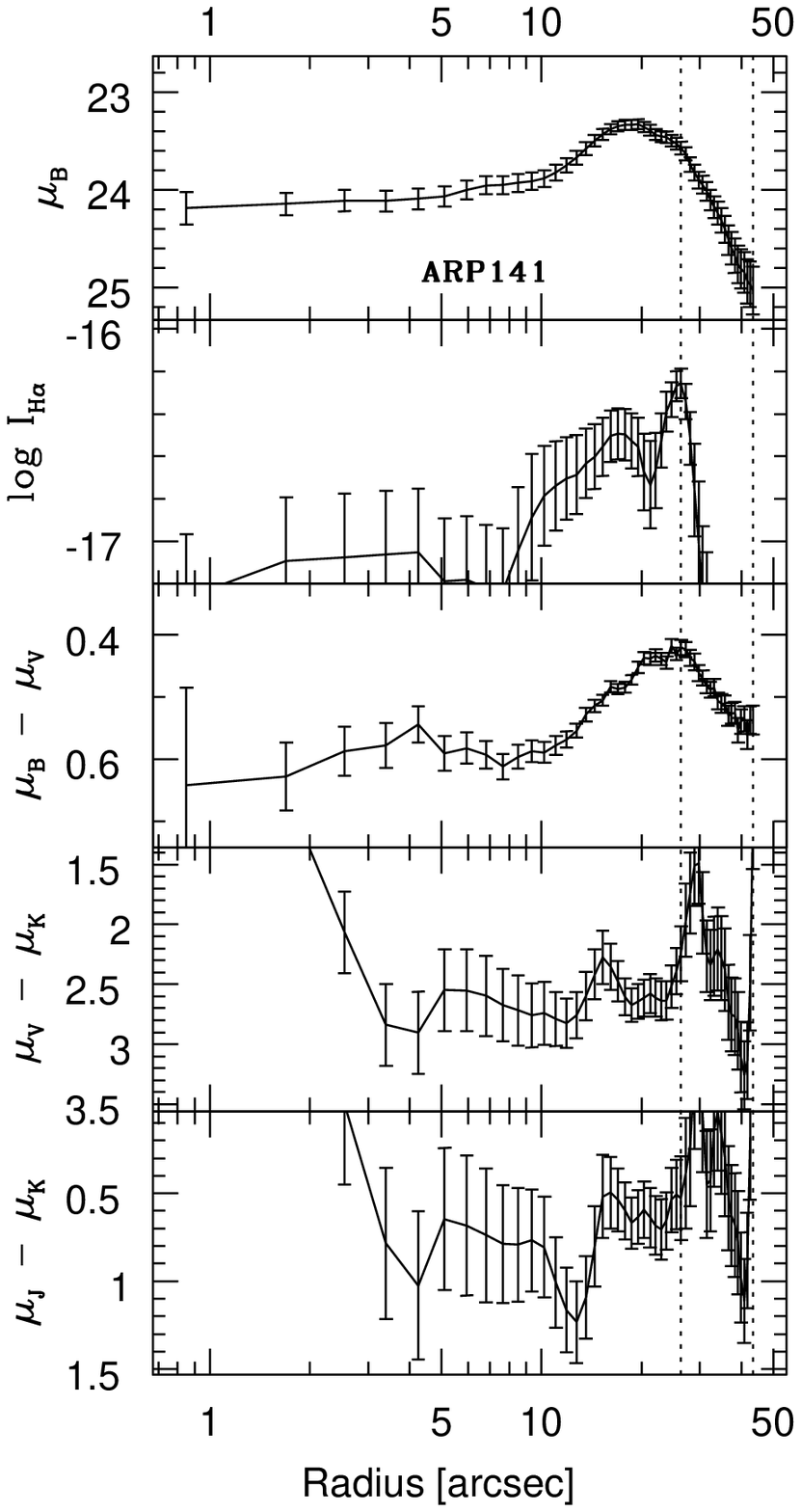}
            \hspace*{-7.5cm}\includegraphics[width=14cm]{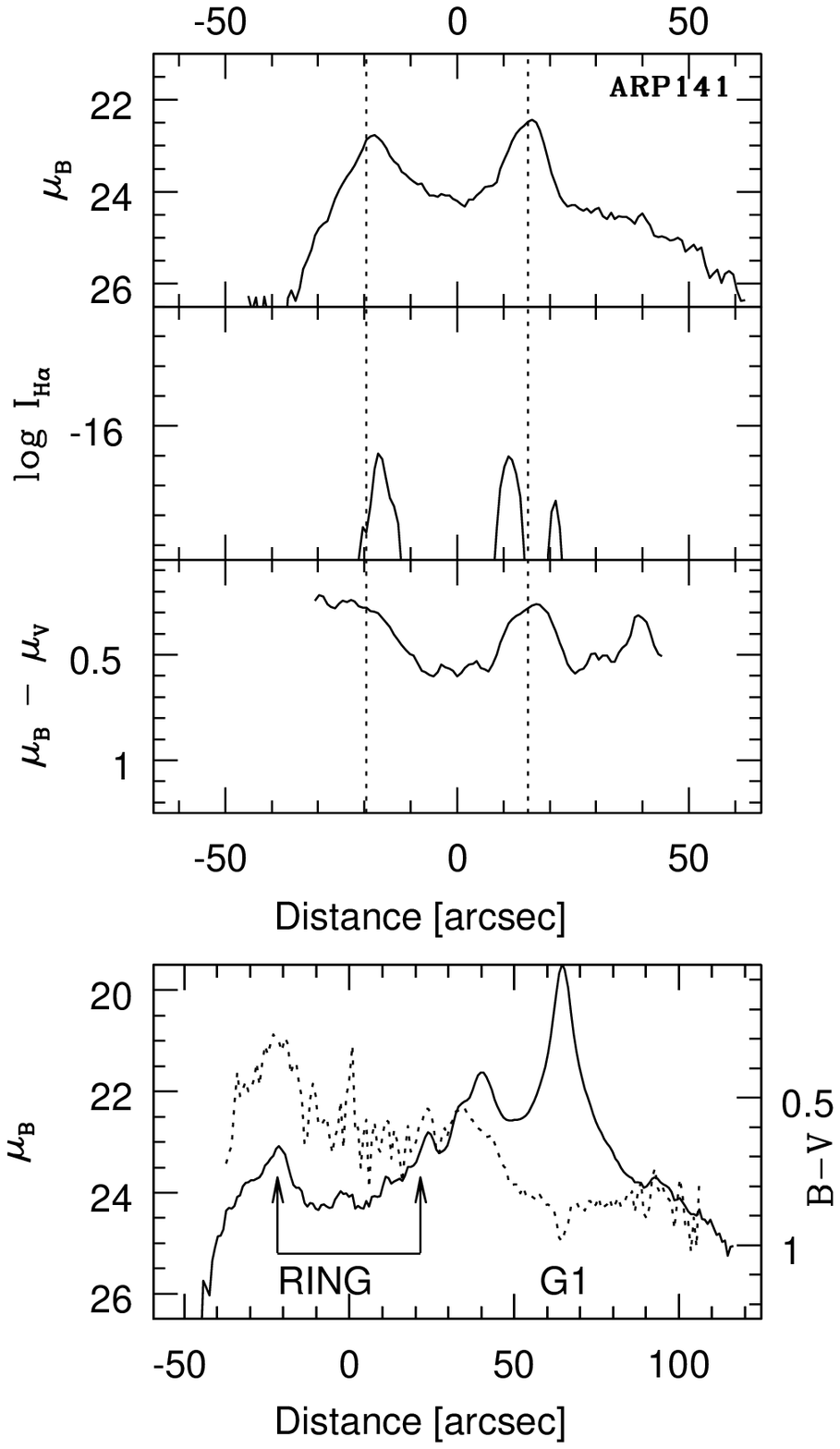}}
\caption{A composite figure to illustrate various morphological features
of Arp141. 
{\it Top:} gray-scale images in \ha\ and $B$-bands on logarithmic 
intensity scale. \ha\ contours are superposed on the $B$-band image. The 
ellipses that best match the ring and the disk are shown by green (inner) 
and red (outer, drawn at 25~mag\,arcsec$^{-2}$) ellipses, respectively.
{\it Bottom left (5 panels):} azimuthally averaged radial profiles of 
intensity (in $B$, and \ha) and colors ($B-V$, $V-K$ and $J-K$). 
{\it Bottom right (top 3 panels):} intensity (in $B$, and \ha) and 
color ($B-V$) cuts across the ring (along the direction marked by solid 
red arrow on the gray-scale image).
{\it Bottom right (bottom most panel):} intensity (in $B$, solid line, 
scale on the left axis) and color ($B-V$, dotted line, scale on the right
axis) cuts along the line joining the ring to the nucleus of the companion
(shown by the dotted blue arrow on the gray-scale image).
See the notes in the appendix for more details on these figures.
}
\end{figure}

\begin{figure}
\epsscale{0.75}
\plotone{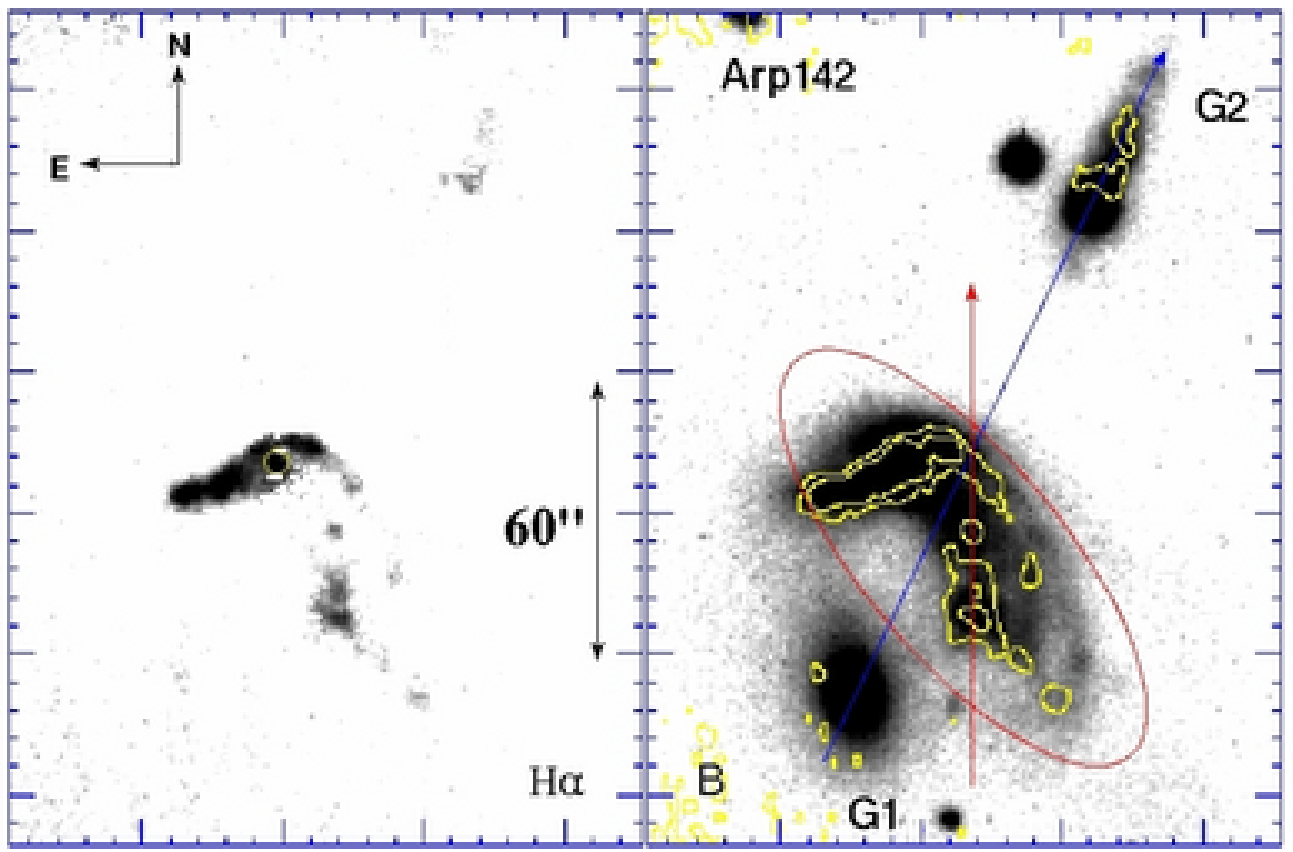}
\centerline{\hspace*{1.5cm}\includegraphics[width=14cm]{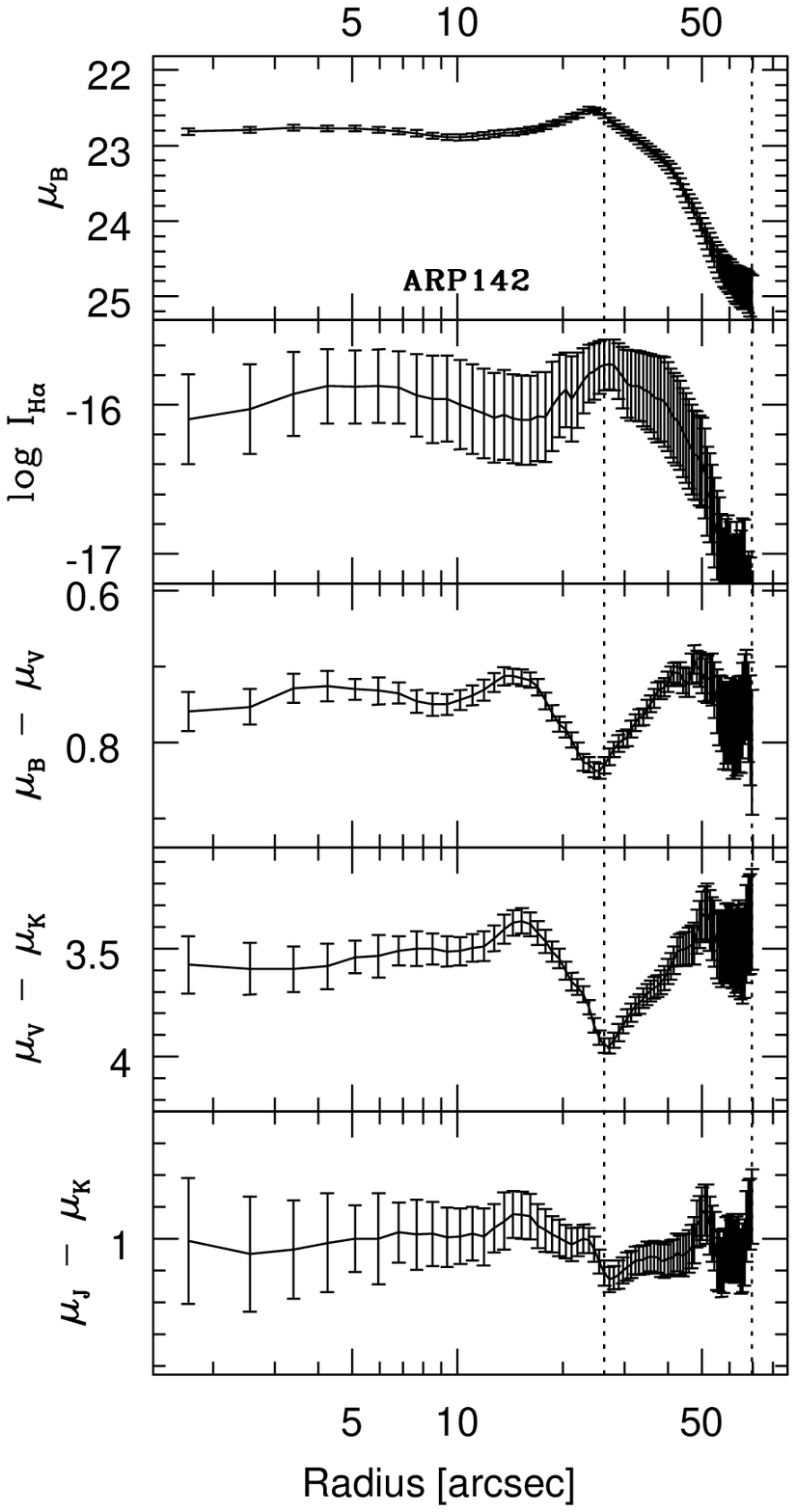}
            \hspace*{-7.5cm}\includegraphics[width=14cm]{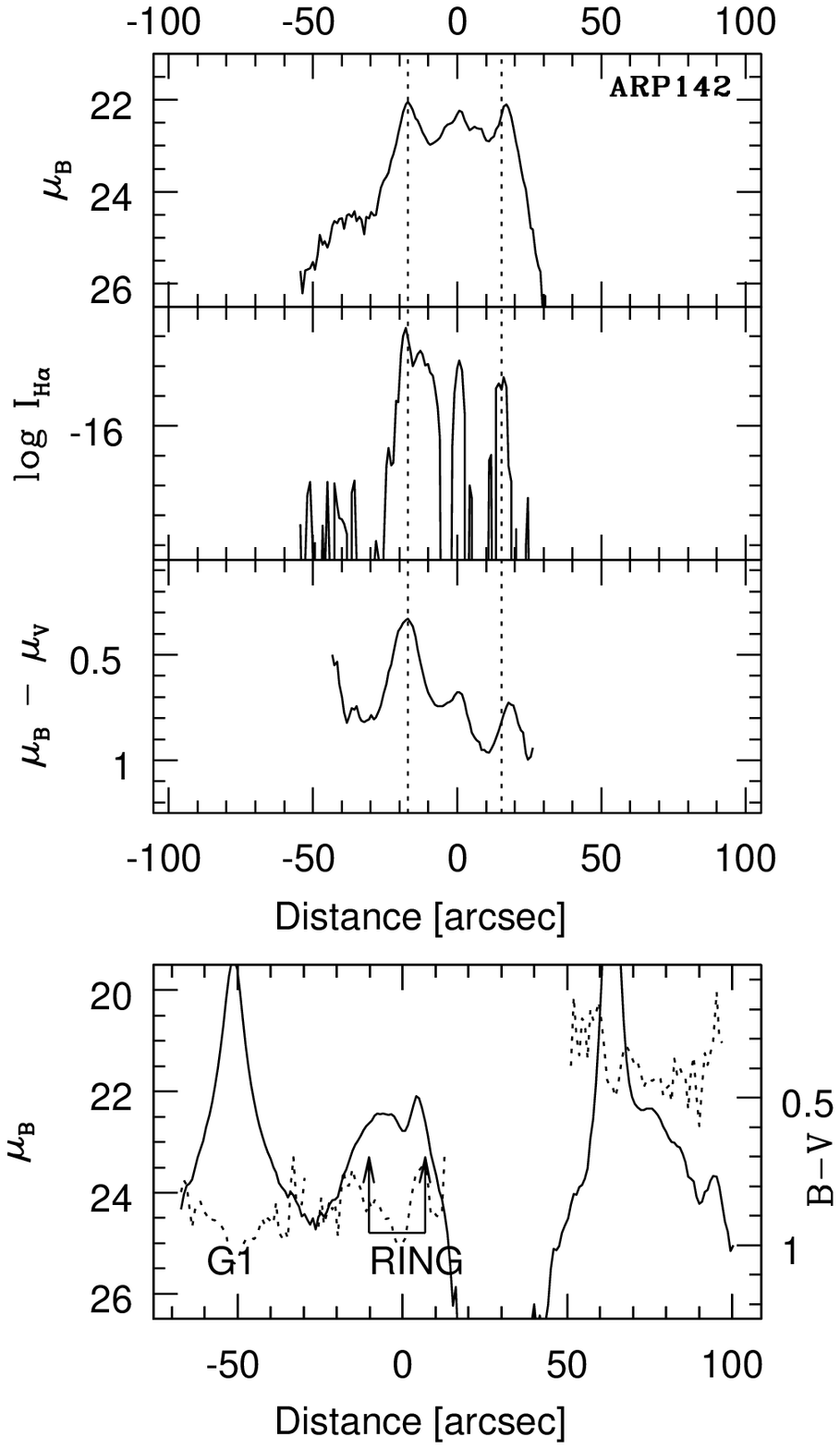}}
\caption{Same as in Figure~7, but for Arp142. There is no ring-like
structure in this galaxy, and the drawn ellipse serves as a guide to
interpret the radial profiles.}
\end{figure}

\begin{figure}
\epsscale{0.75}
\plotone{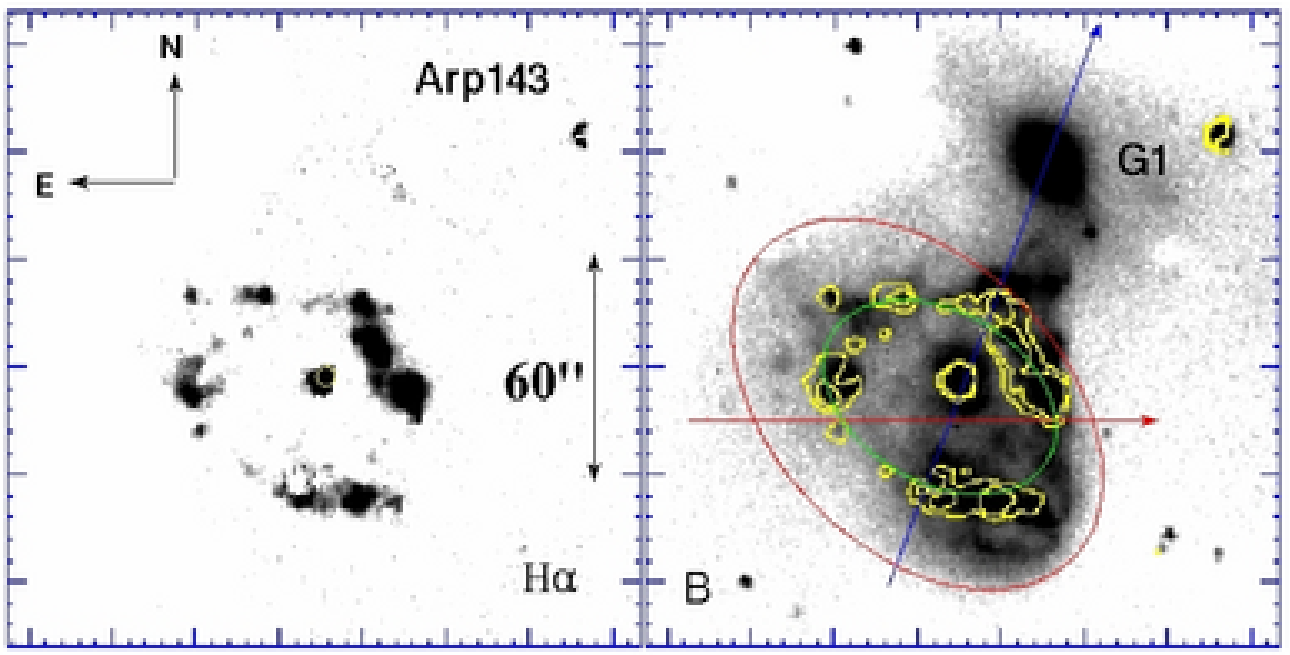}
\centerline{\hspace*{1.5cm}\includegraphics[width=14cm]{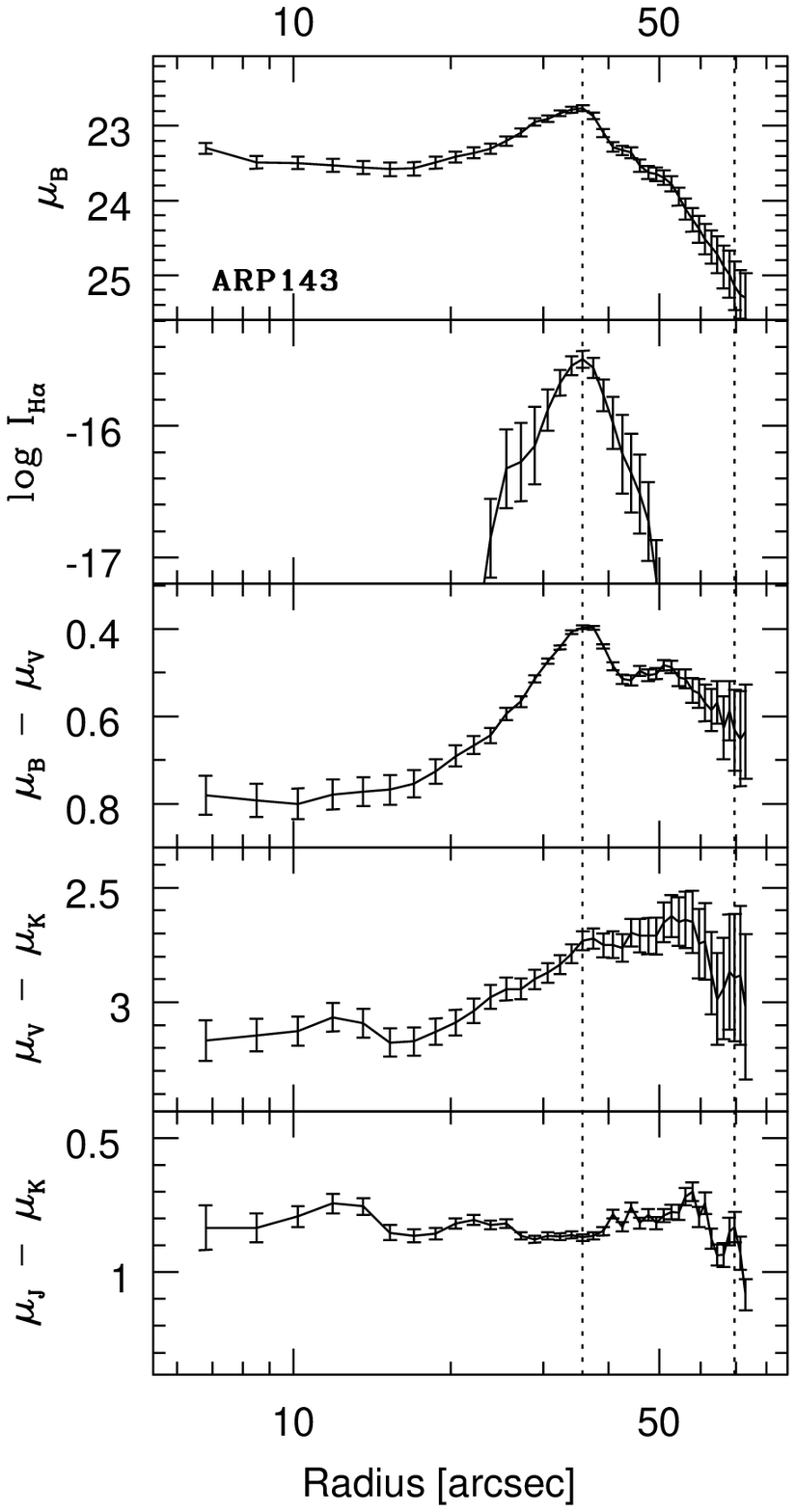}
            \hspace*{-7.5cm}\includegraphics[width=14cm]{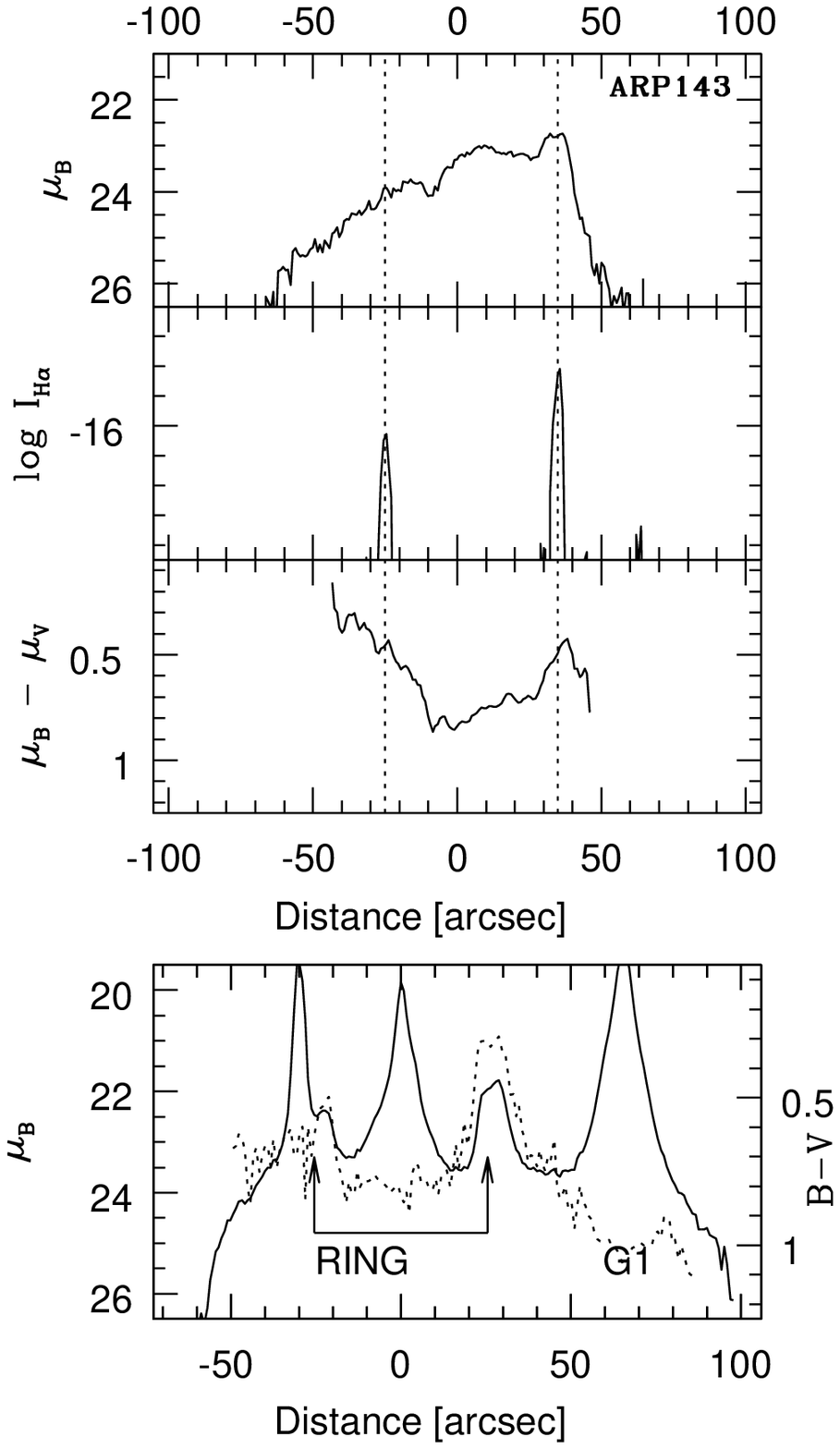}}
\caption{Same as in Figure~7, but for Arp143.}
\end{figure}

\begin{figure}
\epsscale{0.75}
\plotone{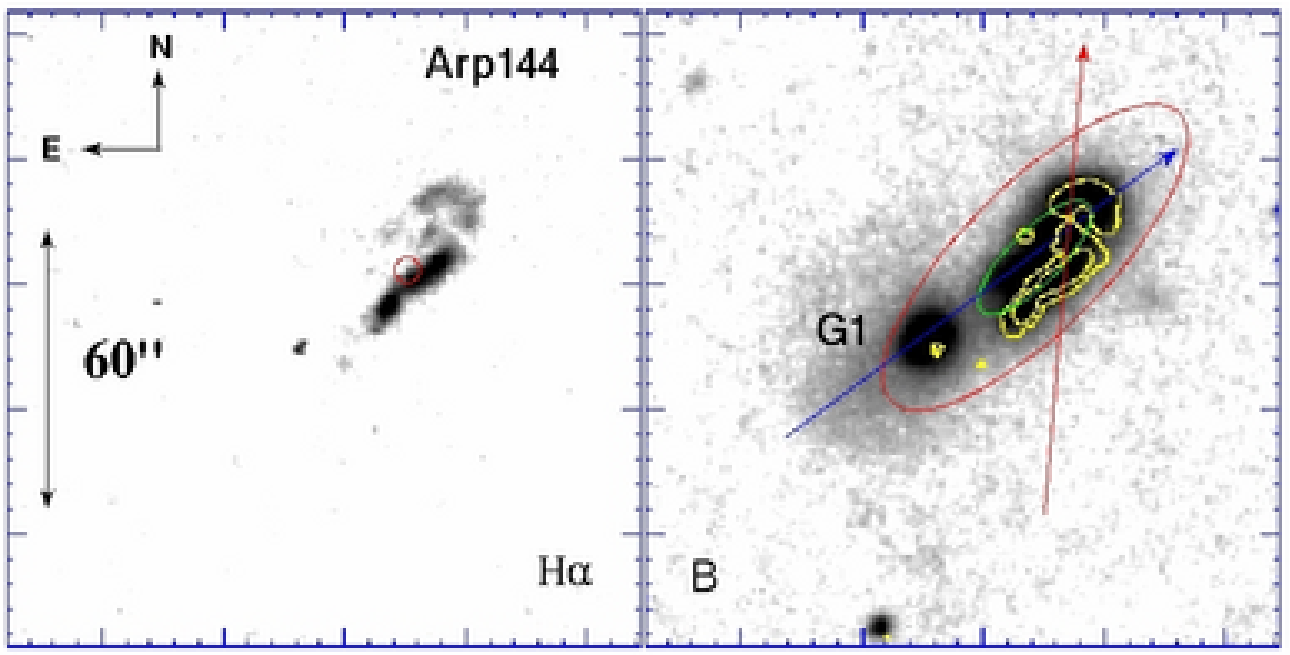}
\centerline{\hspace*{1.5cm}\includegraphics[width=14cm]{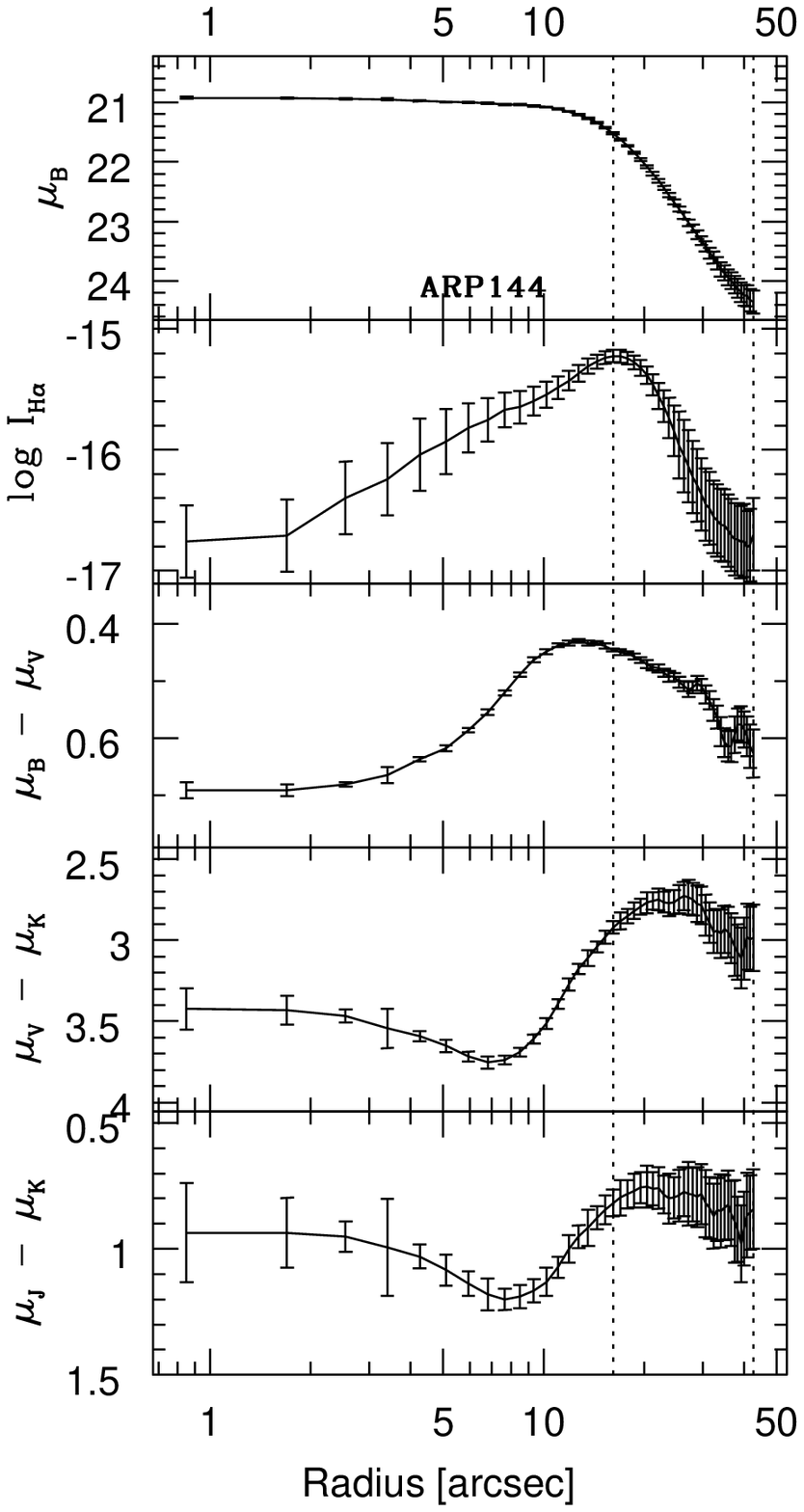}
            \hspace*{-7.5cm}\includegraphics[width=14cm]{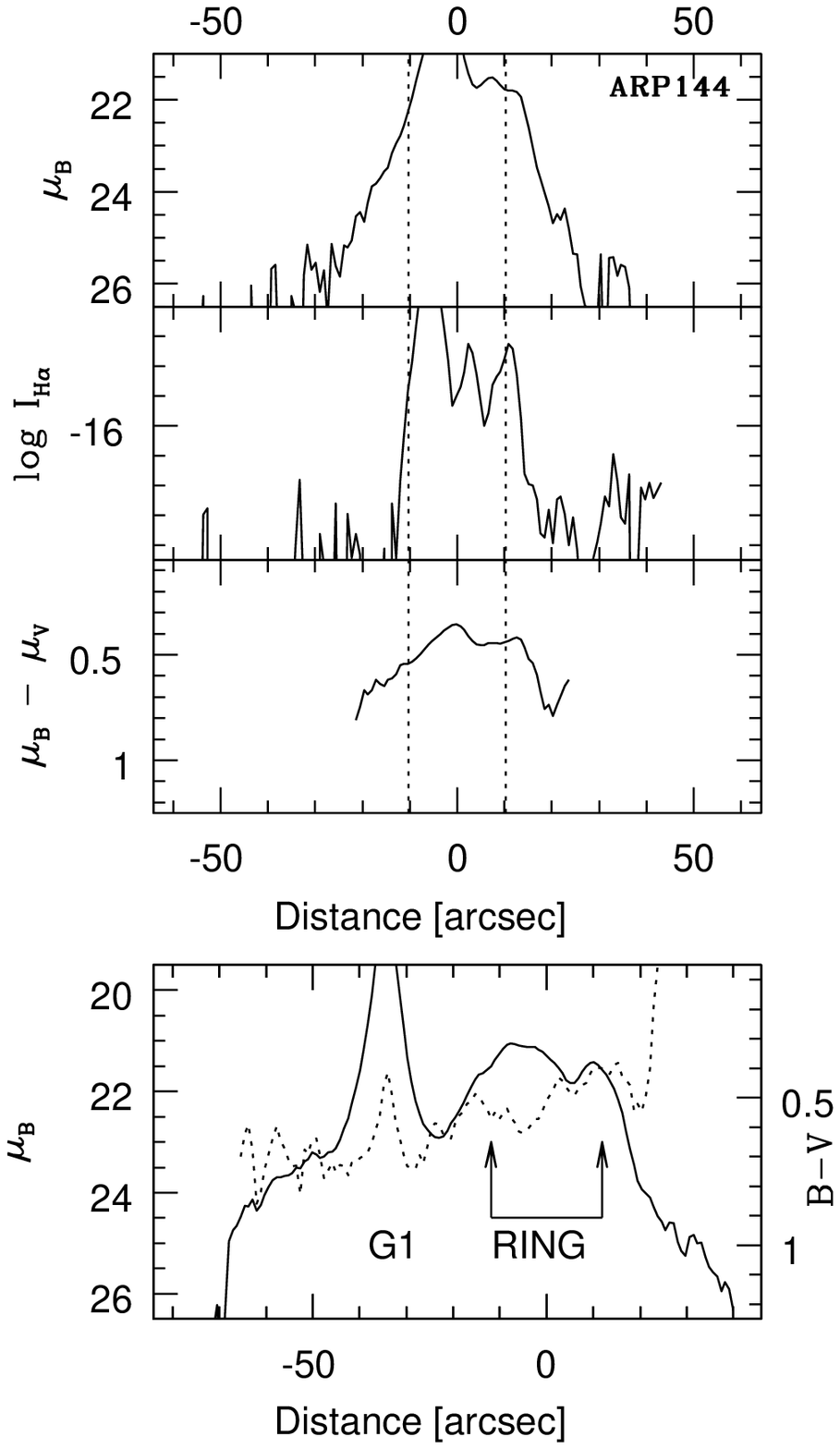}}
\caption{Same as in Figure~7, but for Arp144. This galaxy lacks a complete 
ring. However, the observed patchy \ha\ morphology can be fitted with an 
ellipse, which is shown in green (inner ellipse). The red (outer) 
ellipse is drawn corresponding to 25~mag\,arcsec$^{-1}$, with its 
ellipticity and center fixed to the values of the inner ellipse.} 
\end{figure}

\begin{figure}
\epsscale{0.75}
\plotone{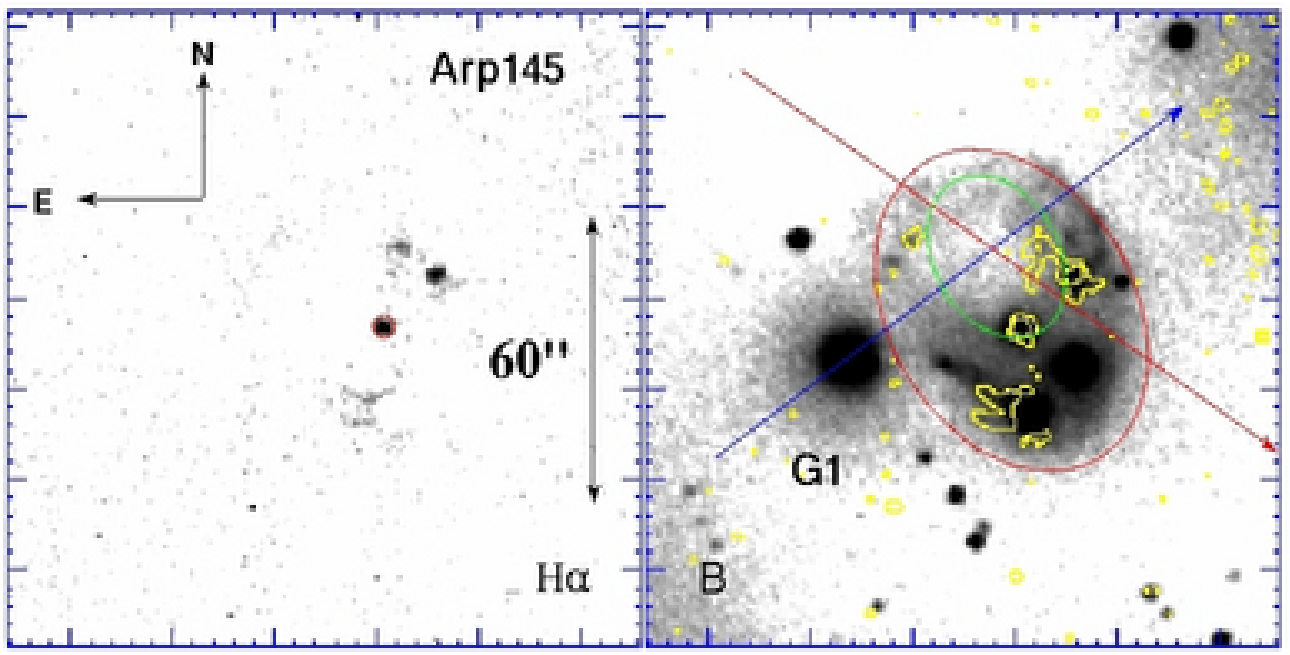}
\centerline{\hspace*{1.5cm}\includegraphics[width=14cm]{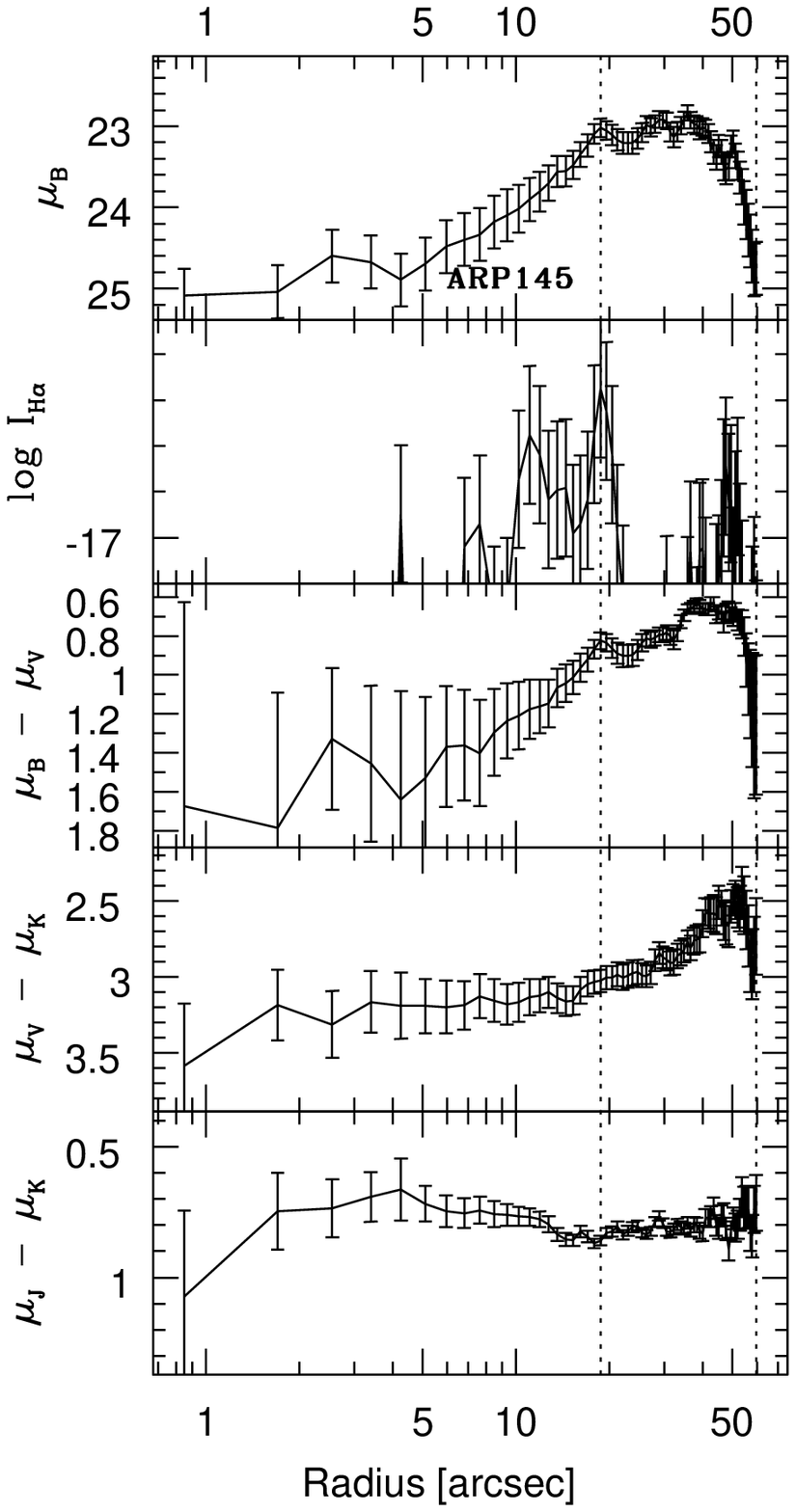}
            \hspace*{-7.5cm}\includegraphics[width=14cm]{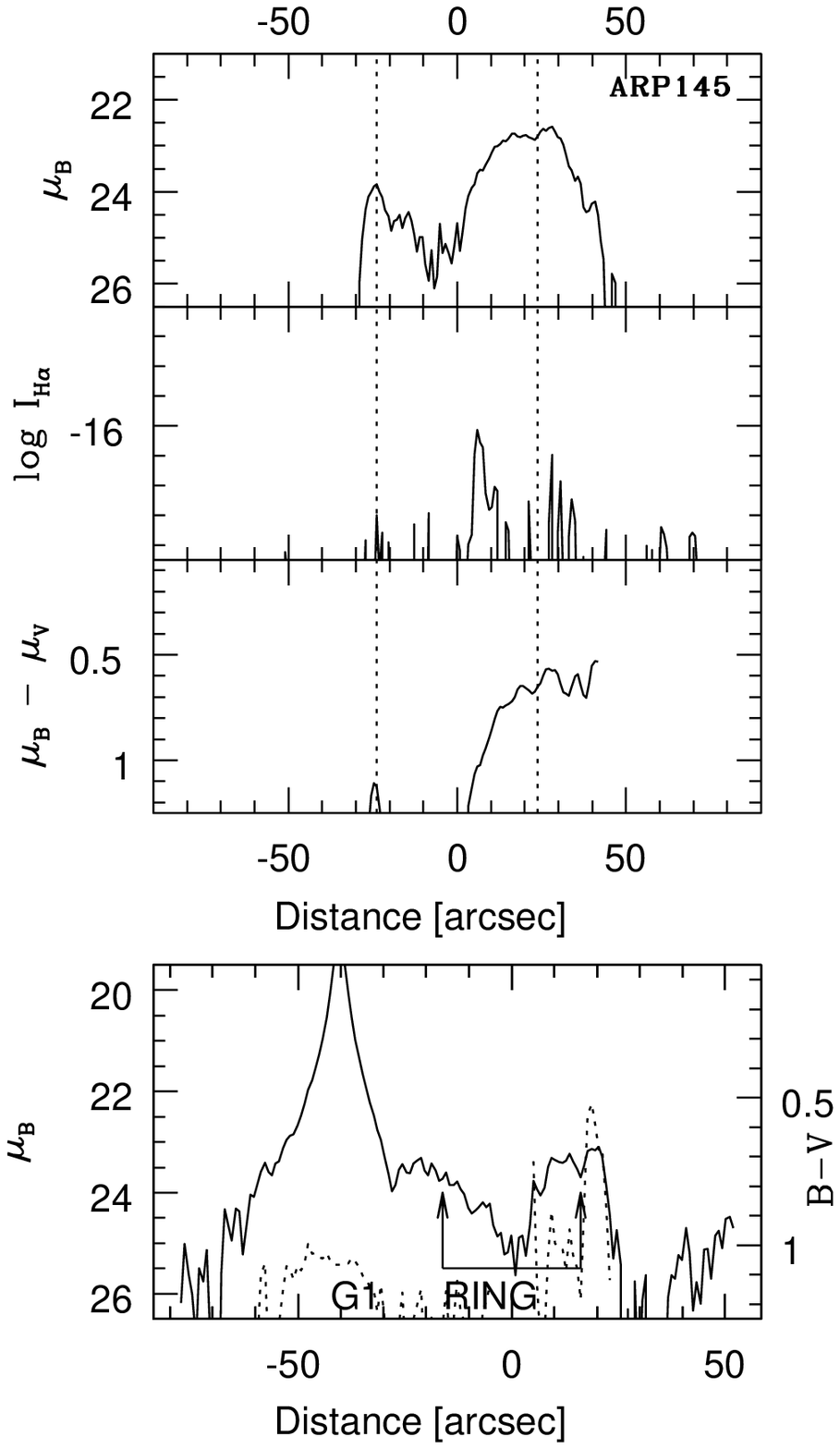}}
\caption{Same as in Figure~7, but for Arp145.}  
\end{figure}

\begin{figure}
\epsscale{0.75}
\plotone{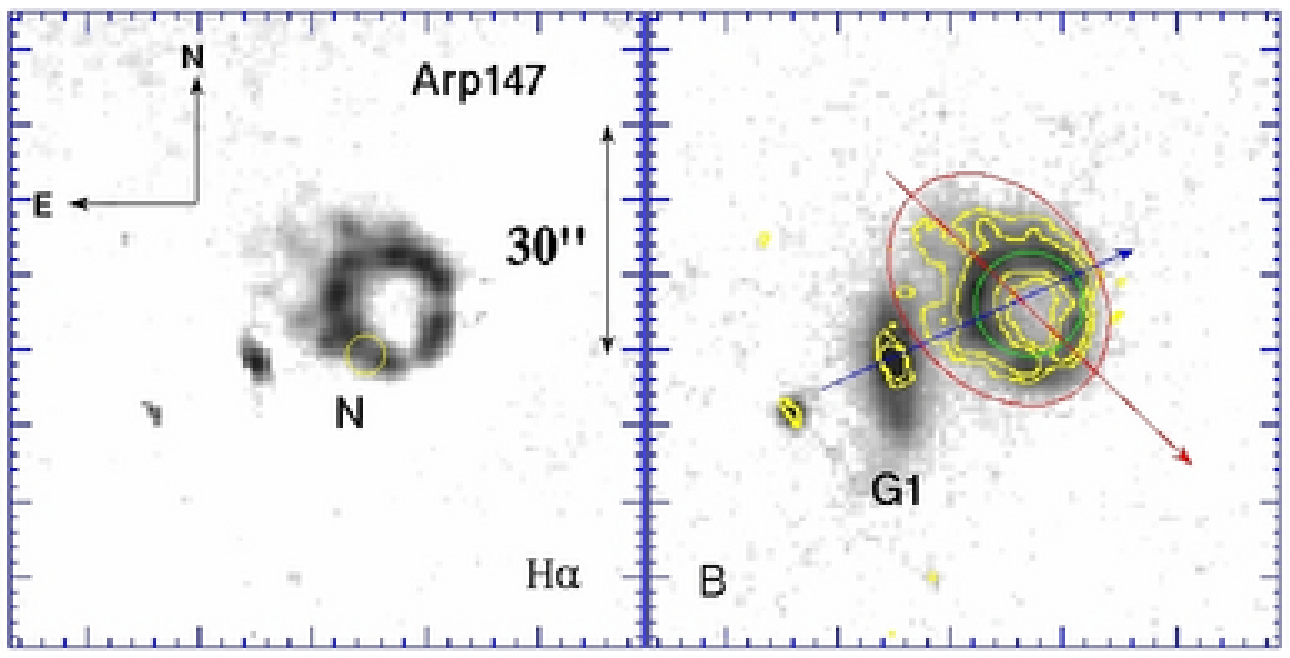}
\centerline{\hspace*{1.5cm,}\includegraphics[width=14cm]{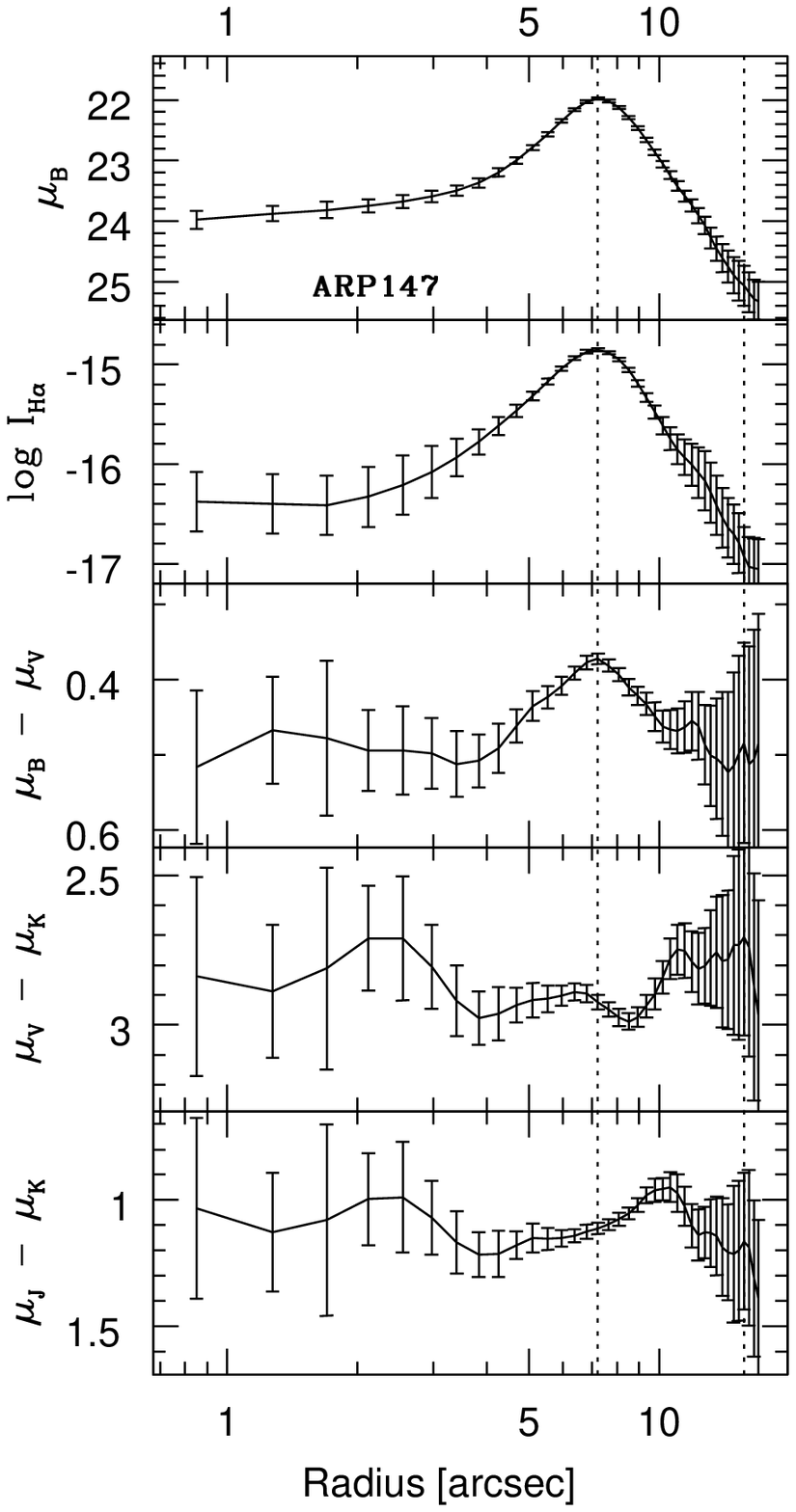}
            \hspace*{-7.5cm}\includegraphics[width=14cm]{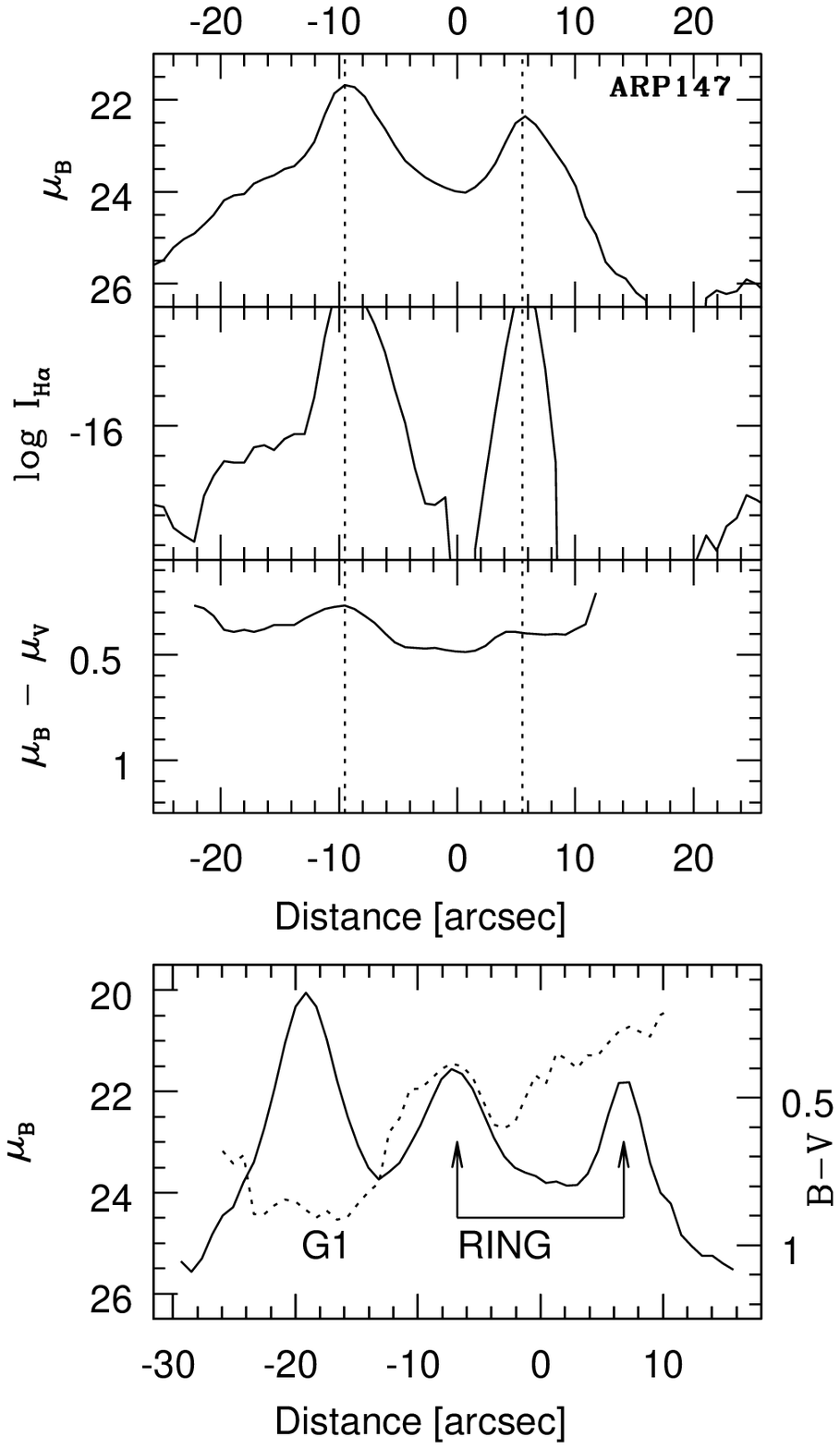}}
\caption{Same as in Figure~7, but for Arp147.} 
\end{figure}

\begin{figure}
\epsscale{0.75}
\plotone{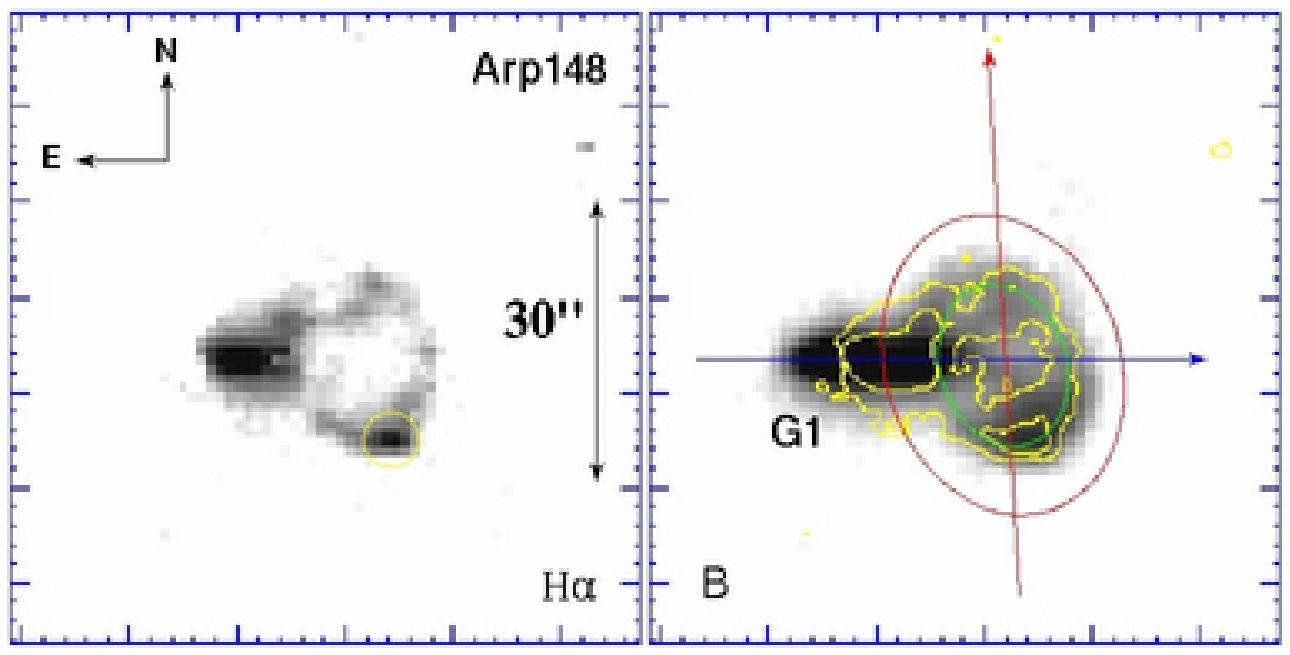}
\centerline{\hspace*{1.5cm}\includegraphics[width=14cm]{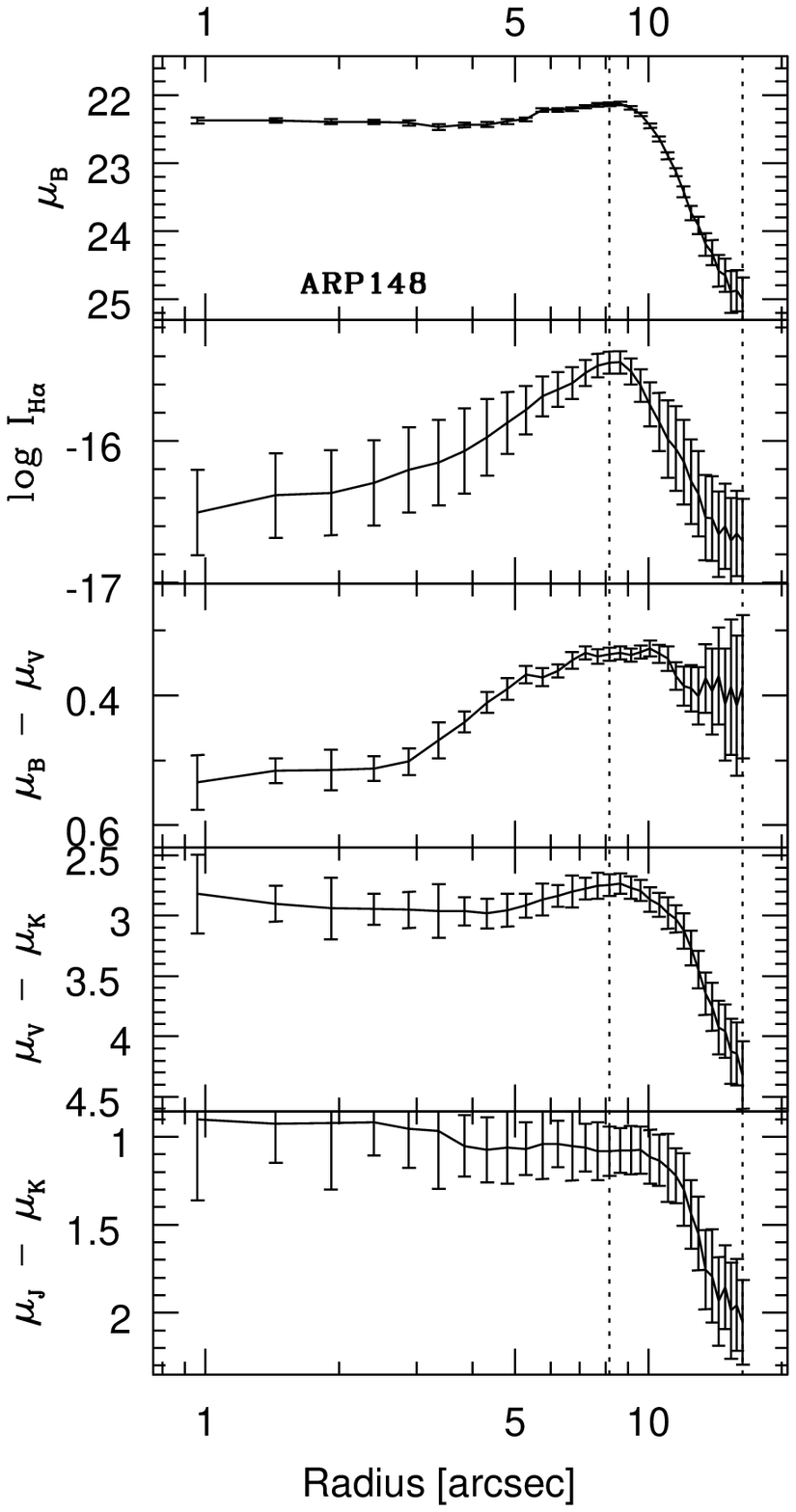}
            \hspace*{-7.5cm}\includegraphics[width=14cm]{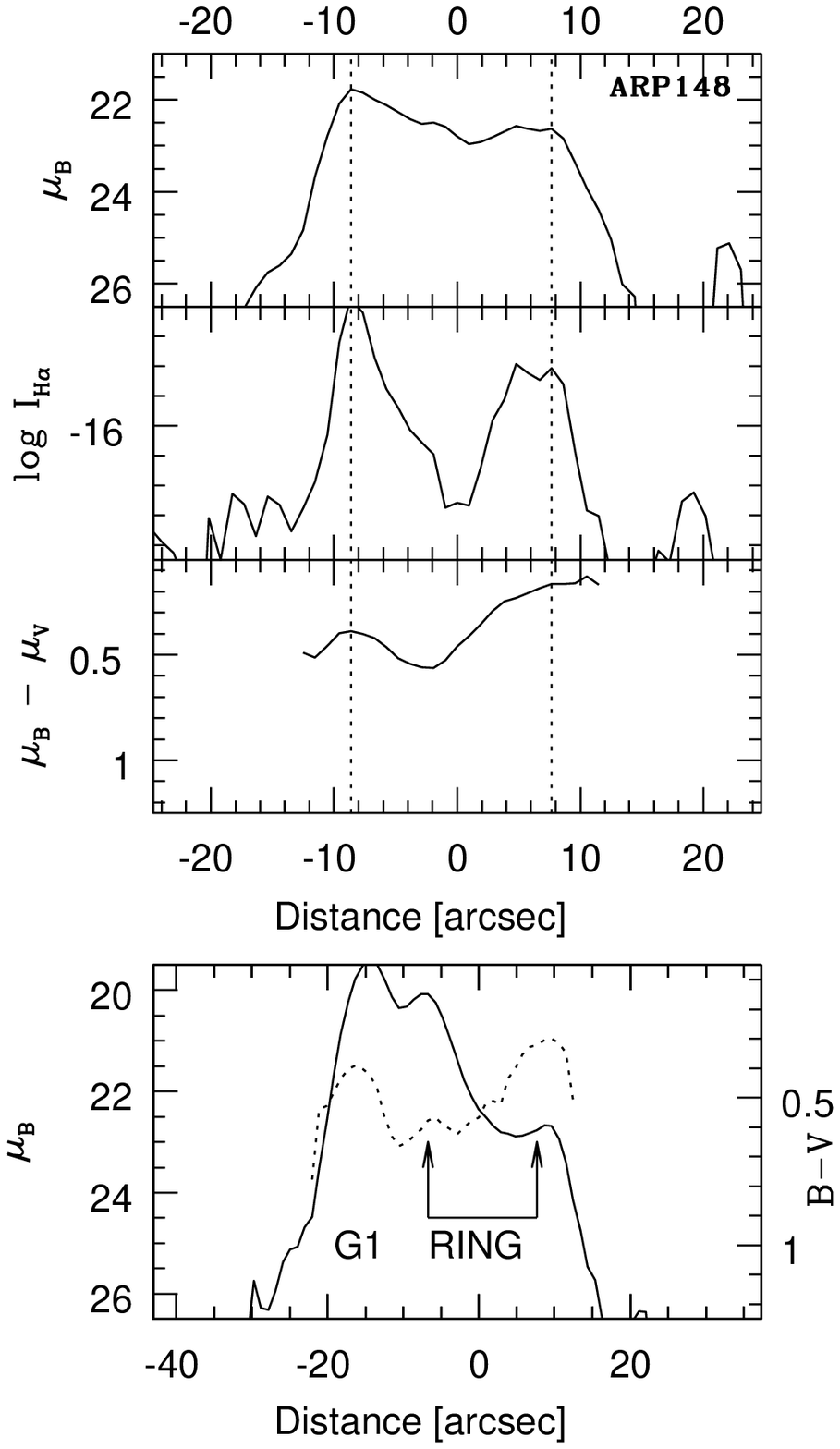}}
\caption{Same as in Figure~7, but for Arp148.} 
\end{figure}

\begin{figure}
\epsscale{0.75}
\plotone{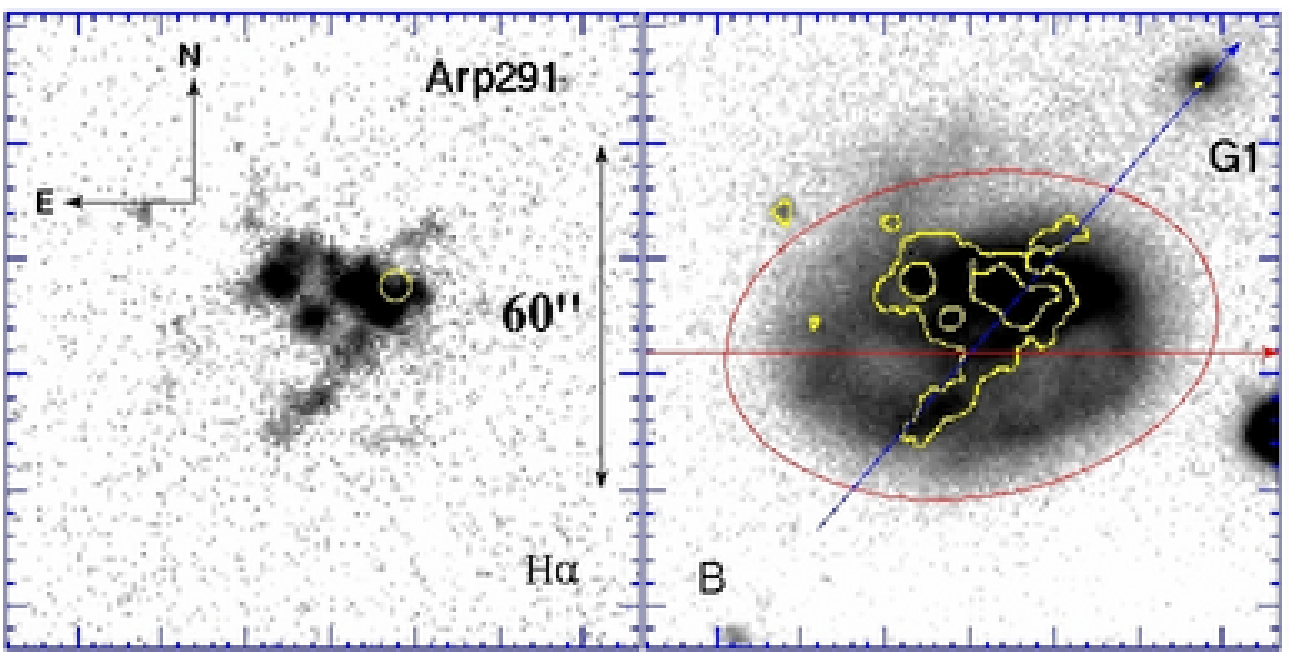}
\centerline{\hspace*{1.5cm}\includegraphics[width=14cm]{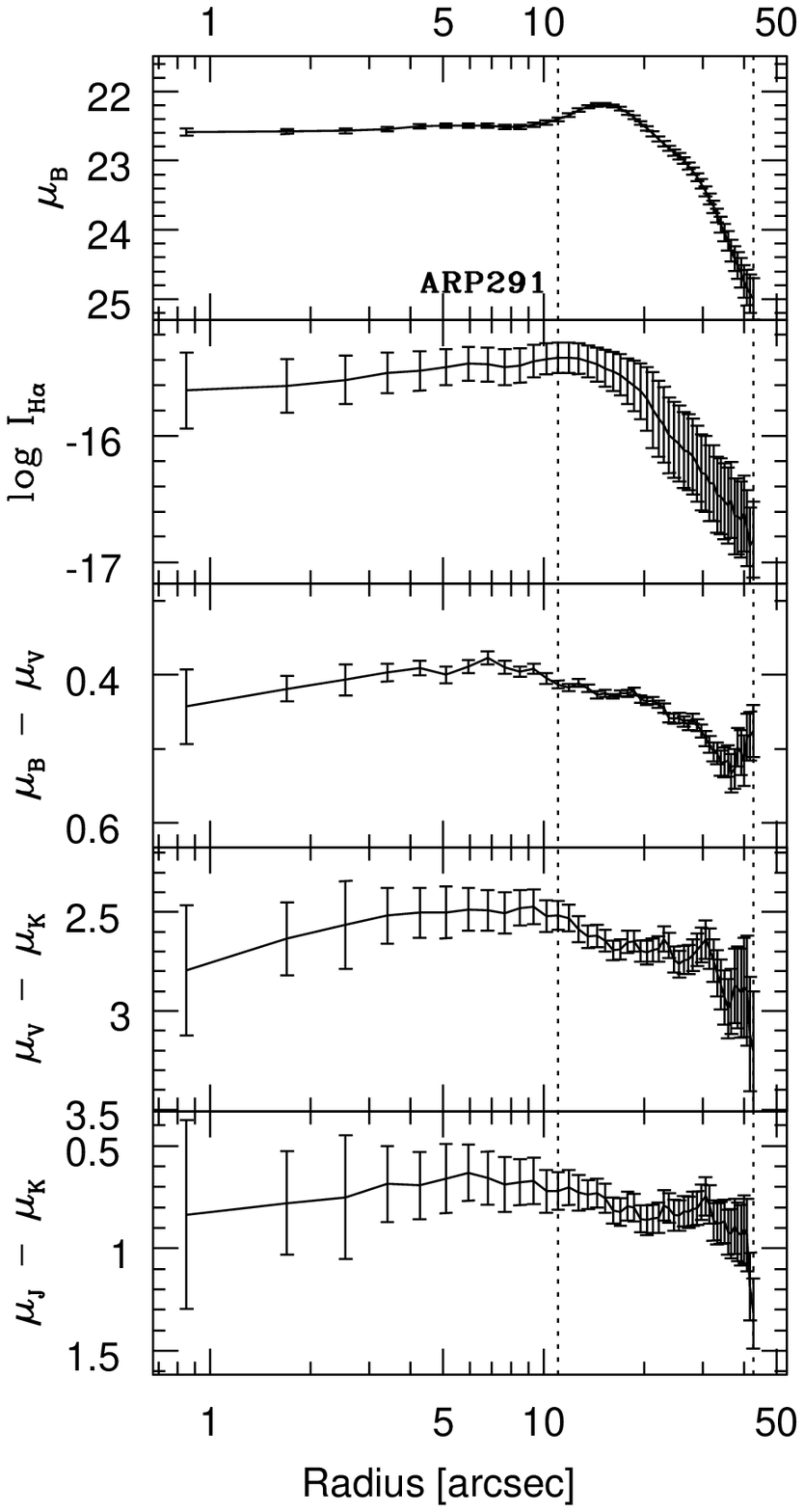}
            \hspace*{-7.5cm}\includegraphics[width=14cm]{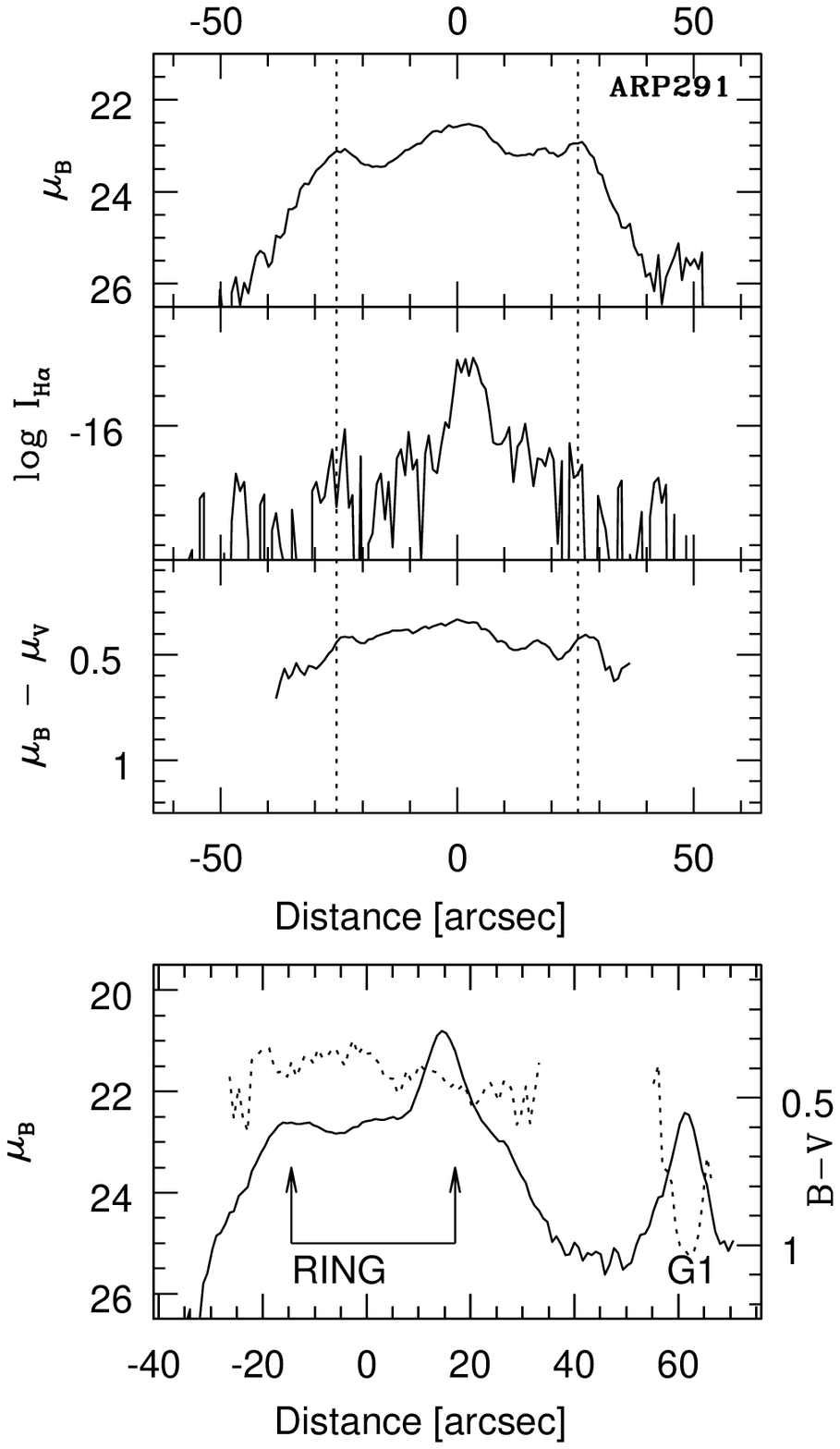}}
\caption{Same as in Figure~7, but for Arp291. Lacks a ring-like structure 
in \ha. The red ellipse is drawn corresponding to 25~mag\,arcsec$^{-1}$, 
with its ellipticity and center fixed to the values of the ring-like
structure seen on the $B$-band image.} 
\end{figure}

\begin{figure}
\epsscale{0.75}
\plotone{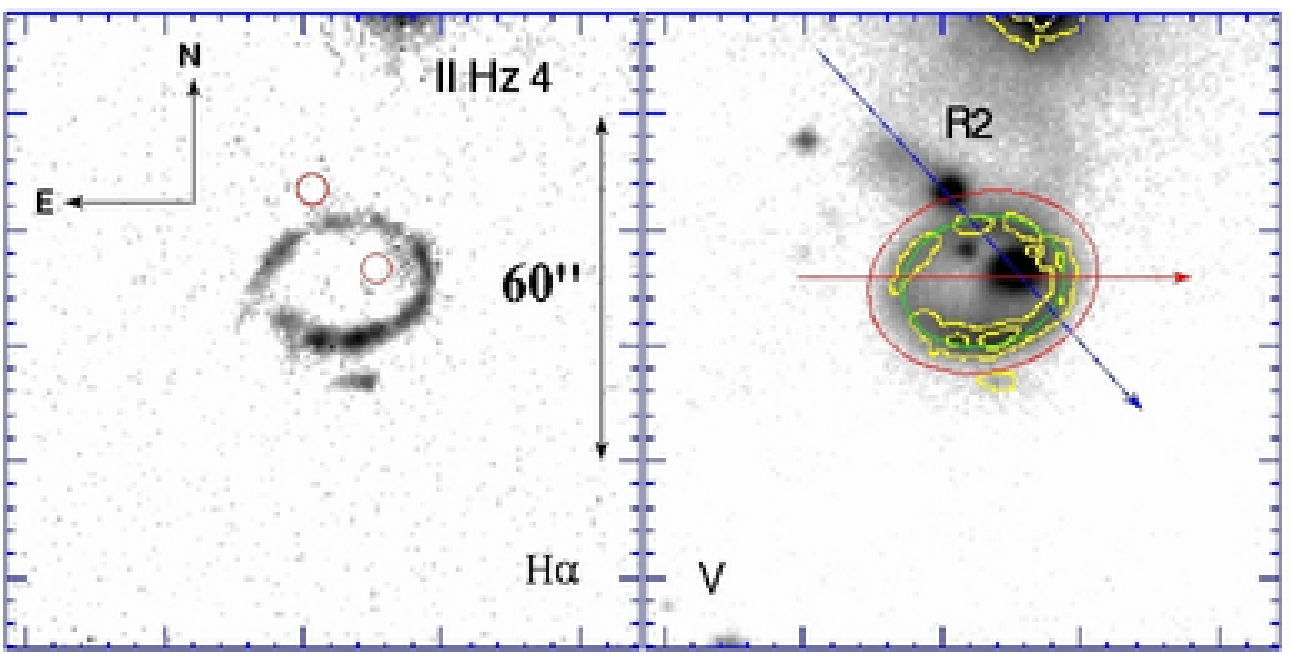}
\centerline{\hspace*{1.5cm}\includegraphics[width=14cm]{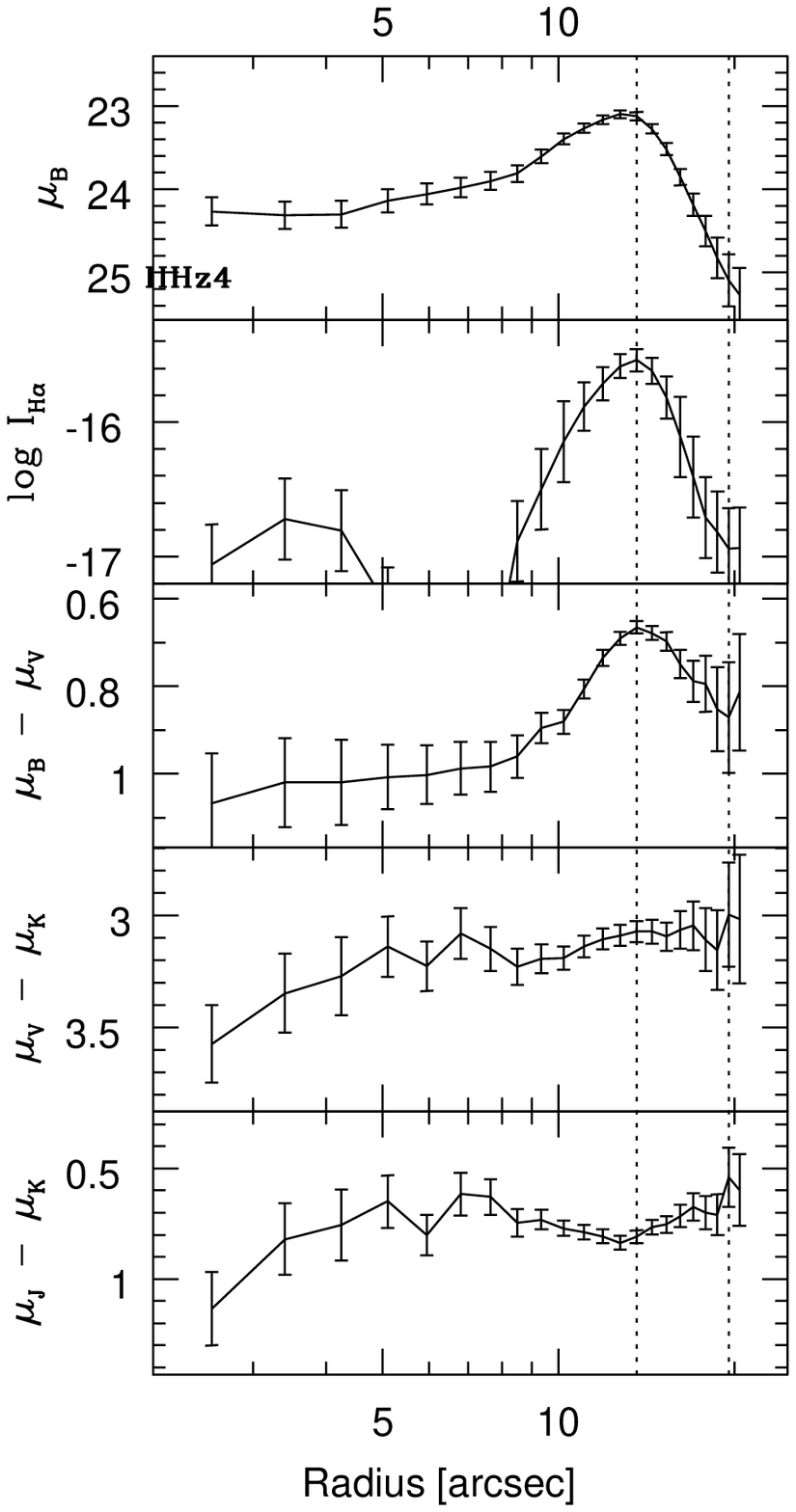}
            \hspace*{-7.5cm}\includegraphics[width=14cm]{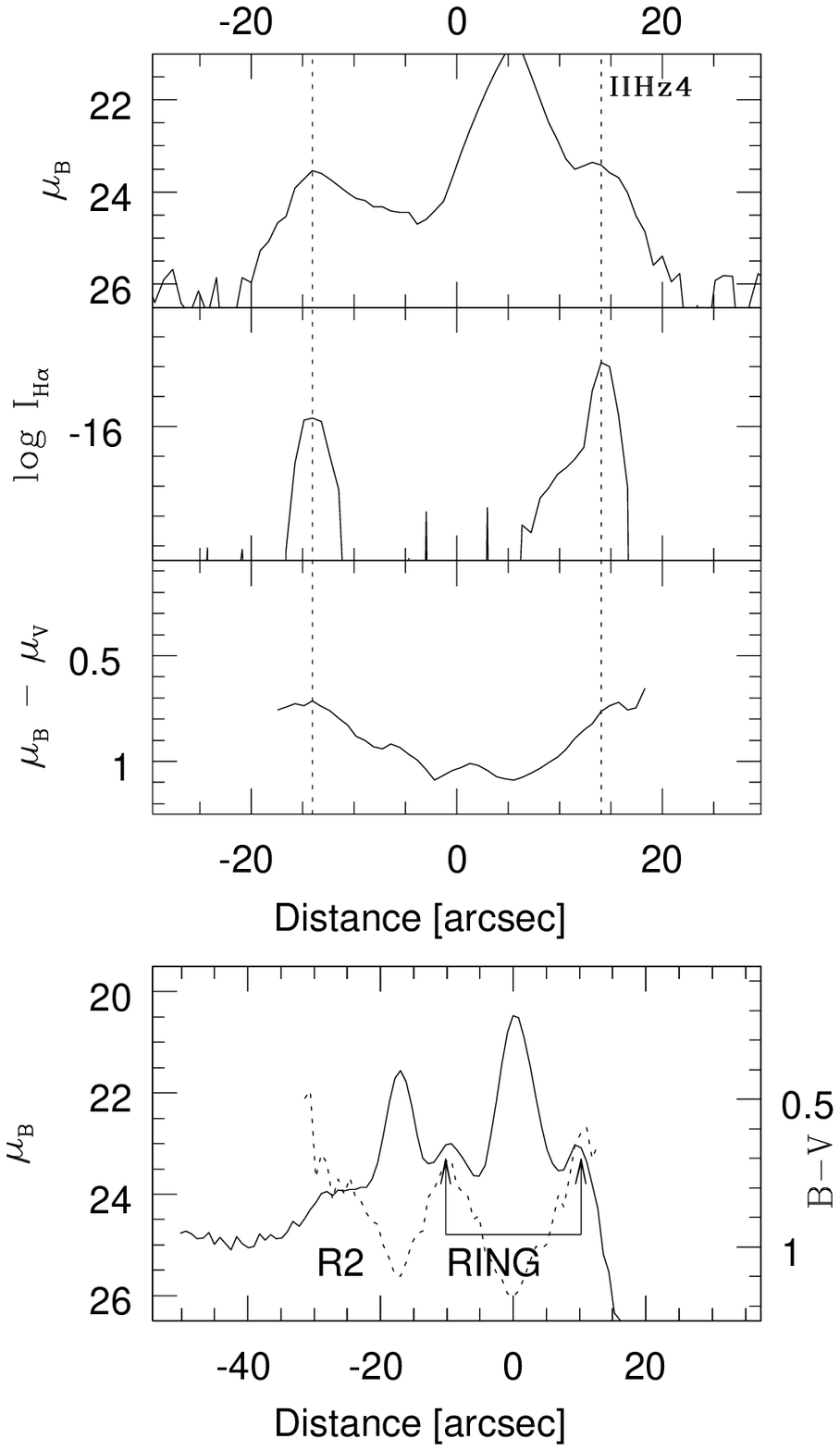}}
\caption{Same as in Figure~7, but for IIHz4.}
\end{figure}

\begin{figure}
\epsscale{0.75}
\plotone{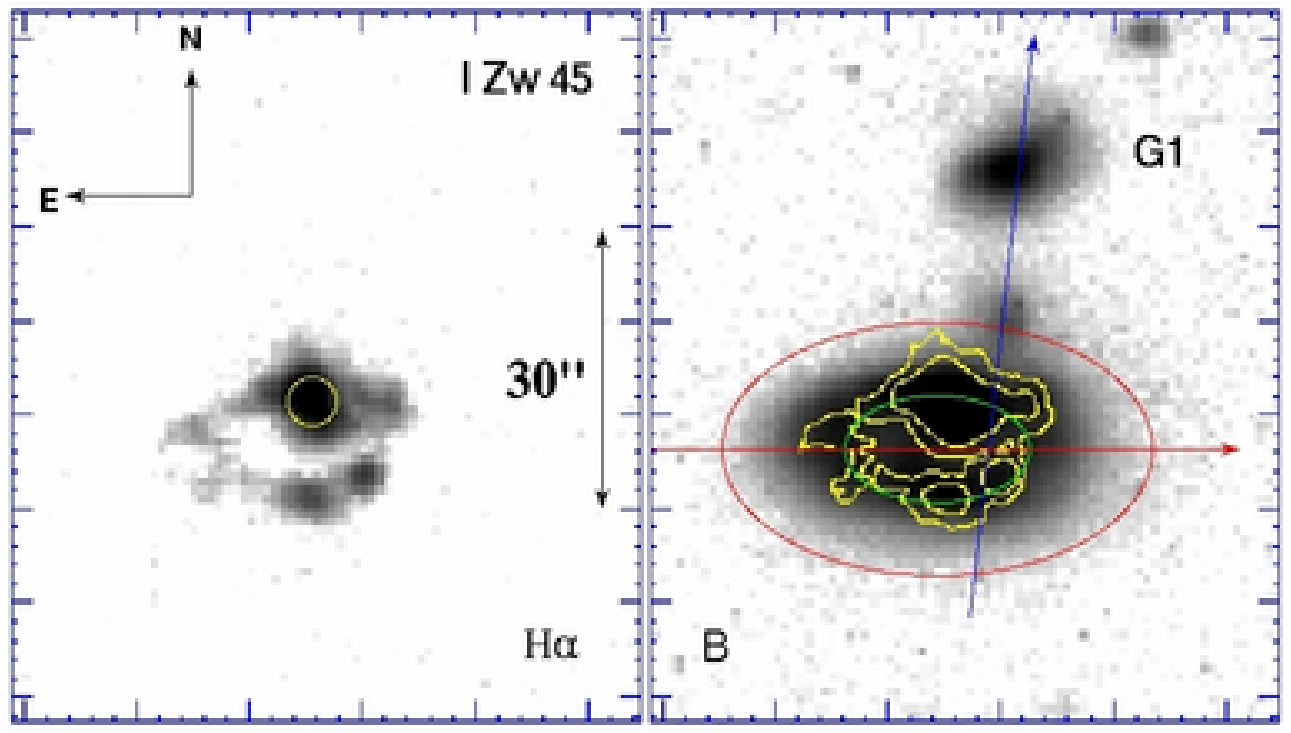}
\centerline{\hspace*{1.5cm}\includegraphics[width=14cm]{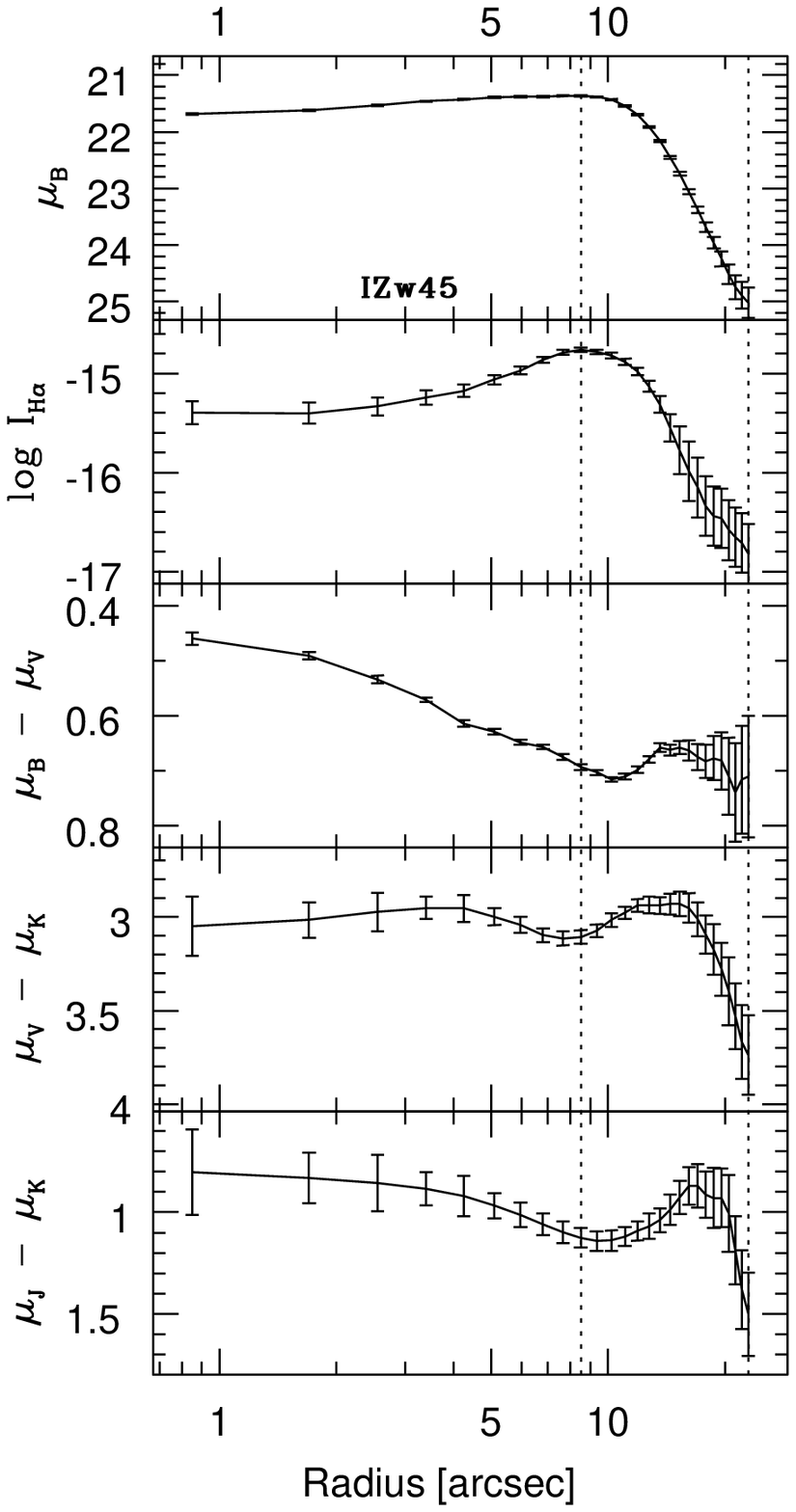}
            \hspace*{-7.5cm}\includegraphics[width=14cm]{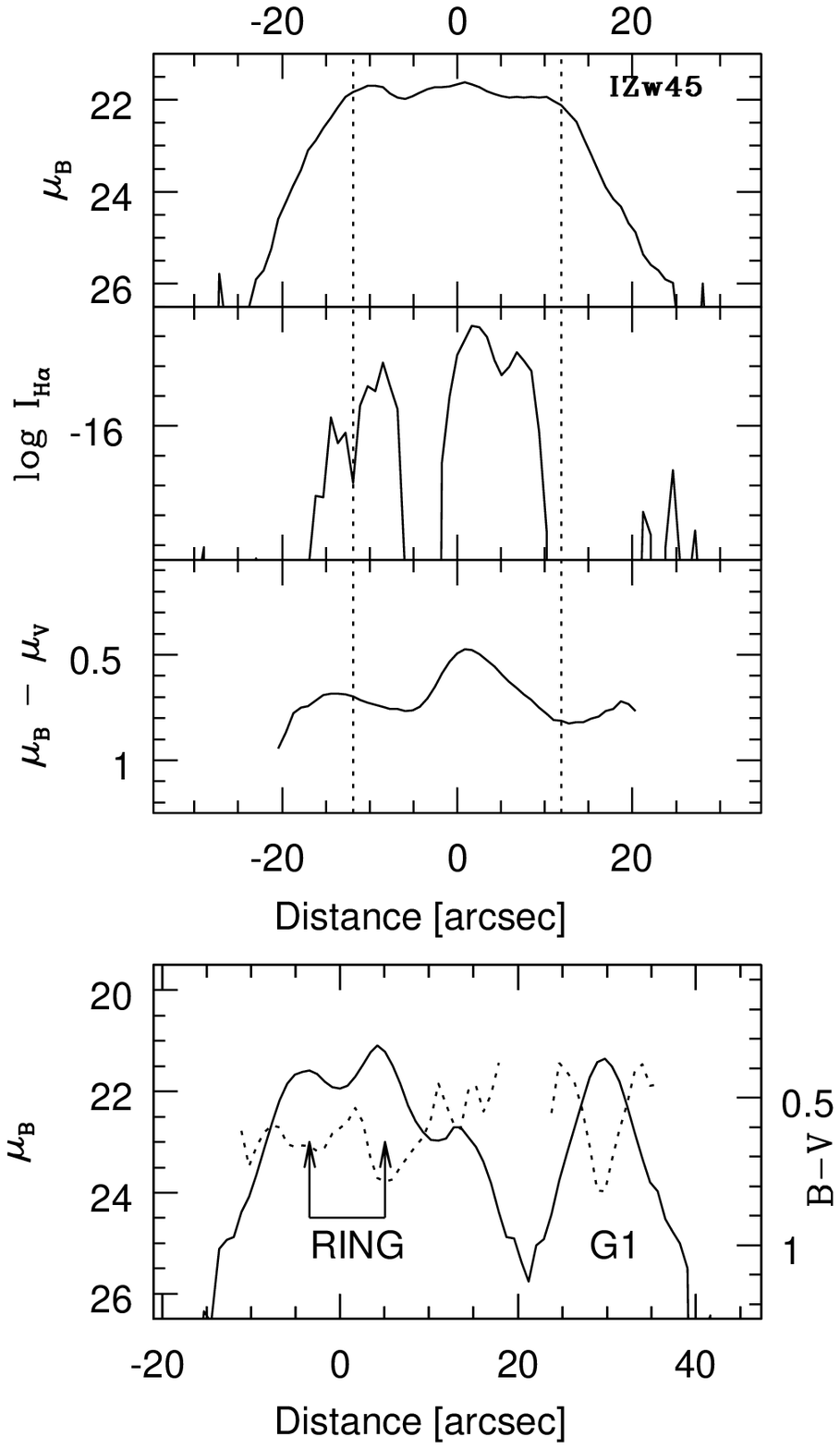}}
\caption{Same as in Figure~7, but for IZw45.} 
\end{figure}

\begin{figure}
\epsscale{0.75}
\plotone{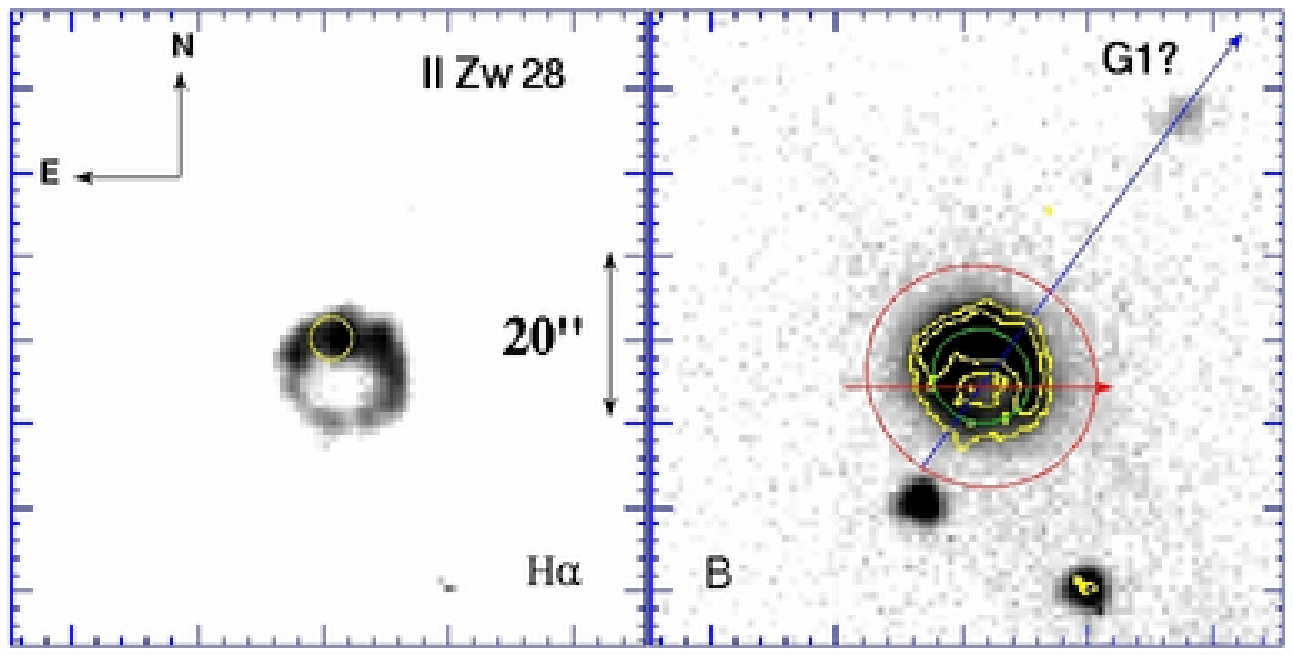}
\centerline{\hspace*{1.5cm}\includegraphics[width=14cm]{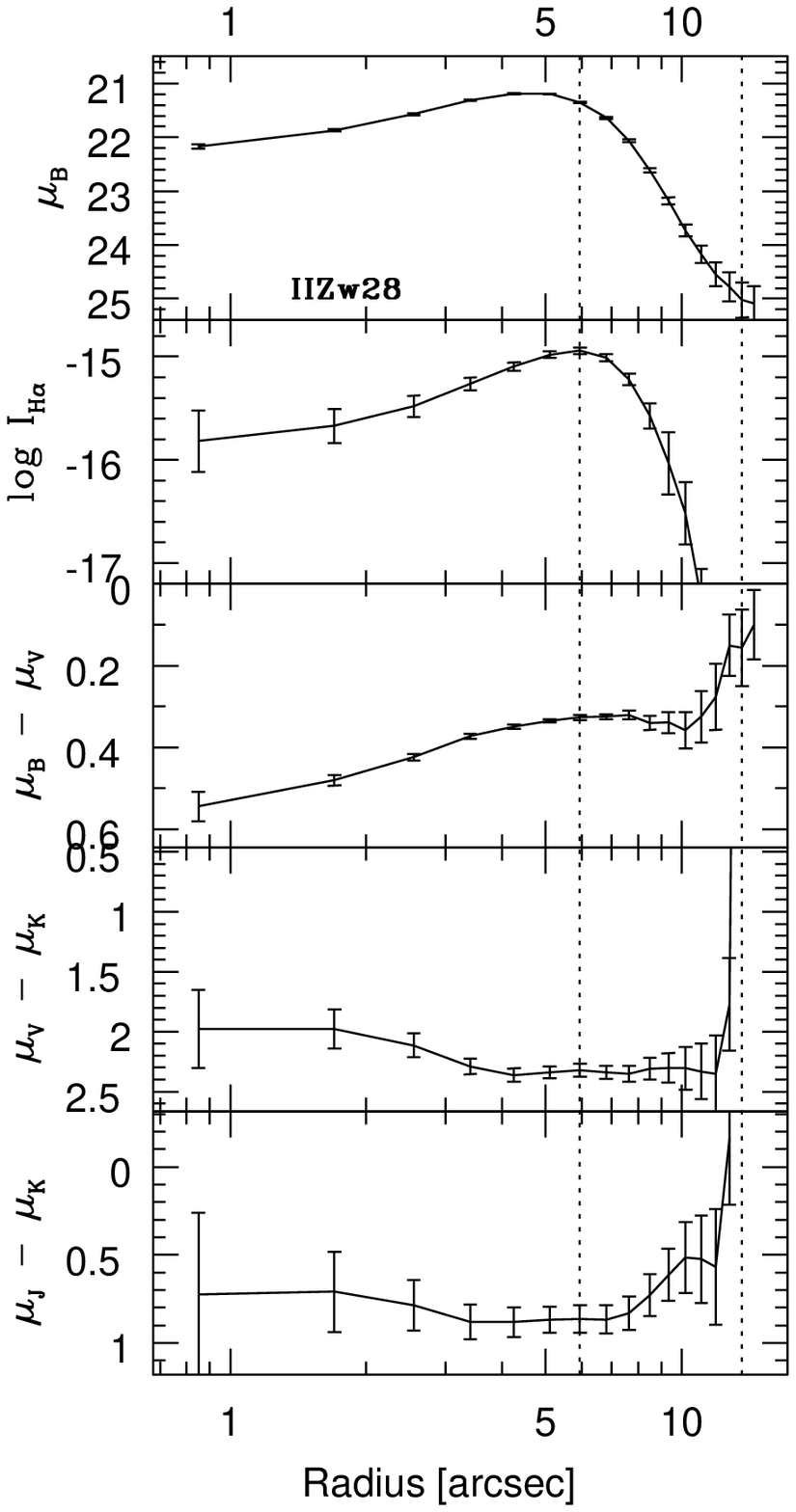}
            \hspace*{-7.5cm}\includegraphics[width=14cm]{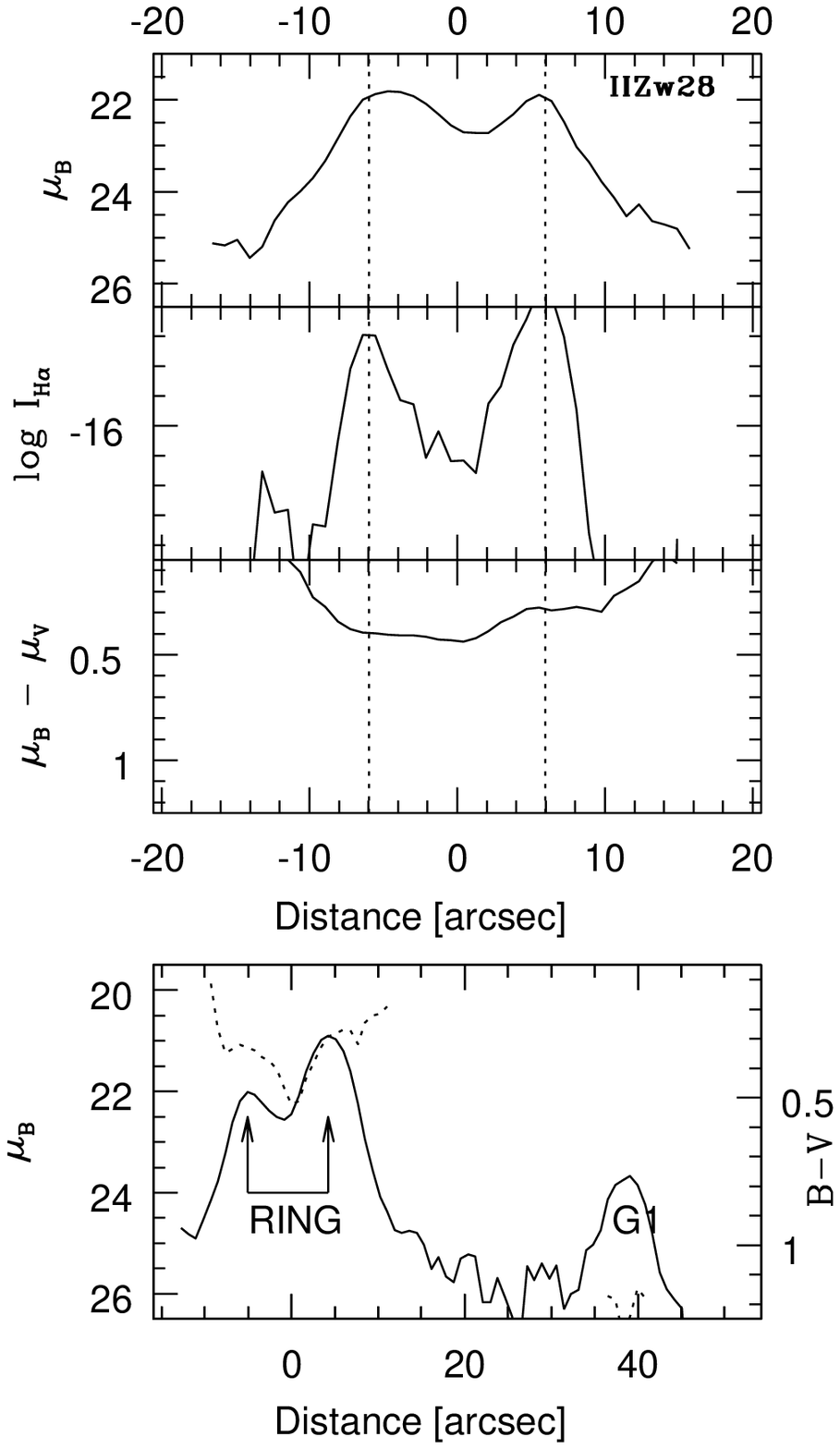}}
\caption{Same as in Figure~7, but for IIZw28. There is no confirmed
companion to this galaxy. G1 is the brightest extended object in the 
neighborhood, and is a likely candidate for the companion.} 
\end{figure}

\clearpage

\begin{figure}
\epsscale{0.75}
\plotone{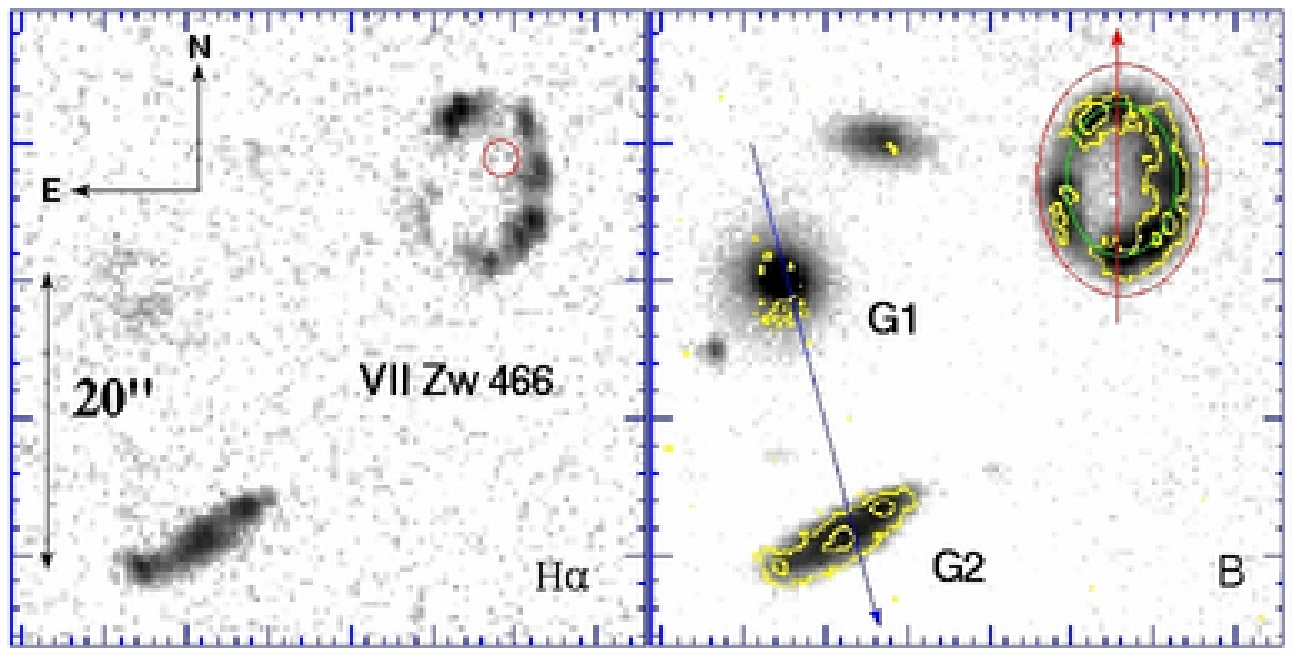}
\centerline{\hspace*{1.5cm}\includegraphics[width=14cm]{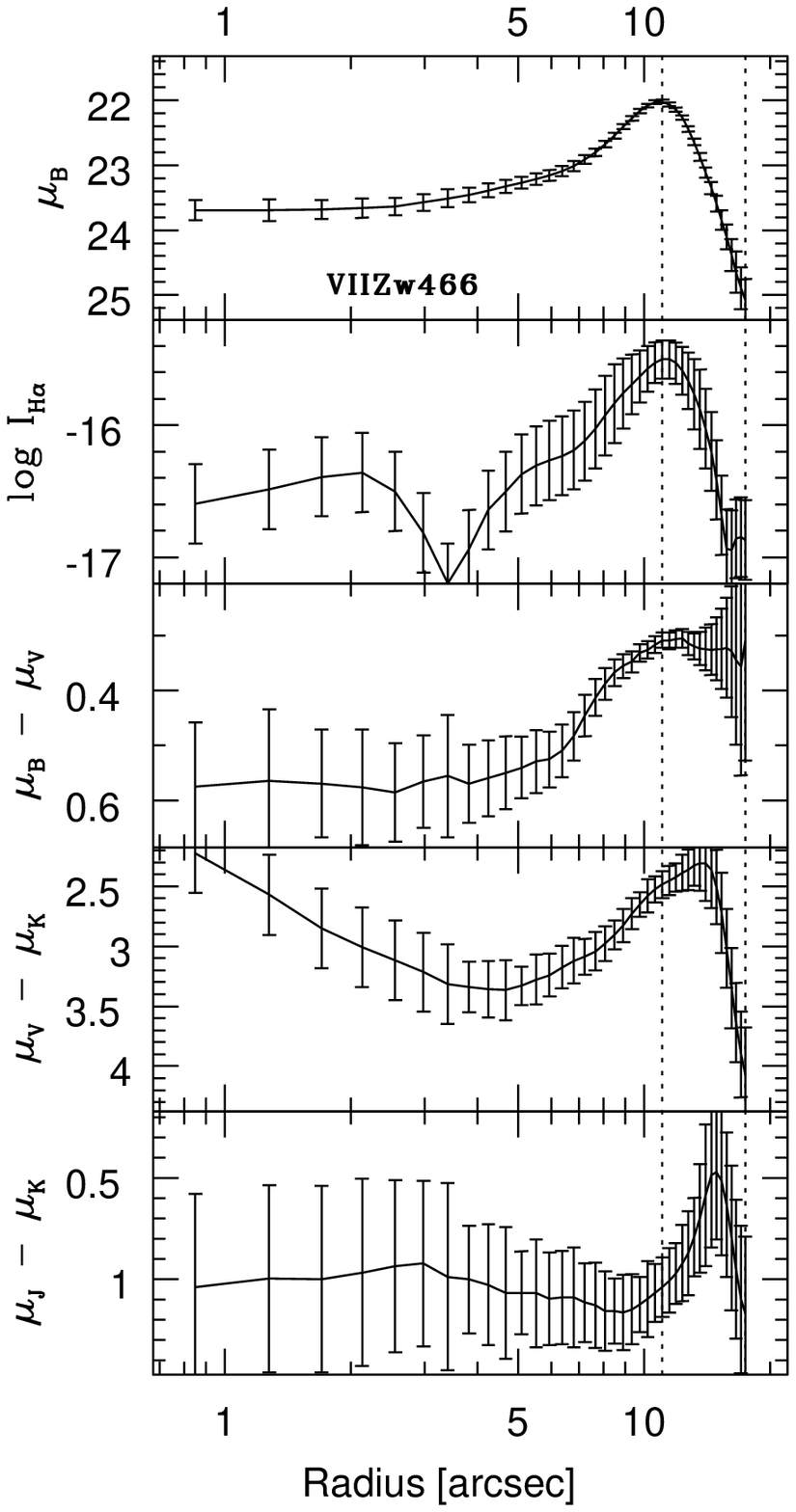}
            \hspace*{-7.5cm}\includegraphics[width=14cm]{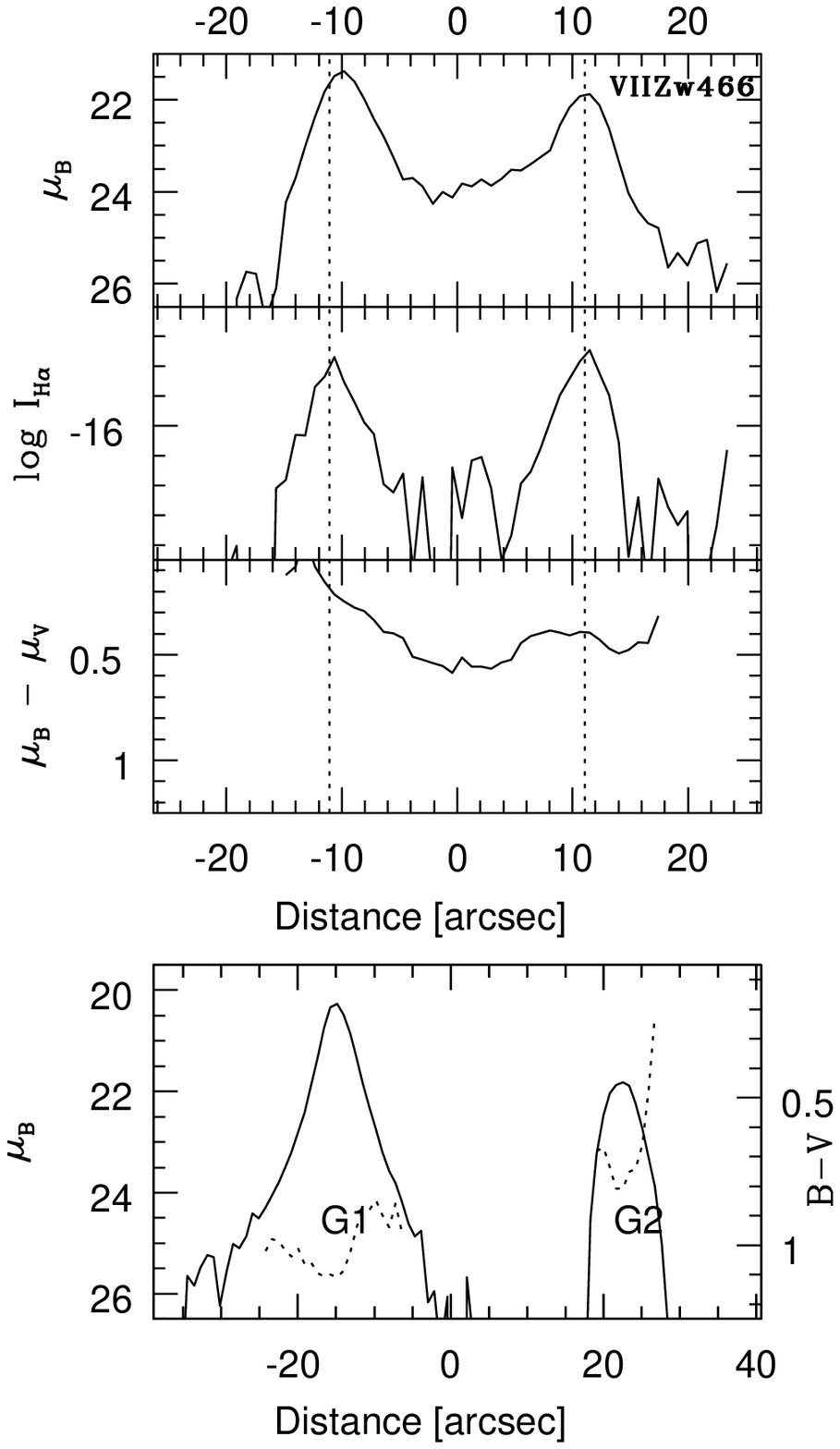}}
\caption{Same as in Figure~7, but for VIIZw466. The second cut 
(last panel on the bottom right) is taken along the line joining
the two candidate companions.} 
\end{figure}

\begin{figure}
\epsscale{0.75}
\plotone{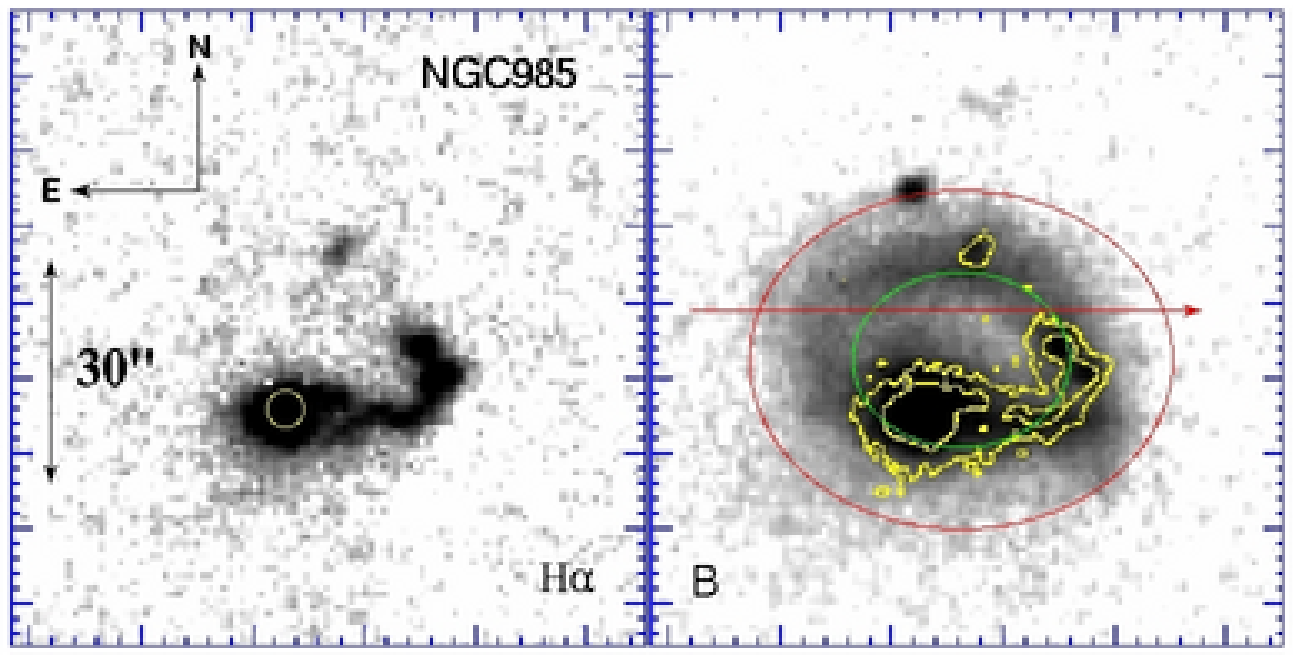}
\centerline{\hspace*{1.5cm}\includegraphics[width=14cm]{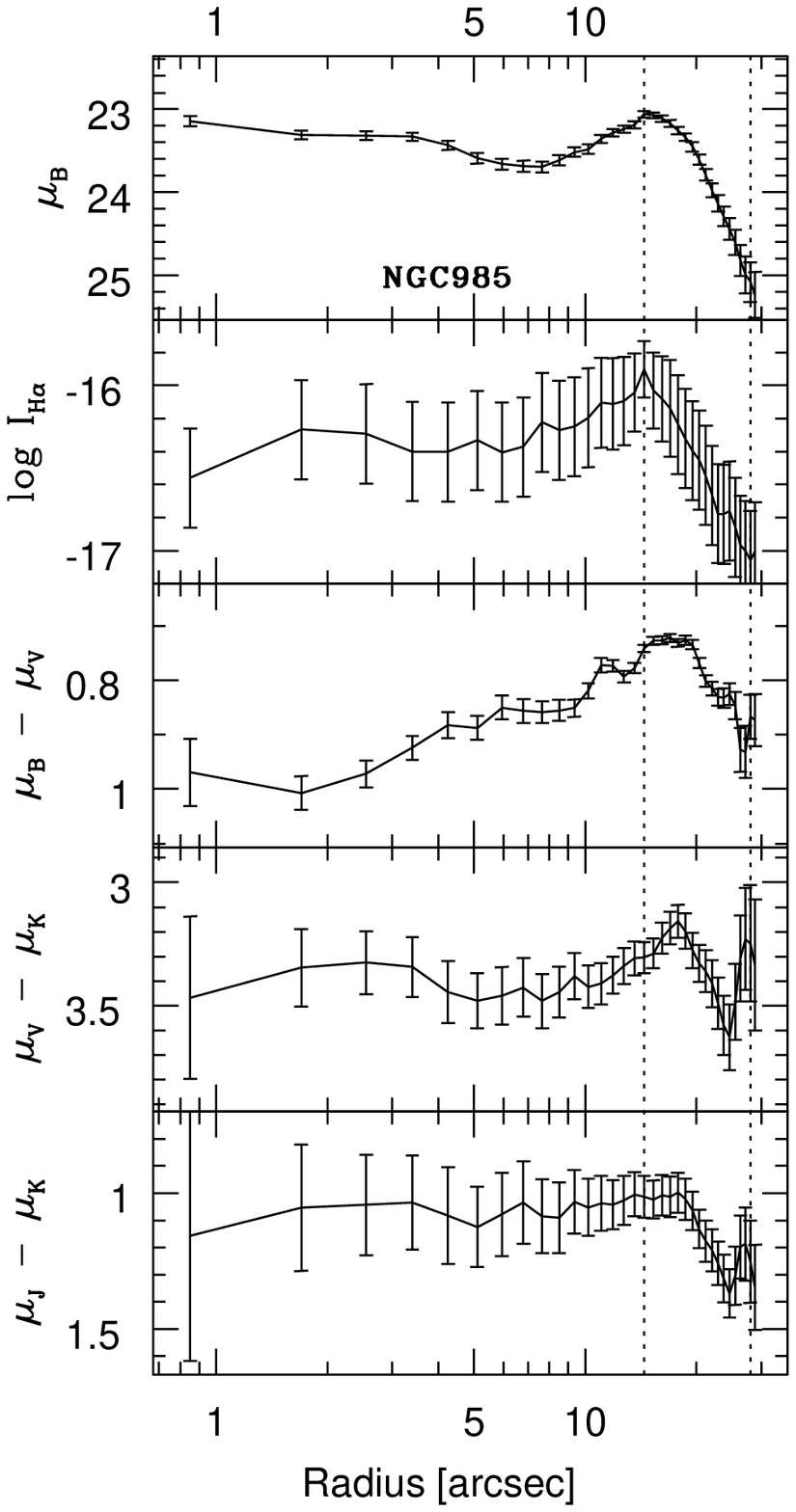}
            \hspace*{-7.5cm}\includegraphics[width=14cm]{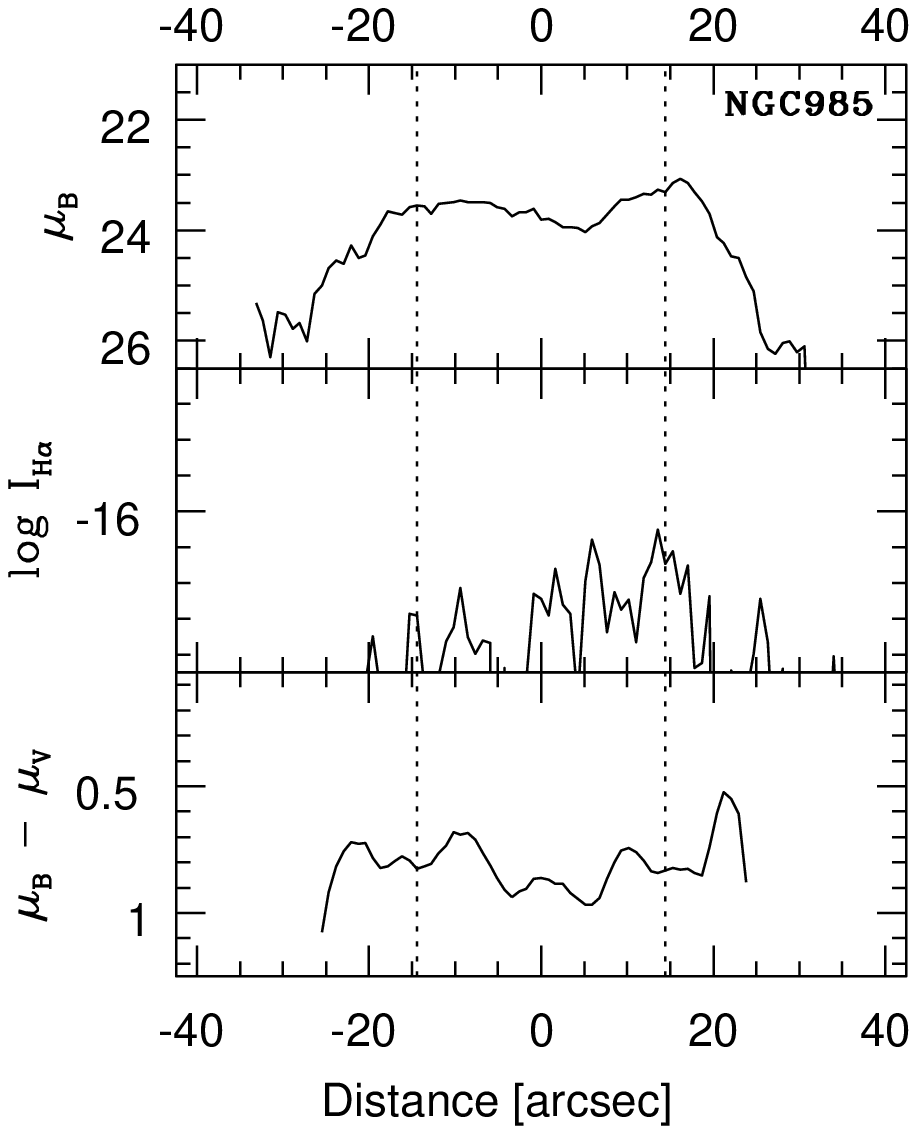}}
\caption{Same as in Figure~7, but for NGC985. As there is not even a 
candidate for the companion, the panel showing the second cut is 
not drawn. } 
\end{figure}

\begin{figure}
\epsscale{0.75}
\plotone{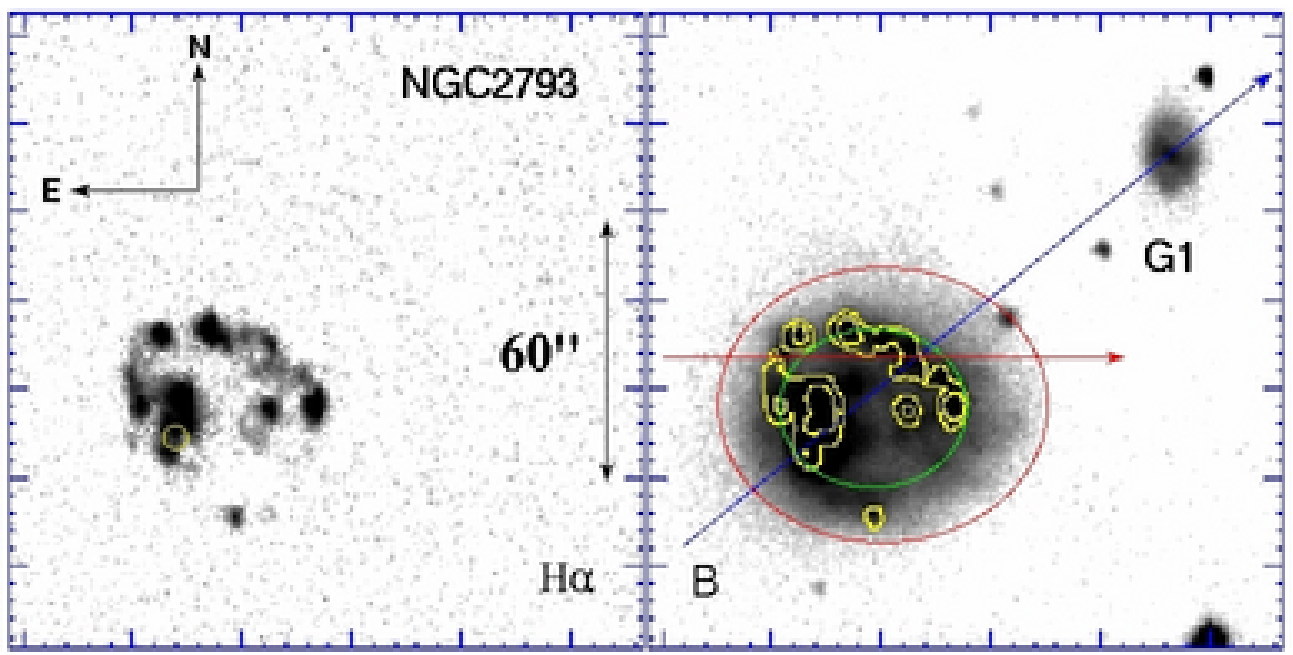}
\centerline{\hspace*{1.5cm}\includegraphics[width=14cm]{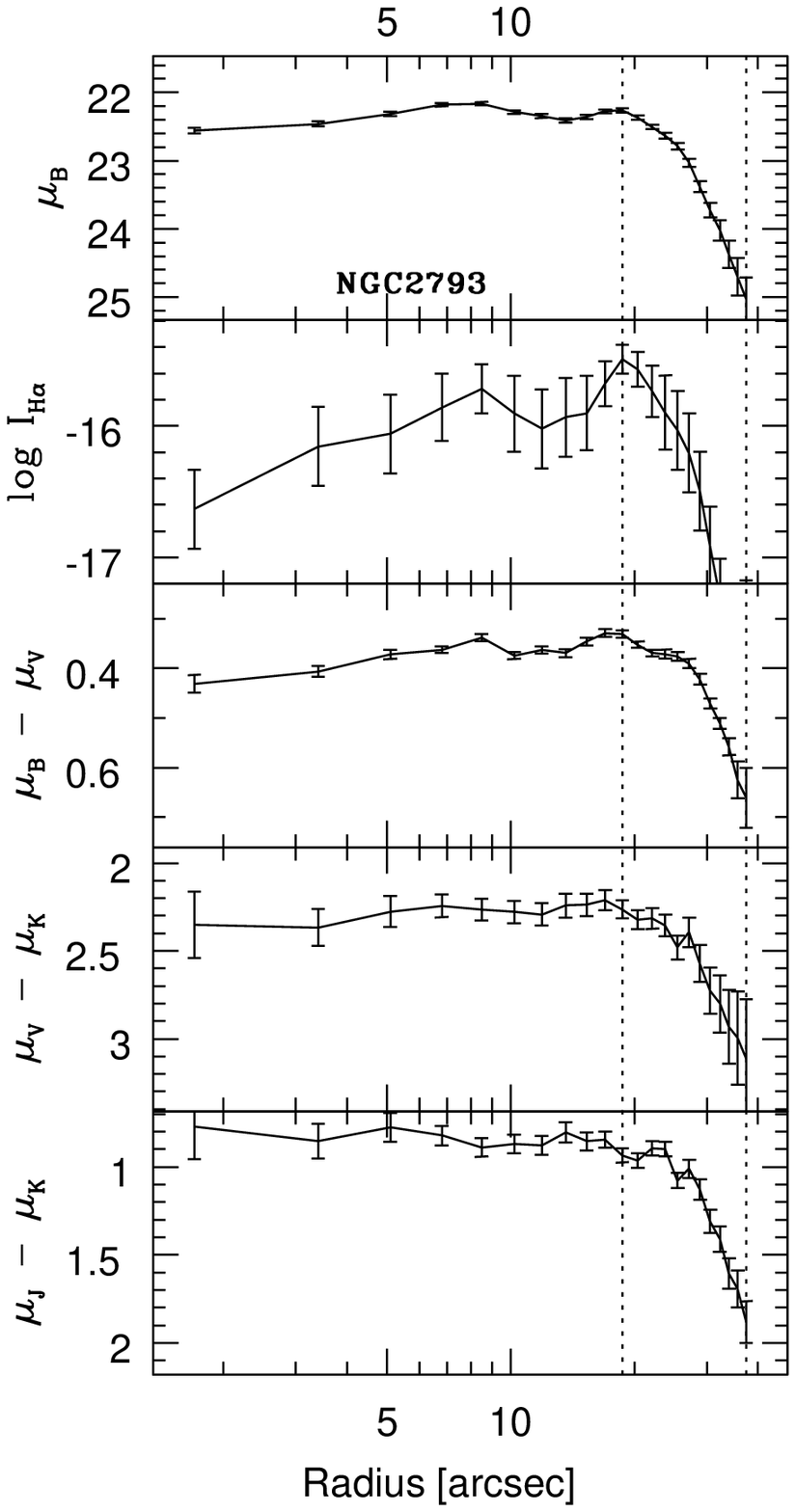}
            \hspace*{-7.5cm}\includegraphics[width=14cm]{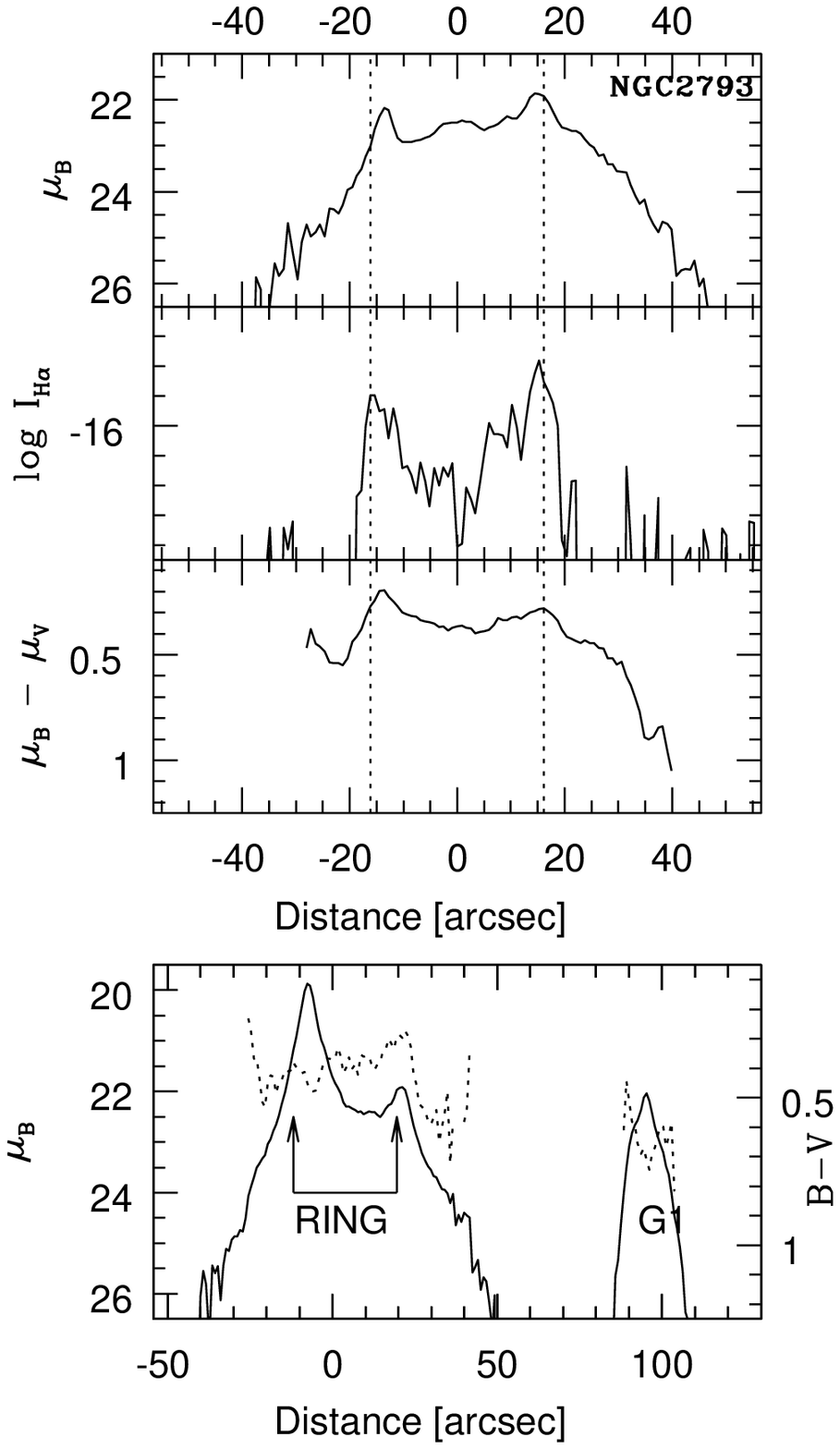}}
\caption{Same as in Figure~7, but for NGC2793.} 
\end{figure}

\begin{figure}
\epsscale{0.75}
\plotone{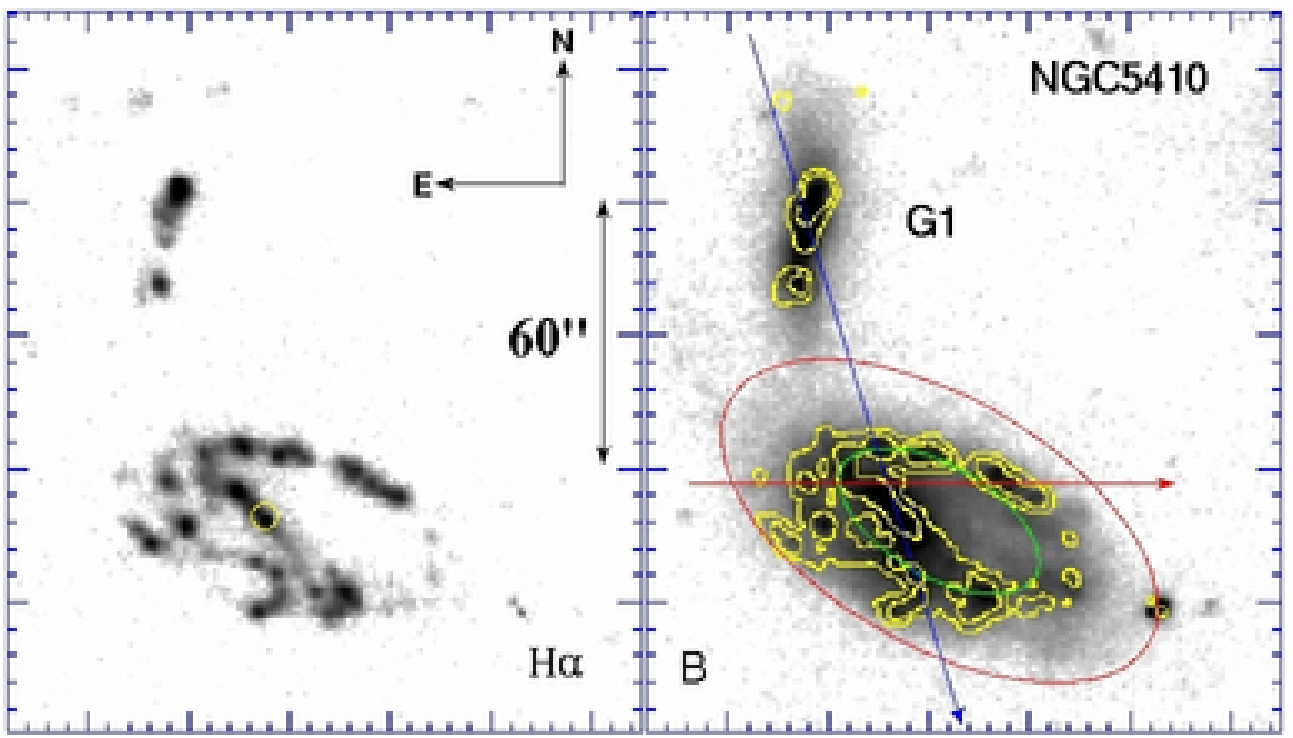}
\centerline{\hspace*{1.5cm}\includegraphics[width=14cm]{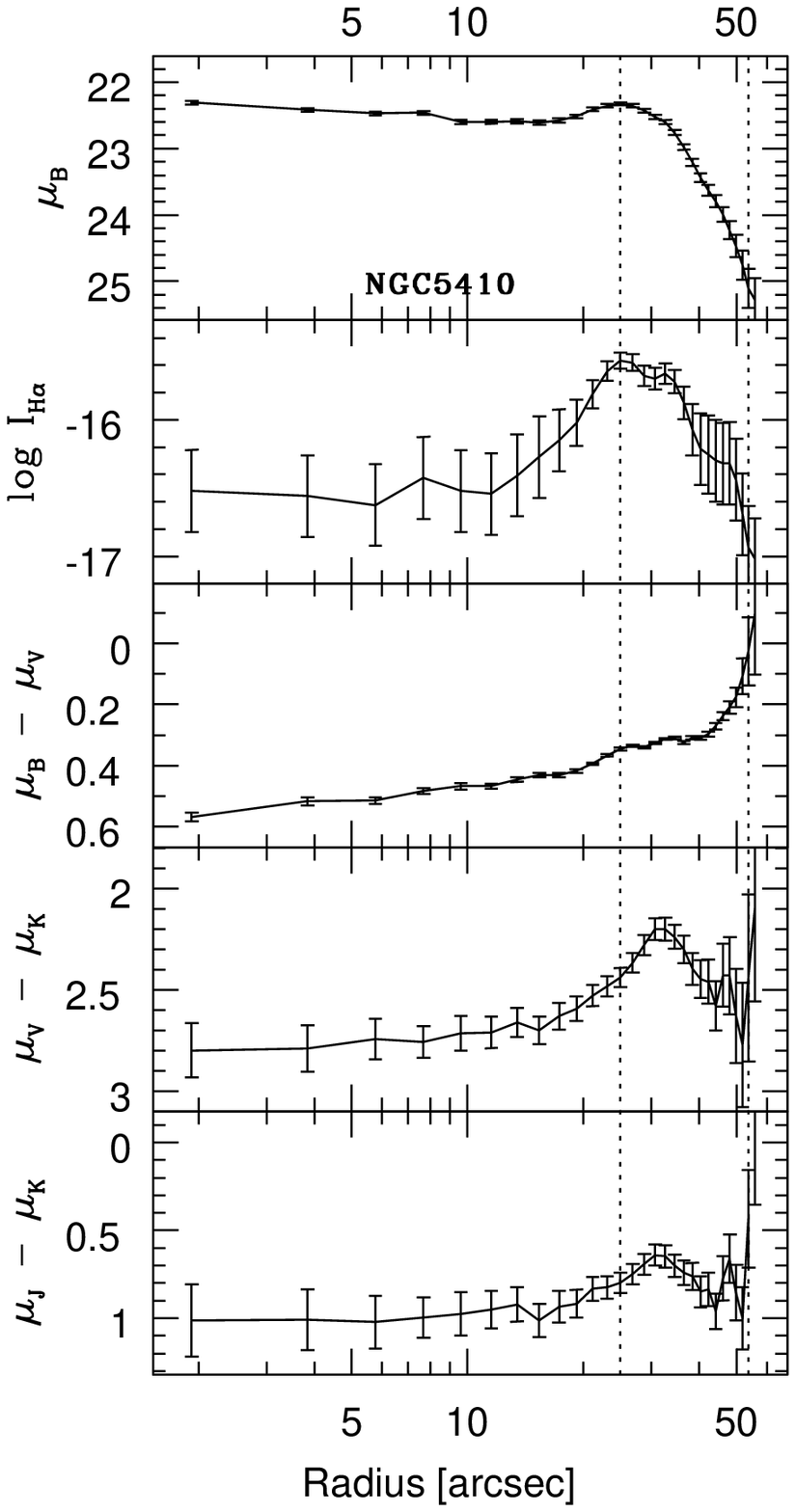}
            \hspace*{-7.5cm}\includegraphics[width=14cm]{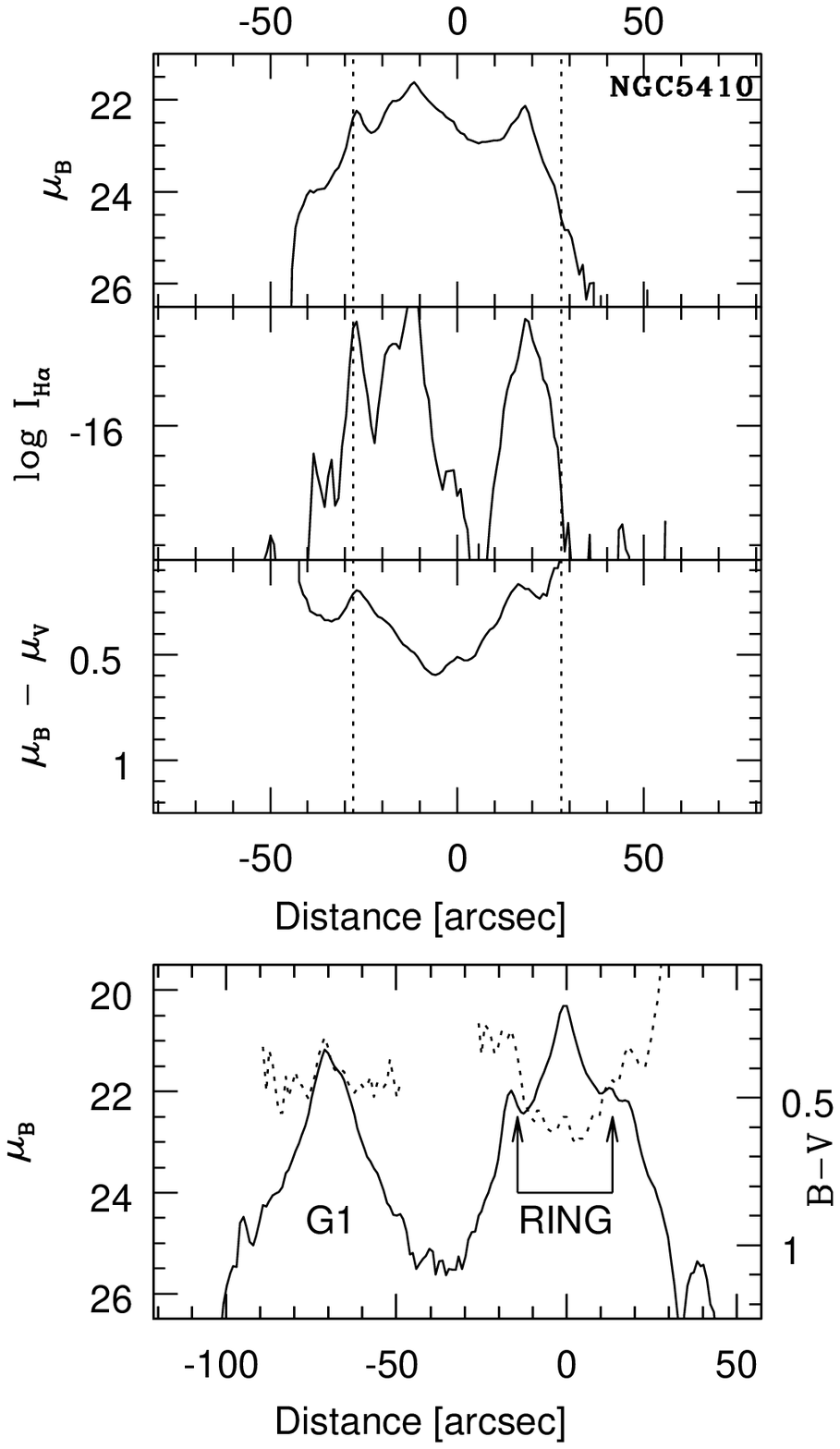}}
\caption{Same as in Figure~7, but for NGC5410.} 
\end{figure}

\begin{figure}
\epsscale{0.75}
\plotone{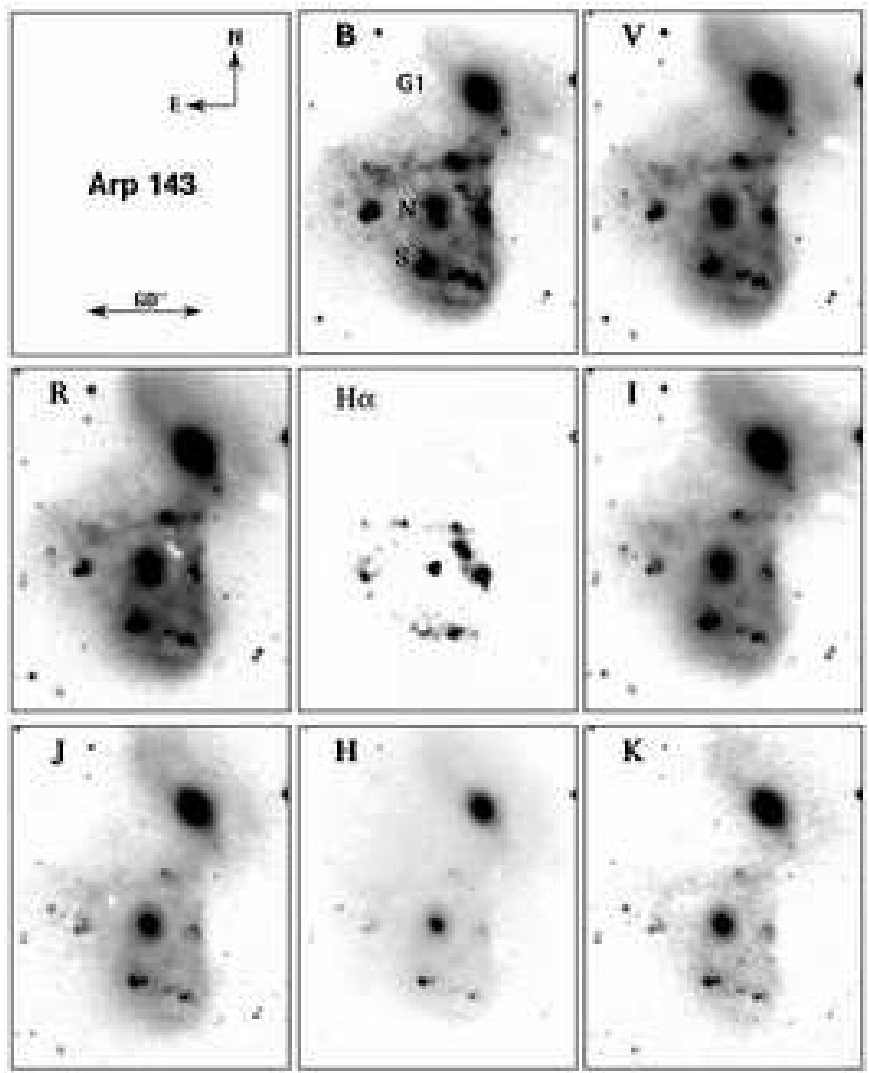}
\caption{$BVR$H$\alpha IJHK$-band gray-scale maps of Arp143. The orientation
and the image scale are shown in the first panel. The intensities are
scaled logarithmically into the gray-scale. Similar maps
for the rest of the galaxies are available 
at the website http://www.inaoep.mx/$\tilde{\ }$ydm/rings/.
}
\end{figure}

\end{document}